\documentclass[twocolumn]{aastex63}
\usepackage{scrextend}
\deffootnote[0.45em]{0.2em}{0.2em}{\textsuperscript{\thefootnotemark}\,}
\usepackage{courier}
\usepackage[T1]{fontenc}
\usepackage{mathtools}
\usepackage{changepage} 
\usepackage{url,multirow}
\usepackage{appendix}
\usepackage{microtype}
\DeclareUnicodeCharacter{2212}{-}
\usepackage{textcomp}
\usepackage{gensymb}

\title{NGC 1358 monitoring}

\def\xmm{{XMM-{\it Newton\/}}}
\def\XMM{{XMM-{\it Newton\/}}}
\def\cha{{\it Chandra}}
\def\lu{{erg\,s$^{-1}$}}
\def\swi{{{\it Swift}-BAT}\/}
\def\xrt{{{\it Swift}-XRT}\/}

\def\flu{{erg\,s$^{-1}$\,cm$^{-2}$}}

\def\nus{{\it NuSTAR}}
\def\nhlos{N$_{\rm H,los}$}
\def\nhtor{N$_{\rm H,tor}$}
\def\NuSTAR{{\it NuSTAR}}

\def\myt{\texttt{MYTorus}}
\def\borus{\texttt{borus02}}
\def\uxcl{\texttt{UXCLUMPY}}

\def\aap{A\&A}

\def\apjl{ApJL}
\def\apjs{ApJS}

\shorttitle{NGC 1358 monitoring}
\shortauthors{Marchesi et al.}

\begin{document}

\title{Compton-Thick AGN in the NuSTAR era VIII: A joint \nus-\xmm\ monitoring of the changing-look Compton-thick AGN NGC 1358}

\author{S. Marchesi}
\affiliation{INAF - Osservatorio di Astrofisica e Scienza dello Spazio di Bologna, Via Piero Gobetti, 93/3, 40129, Bologna, Italy}
\affiliation{Department of Physics and Astronomy, Clemson University,  Kinard Lab of Physics, Clemson, SC 29634, USA}

\author{X. Zhao}
\affiliation{Center for Astrophysics | Harvard \& Smithsonian, 60 Garden Street, Cambridge, MA 02138, USA}
\affiliation{Department of Physics and Astronomy, Clemson University,  Kinard Lab of Physics, Clemson, SC 29634, USA}

\author{N. Torres-Alb\`{a}}
\affiliation{Department of Physics and Astronomy, Clemson University,  Kinard Lab of Physics, Clemson, SC 29634, USA}

\author{M. Ajello}
\affiliation{Department of Physics and Astronomy, Clemson University,  Kinard Lab of Physics, Clemson, SC 29634, USA}

\author{M. Gaspari}
\affiliation{Department of Astrophysical Sciences, Princeton University, Princeton, NJ 08544, USA}

\author{A. Pizzetti}
\affiliation{Department of Physics and Astronomy, Clemson University,  Kinard Lab of Physics, Clemson, SC 29634, USA}

\author{J. Buchner}
\affiliation{Max Planck Institute for Extraterrestrial Physics, Giessenbachstrasse, D-85741 Garching, Germany}

\author{E. Bertola}
\affiliation{Dipartimento di Fisica e Astronomia ``Augusto Righi'', Università di Bologna, via Gobetti 93/2, I-40129 Bologna, Italy}
\affiliation{INAF - Osservatorio di Astrofisica e Scienza dello Spazio di Bologna, Via Piero Gobetti, 93/3, 40129, Bologna, Italy}

\author{A. Comastri}
\affiliation{INAF - Osservatorio di Astrofisica e Scienza dello Spazio di Bologna, Via Piero Gobetti, 93/3, 40129, Bologna, Italy}

\author{A. Feltre}
\affiliation{INAF - Osservatorio di Astrofisica e Scienza dello Spazio di Bologna, Via Piero Gobetti, 93/3, 40129, Bologna, Italy}

\author{R. Gilli}
\affiliation{INAF - Osservatorio di Astrofisica e Scienza dello Spazio di Bologna, Via Piero Gobetti, 93/3, 40129, Bologna, Italy}

\author{G. Lanzuisi}
\affiliation{INAF - Osservatorio di Astrofisica e Scienza dello Spazio di Bologna, Via Piero Gobetti, 93/3, 40129, Bologna, Italy}

\author{G. Matzeu}
\affiliation{Dipartimento di Fisica e Astronomia ``Augusto Righi'', Università di Bologna, via Gobetti 93/2, I-40129 Bologna, Italy}

\author{F. Pozzi}
\affiliation{Dipartimento di Fisica e Astronomia ``Augusto Righi'', Università di Bologna, via Gobetti 93/2, I-40129 Bologna, Italy}
\affiliation{INAF - Osservatorio di Astrofisica e Scienza dello Spazio di Bologna, Via Piero Gobetti, 93/3, 40129, Bologna, Italy}

\author{F. Salvestrini}
\affiliation{INAF – Osservatorio Astrofisico di Arcetri, Largo E. Fermi 5, 50125, Firenze, Italy}

\author{D. Sengupta}
\affiliation{Dipartimento di Fisica e Astronomia ``Augusto Righi'', Università di Bologna, via Gobetti 93/2, I-40129 Bologna, Italy}
\affiliation{INAF - Osservatorio di Astrofisica e Scienza dello Spazio di Bologna, Via Piero Gobetti, 93/3, 40129, Bologna, Italy}

\author{R. Silver}
\affiliation{Department of Physics and Astronomy, Clemson University,  Kinard Lab of Physics, Clemson, SC 29634, USA}

\author{F. Tombesi}
\affiliation{Department of Physics, Tor Vergata University of Rome, Via della Ricerca Scientifica 1, 00133 Rome, Italy}
\affiliation{INAF – Osservatorio Astronomico di Roma, Via Frascati 33, 00078 Monte Porzio Catone, Italy}
\affiliation{Department of Astronomy, University of Maryland, College Park, MD 20742, USA}
\affiliation{NASA Goddard Space Flight Center, Greenbelt, MD 20771, USA}
\affiliation{INFN - Roma Tor Vergata, Via della Ricerca Scientifica 1, 00133 Rome, Italy}

\author{A. Traina}
\affiliation{Dipartimento di Fisica e Astronomia ``Augusto Righi'', Università di Bologna, via Gobetti 93/2, I-40129 Bologna, Italy}
\affiliation{INAF - Osservatorio di Astrofisica e Scienza dello Spazio di Bologna, Via Piero Gobetti, 93/3, 40129, Bologna, Italy}

\author{C. Vignali}
\affiliation{Dipartimento di Fisica e Astronomia ``Augusto Righi'', Università di Bologna, via Gobetti 93/2, I-40129 Bologna, Italy}
\affiliation{INAF - Osservatorio di Astrofisica e Scienza dello Spazio di Bologna, Via Piero Gobetti, 93/3, 40129, Bologna, Italy}

\author{L. Zappacosta}
\affiliation{INAF – Osservatorio Astronomico di Roma, Via Frascati 33, 00078 Monte Porzio Catone, Italy}

\begin{abstract}
We present the multi-epoch monitoring with \nus\ and \xmm\ of NGC 1358, a nearby Seyfert 2 galaxy whose properties made it a promising candidate X-ray changing-look active galactic nucleus (AGN), i.e., a source whose column density could transition from its 2017 Compton-thick (CT--, having line-of-sight Hydrogen column density N$_{H,los}>$10$^{24}$ cm$^{-2}$) state to a Compton-thin (N$_{H,los}<$10$^{24}$ cm$^{-2}$) one. The multi-epoch X-ray monitoring confirmed the presence of significant N$_{H,los}$ variability over time-scales from weeks to years, and allowed us to confirm the ``changing-look'' nature of NGC 1358, which has most recently been observed in a Compton-thin status. 
Multi-epoch monitoring with \nus\ and \xmm\ is demonstrated to be highly effective in simultaneously constraining three otherwise highly degenerate parameters: the torus average column density and covering factor, and the inclination angle between the torus axis and the observer.
We find a tentative anti-correlation between column density and luminosity, which can be understood in the framework of Chaotic Cold Accretion clouds driving recursive AGN feedback. The monitoring campaign of NGC 1358 has proven the efficiency of our newly developed method to select candidate \nhlos--variable, heavily obscured AGN, which we plan to soon extend to a larger sample to better characterize the properties of the obscuring material surrounding accreting supermassive black holes, as well as constrain AGN feeding models.
\end{abstract}

\section{Introduction} 
Obscuration in active galactic nuclei (AGNs) has been largely studied over the electromagnetic spectrum, from the optical \citep[e.g.,][]{lawrence91,simpson05}, to the infrared \citep[e.g.,][]{jaffe04,nenkova08a,feltre12}, and to the X-rays \citep[][]{gilli07,ricci15,hickox18}. It is commonly accepted that the obscuration is mostly caused by a ``dusty torus'', i.e., a distribution of molecular gas and dust located at $\sim$1--10\,pc from the accreting supermassive black hole (SMBH). While the existence of this obscuring material is universally accepted, its geometrical distribution and chemical composition are still a matter of debate. Several works reported observational evidence favoring a ``clumpy torus'' scenario, where the obscuring material is distributed in clumps formed by optically thick clouds \citep[e.g.,][]{jaffe04,nenkova08a,elitzur06,risaliti07,honig07,burtscher13}.  Theoretical/numerical models of accretion onto SMBHs also predict a highly clumpy and chaotic multi--phase medium (\citealt{gaspari13,gaspari17_cca,gaspari20} for a review), in particular within $r<$100\,pc of the AGN, where Chaotic Cold Accretion (CCA) is expected to boost the feeding rates. Such CCA `rain' has been now observationally probed in many systems and bands spanning from X-ray to optical/IR and radio (e.g., \citealt{rose19,gaspari19,maccagni21,mckinley22,olivares22,temi22}).

If the obscuring environment is indeed inhomogeneous, one would expect to observe significant variability in the obscuring material line-of-sight (l.o.s.) column density (\nhlos) and even, in some cases, a transition from a Compton-thick (CT-) scenario (where \nhlos$>$10$^{24}$ cm$^{-2}$) to a Compton-thin one (where \nhlos$<$10$^{24}$ cm$^{-2}$). This transition should occur in a period of time as short as a day and as long as several months, assuming a typical range of obscuring clouds filling factors, velocities and distances from the accreting BH \citep[e.g,][]{nenkova08a}. However, the number of \textit{bona fide} CT-AGN with high-quality X-ray data is limited \citep[$\sim$35 sources, see, e.g.,][]{arevalo14,balokovic14,koss15,masini16,oda17,marchesi18,marchesi19a,torres21,traina21,zhao21}, and only a small fraction of these objects have multi-epoch observations on time-scales that vary from weeks to years, which are key to properly assess any variation in \nhlos\ and/or flux. Consequently, only a few sources have been observed to transition from Compton-thick to Compton-thin: NGC 1365 \citep{risaliti05}; NGC 7582 \citep{bianchi09,rivers15}; Mrk 3 \citep{guainazzi12}; NGC 454 \citep{marchese12}, ESO 323-G77 \citep{miniutti14}; IC 751 \citep{ricci16}. 

This class of ``X-ray changing-look sources'' is the ideal \nhlos--variable sample to study the properties of the obscuring material in a complete, self-consistent way. In fact, Compton-thick to Compton-thick \nhlos\ transitions are difficult to reliably measure with small enough uncertainties to enable the estimate of the SMBH-cloud distance from $\Delta$\nhlos, since at column densities above $\sim$2--3$\times$10$^{24}$\,cm$^{-2}$ almost all photons at energies $<$10--20\,keV are absorbed by the obscuring material \citep[see, e.g.,][]{koss16}. In less obscured AGN (\nhlos$\lesssim$10$^{23}$\,cm$^{-2}$), instead, the \nhlos\ variability can be measured with excellent precision. However, in this class of objects the overall X-ray emission is dominated by the transmitted main component: consequently, it is difficult to accurately measure the obscuring material properties linked to the reprocessed emission, such as the covering factor and the average column density. 
Consequently, the limited sample size of currently available X-ray changing-look AGNs prevents us from getting a complete picture of the properties of the obscuring material surrounding accreting SMBHs.

In this paper, we present the result of a multi-epoch monitoring of the nearby CT-AGN NGC 1358, a Seyfert 2 galaxy whose properties make it a promising ``changing-look'' candidate and an ideal pilot source to start the X-ray characterization of the obscuring material in nearby accreting SMBHs. The work is organized as follows: in Section~\ref{sec:ngc1358} we present the source, with a particular focus on previous X-ray works. In Section~\ref{sec:new_data} we present the data analysis and results of the joint spectral fitting for the new and old \nus\ and \xmm\ observations. We then discuss in Section~\ref{sec:discussion} how the results of the X-ray monitoring can be explained in the framework of a ``clumpy obscuration'' model. Finally, we summarize the results of our work in Section~\ref{sec:conclusions}.

Through the rest of the work, we assume a flat $\Lambda$CDM cosmology with H$_0$=69.6\,km\,s$^{-1}$\,Mpc$^{-1}$, $\Omega_m$=0.29 and $\Omega_\Lambda$=0.71 \citep{bennett14}. Errors are quoted at the 90\,\% confidence level, unless otherwise stated.

\section{NGC 1358}\label{sec:ngc1358}
\begin{figure*} 
\begin{minipage}{0.49\textwidth} 
 \centering 
 \includegraphics[width=1.0\textwidth]{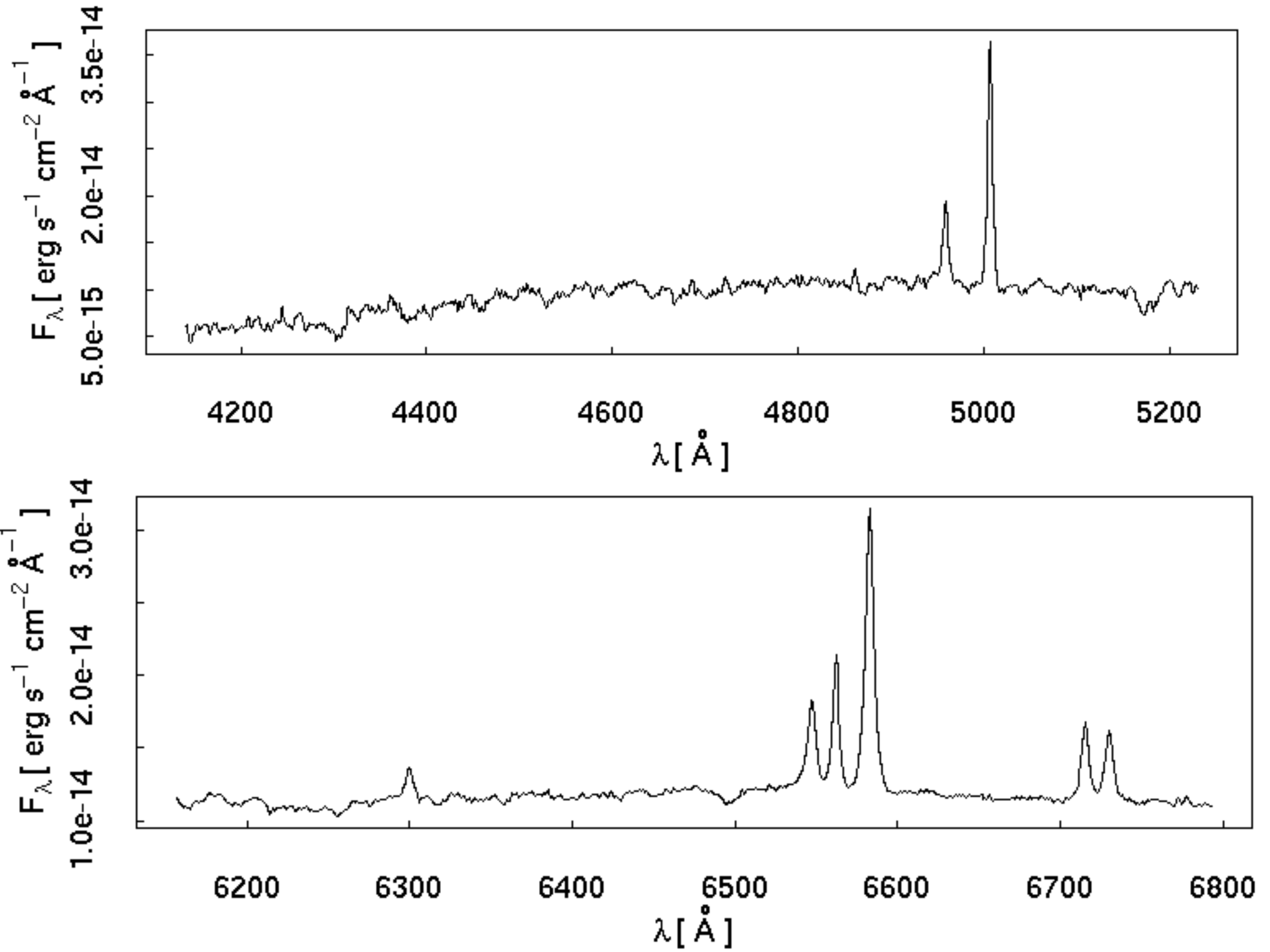} 
 \end{minipage} 
\begin{minipage}{0.49\textwidth} 
 \centering 
 \includegraphics[width=1\textwidth]{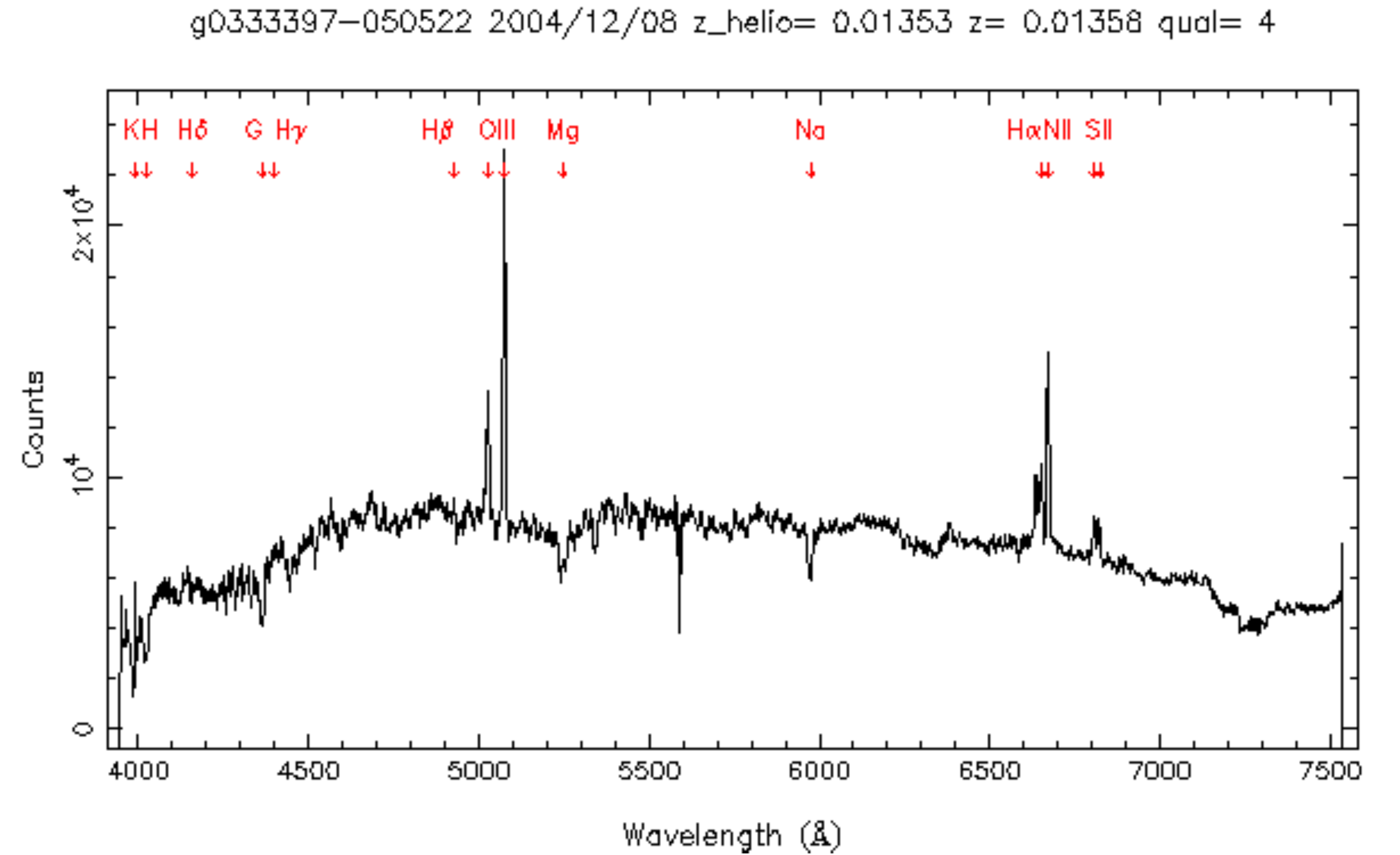}
 \end{minipage}
 \begin{minipage}{1.0\textwidth} 
 \centering 
 \includegraphics[width=0.55\textwidth]{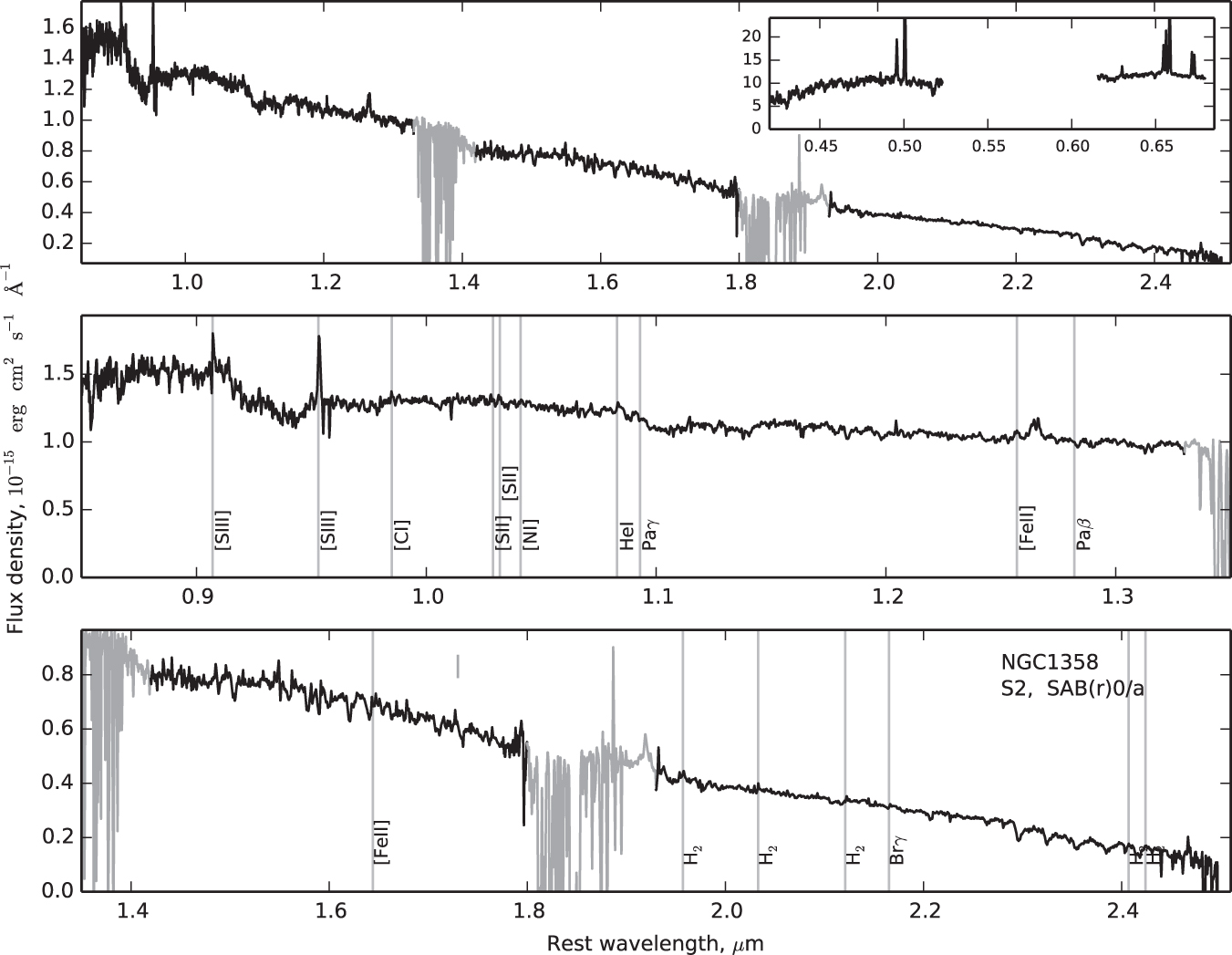}
 \end{minipage}
\caption{\normalsize 
Top panels: optical spectra of NGC 1358 obtained using the Double Spectrograph mounted on the Palomar 200-inch Hale Telescope \citep[][left; spectrum taken in 1985]{filippenko85} and using the multi-object spectrograph mounted on the 1.2\,m UK Schmidt Telescope \citep[][right; spectrum taken in 2004]{jones09}.
Bottom panel: near-infrared spectrum of NGC 1358 obtained using the Gemini Near-IR Spectrograph on the Gemini North telescope \citep[][spectrum taken in 2011]{mason15}.
}\label{fig:optical_spectra}
\end{figure*}

NGC 1358 is a nearby \citep[$z$=0.01344][]{theureau98}, X-ray bright (having 15--150\,keV observed flux f$_{\rm 15-150\,keV}\geq$5$\times$10$^{-12}$\,\flu) Seyfert 2 galaxy. The source was originally classified as a narrow-line (NL) Seyfert 2 source in \citet{filippenko85} using the Double Spectrograph mounted on the Palomar 200-inch Hale Telescope. A new optical spectrum of NGC 1358 was then taken in 2004 within the 6dF Galaxy Survey, using the multi-object spectrograph mounted on the 1.2\,m UK Schmidt Telescope \citep{jones09}, and no evidence for optical variability with respect to the \citet{filippenko85} spectrum is observed, thus confirming the NL nature of the source. We report both optical spectra in Figure~\ref{fig:optical_spectra}. More recently, \citet{mason15} reported that the NIR spectrum of NGC 1358, obtained using the GNIRS spectrograph mounted on the Gemini North 8m telescope ``only contain[ed] a handful of weak emission lines'' (see Fig.~Figure~\ref{fig:optical_spectra}, bottom panel). No evidence for a significant optical ``changing-look'' behavior has therefore ever been observed in NGC 1358.

In the X-rays, NGC 1358 is detected in the \textit{Swift} \citep[][]{gehrels04} Burst Alert Telescope \citep[BAT][]{barthelmy05} 150-month catalog (K. Iman et al. in prep\footnote{An online version of the catalog is available at \url{https://science.clemson.edu/ctagn/bat-150-month-catalog/}.}) and has been targeted several times by X-ray telescopes\footnote{While NGC 1358 has been observed multiple times by \textit{Swift-XRT}, none of the observations has enough counts to perform a spectral fit. We therefore do not include the \textit{Swift-XRT} observations in this work.}. A first 10\,ks \xmm\ observation was taken in 2005 and analyzed in \citet[][]{marinucci12}. They determined that NGC 1358 is heavily obscured, and potentially a Compton-thick AGN, but the low data quality made their l.o.s. column density poorly constrained (\nhlos=1.3$^{+8.5}_{-0.6}$ $\times$10$^{24}$\,cm$^{-2}$). A second observation was performed by \cha\ in November 2015: the joint fit of the \cha\ spectrum with the \swi\ 100-month one is reported in \citet{marchesi17a}. The source was once again found to be heavily obscured, having \nhlos=1.0$^{+0.4}_{-0.6}$ $\times$10$^{24}$\,cm$^{-2}$. However, while a physically self-consistent spectral model such as \myt\ \citep[][]{murphy09} was used to perform the fit, the limited count statistic ($<$100 net counts in the 0.5--7\,keV band) of the \cha\ spectrum did not allow for a reliable characterization of the obscuring material's physical and geometrical properties.

For this reason, NGC 1358 was subsequently targeted by a joint \nus\ and \xmm\ observation performed in August 2017, whose results are reported in \citet{zhao19a}. The high count statistic in the 0.5--70\,keV band obtained in this deep observation ($>$4500 net counts, $\sim$50\,\% of which detected by \nus\ in the 3--50\,keV band) made it possible to use models that self-consistently characterize AGN obscuration in X-ray spectra, such as \myt\ \citep{murphy09,yaqoob12,yaqoob15} and \texttt{borus02} \citep[][]{balokovic18}. In particular, \texttt{borus02} measures important physical and geometric parameters such as the obscuring material l.o.s. column density, its average column density (see Section~\ref{sec:model_uniform} for more details on this quantity), and its covering factor, among others. NGC 1358 was found to have: $i$) l.o.s. column density, \nhlos=2.4$^{+0.4}_{-0.1}\times$10$^{24}$\,cm$^{-2}$, well above the CT threshold, at a $>$3\,$\sigma$ confidence level; $ii$) Compton-thin average column density, \nhtor=6.5$^{+0.5}_{-1.6}\times$10$^{23}$\,cm$^{-2}$, i.e.,  $\sim$4 times smaller than the l.o.s. column density; $iii$) low covering factor ($f_c<$0.15). The large $\Delta$N$_{\rm H}$ ($\log$(\nhlos)-$\log$(\nhtor)$\sim$0.6) measured in NGC 1358, combined with its low $f_c$ suggests that this source is a promising candidate ``clumpy-torus'' CT-AGN, where the obscuring material is distributed in clumps at the micro- to meso-scale \citep[i.e., within a few pc from the SMBH,][]{gaspari20}, rather than uniformly. Notably, the large $\Delta$N$_{\rm H}$ measured using \borus\ was independently confirmed using \myt\ in its ``decoupled'' configuration \citep[][]{yaqoob12,yaqoob15}, which allows one to independently measure \nhlos\ and \nhtor. 

\begin{figure}[htbp]
 \centering 
 \includegraphics[width=0.47\textwidth]{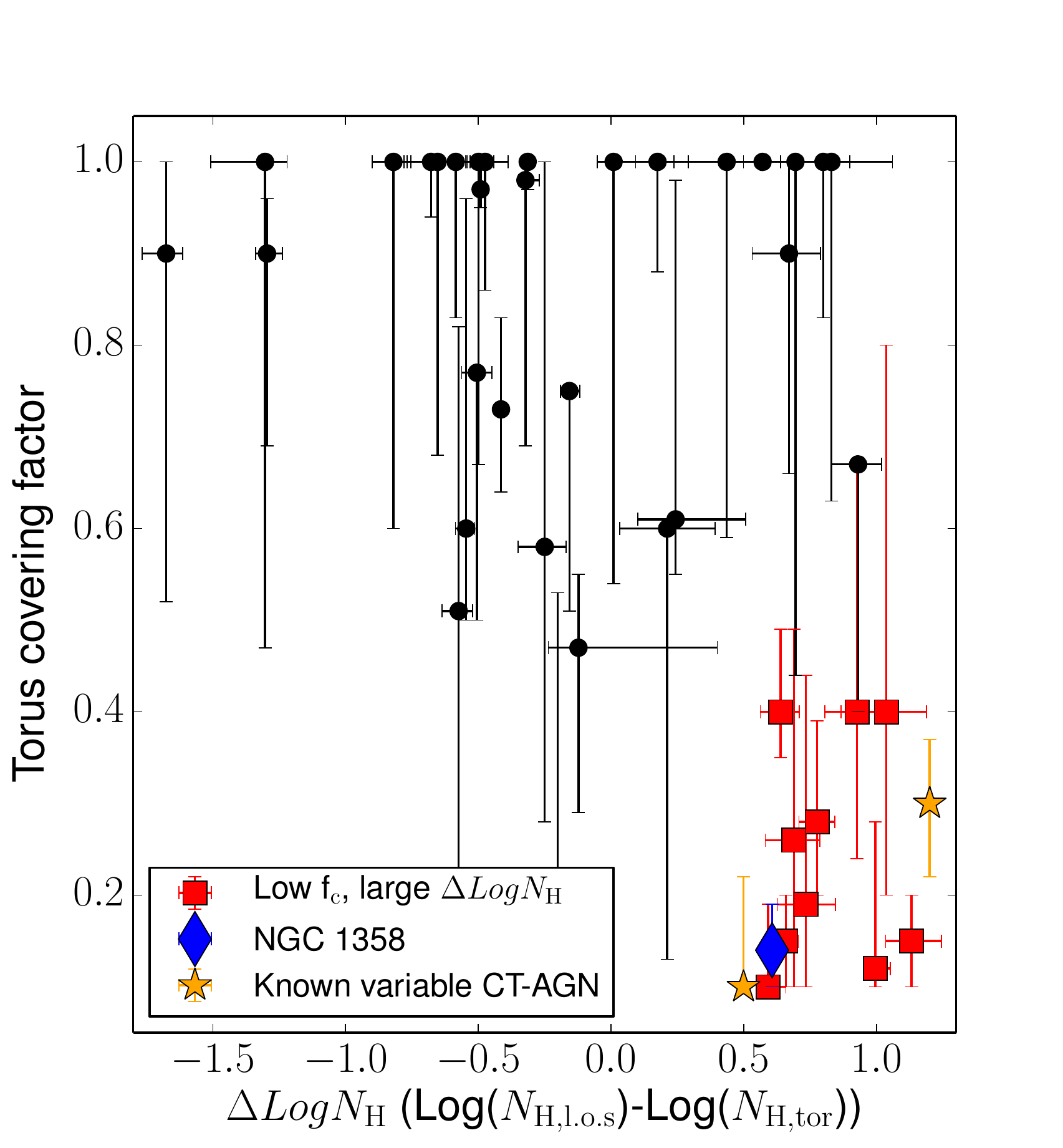} 
\caption{\normalsize 
Obscuring material covering factor ($f_c$) as a function of the difference ($\Delta$Log$N_{\rm H}$) between the logarithms of the line-of-sight and torus average column density, for a sample of nearby CT-AGNs observed with \nus\ and a 0.5--10\,keV facility \citep[\xmm, \cha\ or \xrt; from][]{marchesi19a,torres21}. Sources with $f_c<$0.4 and $\Delta$Log$N_{\rm H}>$0.5 are plotted as red squares. NGC 1358 is shown as a blue diamond: its large $\Delta N_{\rm H}$ and low $f_c$ make it an excellent candidate ``clumpy environment'' CT-AGN. MRK 3 and NGC 4945, both known variable CT-AGN also having large $\Delta N_{\rm H}$ and low $f_c$, are plotted as orange stars.
}\label{fig:NGC_previous_info}
\end{figure}

Based on the above-mentioned observational evidence, NGC 1358 is likely to have been observed through an over-dense region embedded in a significantly less dense environment. In such a scenario, the small covering factor measured with \texttt{borus02} would imply that the overall cloud volume filling factor is small, and the obscuring clouds occupy only a fractional part of the pc-scale region that surrounds the accreting SMBH where the obscuration is expected to take place. Sources with this type of ``low covering factor'' obscurer are ideal candidate X-ray changing-look AGN. In objects with large $f_c$, instead, the number of clouds between the observer and the SMBH is expected to always be $>>$1, thus significantly reducing the chance of observing a significant change in \nhlos. 
Notably, at least two CT-AGN fulfill the proposed selection criteria and are known to be variable. NGC 4945 has been shown to vary significantly even above 10\,keV, likely because of a combination of intrinsic luminosity and l.o.s. column density variability \citep{puccetti14}, while the l.o.s. column density of the material surrounding MRK 3 has been measured at both CT and Compton-thin levels in the past years \citep[see, e.g., ][]{guainazzi12,yaqoob15}. We highlight both these sources in Figure~\ref{fig:NGC_previous_info}.

To confirm the clumpy nature of its circumnuclear material,  NGC 1358 has been selected for a long-term monitoring campaign with \nus\ and \xmm, aimed at detecting significant flux and $N_{\rm H,l.o.s.}$ variability. A joint \nus--\xmm\ monitoring campaign is the best (if not the only) possible approach to constrain the properties of the obscuring material surrounding accreting SMBHs. \xmm\ alone would not be able to detect potential variability above 10\,keV, which can be linked to a variation in the covering factor \citep[see, e.g.,][on the variability above 10\,keV observed in the nearby CT sources NGC 4945 and NGC 1068, respectively]{puccetti14,zaino20}. Furthermore, high-quality data at E$>$10\,keV are key to break the \nhlos-photon index degeneracy in heavily obscured sources \citep[see, e.g.,][]{marchesi19a}. \nus, however, has a $\sim$4 times lower energy resolution than \xmm\ at 6.4\,keV, around the Fe K line region, and does not cover the energy range $<$3\,keV which is required to tightly constrain N$_{\rm H, l.o.s.}$ and, consequently, the AGN intrinsic luminosity.

Among the promising changing-look candidates with low-$f_c$ and large $\Delta N_{\rm H}$ reported in \citet[][see also the red points in Figure~\ref{fig:NGC_previous_info}]{marchesi19a}, the tentative evidence for l.o.s. column density variability between the 2015 and 2017 observations further strengthens a ``clumpy obscuration'' scenario for NGC 1358. This makes it an ideal pilot source to start the X-ray characterization of a whole class of ``clumpy obscuration'' CT-AGNs. 

\begingroup
\renewcommand*{\arraystretch}{1.5}
\begin{table*}
\scalebox{0.9}{
\vspace{.1cm}
 \begin{tabular}{cccccc}
 \hline
 \hline
 Instrument & Sequence & Start Time & End Time & Exposure & Net count rate  \\
& ObsID & (UTC) & (UTC) & ks & 10$^{-2}$ counts s$^{-1}$\\
  \hline    
 \xmm & 0795680101  & 2017-08-01T17:05:27 & 2017-08-02T06:03:10 & 48; 48; 48 & 0.98$\pm$0.05; 0.91$\pm$0.05; 3.68$\pm$0.15 \\
 \nus & 60301026002 & 2017-08-01T03:41:09 & 2017-08-02T06:36:09	& 50 & 2.32$\pm$0.07; 2.28$\pm$0.07 \\
 \xmm & 0862980101  & 2021-02-25T00:25:39 & 2021-02-25T10:30:39 & 33; 33; 24 & 1.26$\pm$0.06; 1.49$\pm$0.07; 7.34$\pm$0.18 \\
 \xmm & 0890700101  & 2021-08-02T17:10:55 & 2021-08-03T01:19:15 & 24; 24; 17 & 1.80$\pm$0.09; 1.51$\pm$0.08; 9.97$\pm$0.25 \\
 \nus & 60702044002 & 2021-08-02T16:21:09 & 2021-08-03T09:31:09	& 31 & 12.76$\pm$0.22; 11.78$\pm$0.22 \\
 \xmm & 0890700201 & 2022-01-21T05:11:06 & 2022-01-21T14:06:32 & 32; 32; 26 & 2.39$\pm$0.09; 2.52$\pm$0.09; 11.64$\pm$0.22 \\
 \nus & 60702044004 & 2022-01-21T06:46:09 & 2022-01-22T00:16:09 & 32 & 16.56$\pm$0.24; 15.76$\pm$0.23 \\
 \xmm & 0890700301 & 2022-02-04T10:20:38 & 2022-02-04T18:07:18 & 25; 25; 18 & 1.91$\pm$0.09; 2.08$\pm$0.09; 9.43$\pm$0.23 \\
 \nus & 60702044006 & 2022-02-03T07:21:09 & 2022-02-03T21:31:09 & 28 & 16.72$\pm$0.26; 16.24$\pm$0.26 \\
  \hline
	\hline
\end{tabular}}
	\caption{\normalsize Summary of the \nus\ and/or \xmm\ observations of NGC 1358 used in this work. All observations taken in 2021 and 2022 are analyzed here for the first time, while the 2017 observations were first studied by \citet{zhao19a}.  The \xmm\ count rates are computed in the 0.6--10\,keV band, while the \nus\ ones are computed in the 3--70\,keV band. Exposures are computed after removing high-background periods.
	}
\label{tab:obs_summary}
\end{table*}
\endgroup

\section{Data analysis and spectral fitting results}\label{sec:new_data}
In this work, we analyze four \xmm\ and three \nus\ observations that were taken between February 2021 and February 2022. The first \xmm\ observation (nominal length: 36\,ks) was taken as part of the \xmm\ proposal 086298 (PI: S. Marchesi), while the remaining three \xmm\ observations (nominal length: 30\,ks) and the \nus\ ones (30\,ks) are part of a \nus\ observing program (proposal ID: 07192; PI: S. Marchesi).  We report a summary of these observations in Table~\ref{tab:obs_summary}. In the rest of the paper, all the quoted errors are computed at the 90\,\% confidence level for a single parameter of interest.

\subsection{Data reduction}
The \nus\ data are obtained from both focal plane modules, FPMA and FPMB. We calibrated, cleaned, and screened the raw files using the \nus\ \texttt{nupipeline} script version 2.1.1. The \nus\ calibration database (CALDB) used in this work is the version 20210210. We then generated the ARF, RMF, and light-curve files with the \texttt{nuproducts} script. Both source and background spectra were extracted from a 60$^{\prime\prime}$--radius circle: this choice of radius was found to maximize the signal-to-noise ratio in the source spectra. The background spectra are extracted from a region nearby the source which is not affected by emission from NGC 1358 or other bright objects. Finally, the spectra are binned with a minimum of 15 counts per bin using the \texttt{grppha} task.

The \xmm\ observations taken in August 2021 and January 2022 were performed quasi-simultaneously to the \nus\ ones, the start-- and end--times of each pair observations being always less than 12\,hours apart (see Table~\ref{tab:obs_summary}). The February 2022 one has instead been taken $\sim$1 day after the \nus\ one, thus allowing us to perform a further variability study (as discussed in Section~\ref{sec:feb22_obs}). We reduced the \xmm\ data using the Science Analysis System \citep[SAS][]{jansen01} version 19.1. The August 2021 and February 2022 observations were affected by strong flares, so the net exposure time is 10--20\,\% (30--50\,\%) shorter than the nominal one for the MOS (pn) cameras. The MOS (pn) source spectra were extracted from a 10$^{\prime\prime}$ (15$^{\prime\prime}$) --radius circle, while the background spectra are extracted from a 45$^{\prime\prime}$--radius circle located nearby NGC 1358 and in a region with no significant contamination from other sources.

\subsection{Spectral modeling}\label{sec:spec_models}
To avoid possible model--dependent effects, we analyze the \nus\ and \xmm\ spectra using three different physically motivated models which have been developed specifically to treat the X-ray spectra of heavily obscured AGN: we report a summary of their properties in Table~\ref{tab:models_summary}. Two of these models assume that the obscuring material is uniformly distributed in a toroidal shape, while the third one works under the assumption that the obscuring material is distributed in clumps.

\begingroup
\renewcommand*{\arraystretch}{1}
\begin{table*}
\scalebox{0.95}{
\vspace{.1cm}
 \begin{tabular}{ccccc}
 \hline
 \hline
Model & Reference(s) & Material distribution & Morphology &  Free parameters  \\
  \hline    
  \hline
\multirow{3}{*}{\myt} & \citet{murphy09} & \multirow{3}{*}{Uniform} & \multirow{3}{*}{Toroidal} & \multirow{3}{*}{$\Gamma$, \nhlos\, \nhtor, $A_{\rm S}$} \ \\
     & \citet{yaqoob12} &         &          &  \\
     & \citet{yaqoob15} &         &          &  \\
\hline
\multirow{2}{*}{\borus} & \multirow{2}{*}{\citet{balokovic18}} & \multirow{2}{*}{Uniform} & Sphere with & \multirow{2}{*}{$\Gamma$, \nhlos\, $\theta\rm _{i}$, \nhtor, $f_{\rm c}$} \\
       &                     &         & biconical cutout &  \\
\hline
\multirow{2}{*}{\uxcl} & \multirow{2}{*}{\citet{buchner19}} & \multirow{2}{*}{Clumpy} & As proposed in & \multirow{2}{*}{$\Gamma$, \nhlos\, $\theta\rm _{i}$, TOR$\sigma$, CTK}\\
      &                   &        & \citet{nenkova08a} & \\
  \hline
	\hline
\end{tabular}}
	\caption{\normalsize Summary of the properties of the X-ray spectral models used in this work. \myt\ is used in its ``decoupled'' configuration (see the text for more details). $\Gamma$ is the power-law photon index; \nhlos\ is the Hydrogen column density between the observer and the accreting SMBH; \nhtor\ is the (volume)-averaged column density of the obscuring material; $A_{\rm S}$ is the intensity of the reprocessed component with respect to the main one;  $\theta\rm _{i}$ is the angle between the torus axis and the observer; $f_{\rm c}$ is the obscuring material covering factor; TOR$\sigma$ is the angular width parameter, and models the vertical extent of the cloud population; finally, CTK is the covering factor of the inner obscuring ring.
	The free parameters reported in the Table are those that were actually left free to vary in our analysis: further details are available in \citet{saha22}.
	}
\label{tab:models_summary}
\end{table*}
\endgroup

\subsubsection{Uniform torus models}\label{sec:model_uniform}
The first model we use in our analysis is \myt\ \citep[][]{murphy09,yaqoob12,yaqoob15}, which we use in its so-called ``decoupled'' configuration, where the line-of-sight column density, \nhlos, can in principle be different from the (volume) average column density, \nhtor. \myt\ works under the assumption that the obscuration in AGN is caused by a torus with circular cross section, having half-opening angle $\theta_{\rm T}$=60$\degree$, where $\theta_{\rm T}$ is computed starting from the torus axis. This means that in \myt\ the torus covering factor is not a free parameter and is fixed to $f_c$=cos($\theta_{\rm T}$)=0.5.

\myt\ is made of three separate components. In \texttt{XSpec} \citep{arnaud96} the model is written as follows:
\begin{equation}\label{eq:myt}
\begin{aligned}
C_{\rm NuS}*pha*(zpo1*MYTZ + A_{\rm S}*MYTS \\
+  A_{\rm S}*MYTL +  f_{\rm S}*zpo2+mekal),
\end{aligned}
\end{equation}
where $pha$ is the absorption due to our own Galaxy, $N_{\rm H,Gal}$=3.83$\times$10$^{20}$\,cm$^{-2}$ \citep{kalberla05}. The first \myt\ component, $MYTZ$, is an absorber applied to the direct continuum (modeled with a power law, $zpo1$) and is used to model the line-of-sight absorption \nhlos, the one caused by the material between the observer and the accreting SMBH. The second component, $MYTS$, models the so-called reprocessed (or scattered) emission, those photons that end up in the observer line of sight after being up-scattered by the gas surrounding the AGN. Consequently, the column density of this component can be treated as a good approximation of the average torus column density, \nhtor. Finally, the third component, $MYTL$, models two typical fluorescence lines observed in AGN spectra, the Iron K$\alpha$ and K$\beta$ lines at 6.4\,keV and 7.06\,keV, respectively. The relative intensity of the reprocessed component and of the fluorescence lines with respect to the main power law is described by a constant, A$_S$. This constant takes into account the time delay between the main component intrinsic emission and the reprocessed one, which can therefore vary in intensity due to the well known AGN variability. Furthermore, A$_S$ can give some loose indication on the actual torus covering factor, since at higher covering factors corresponds a larger intensity of the reprocessed component at energies $>$6\,keV \citep[see, e.g., Figure~A1 in][]{zhao20}.

In \myt\ decoupled the inclination angle of the reprocessed component and the fluorescence lines can be fixed to one of two values: 90$^\degree$ or 0$^\degree$. The 90$^\degree$ scenario is one where most of the reprocessed emission comes from material which is located between the accreting SMBH and the observer, while the 0$^\degree$ scenario is a ``back--reflection'' one, where most of the reprocessed emission is coming from the material located on the back side of the torus with respect to the observer perspective. In this paper, we test both the configurations separately.

The second model we adopt for our analysis is \borus\ \citep[][]{balokovic18}. \borus\ works under the assumption that the shape of the obscuring material responsible for the reprocessed emission (including the Iron K$\alpha$ line) is a uniform-density sphere with two conical polar cutouts. The opening angle of these cutouts is a free parameter of the model.

In \texttt{XSpec} the model is written as follows:
\begin{equation}\label{eq:borus}
\begin{aligned}
C_{\rm NuS}*pha*(borus02 + zphabs*cabs*zpo1\\
+ f_{\rm S}*zpo2+mekal).
\end{aligned}
\end{equation}
The torus covering factor varies in the range $f_c$=[0.1--1] (i.e., in a range of opening angles $\theta_{\rm T}$=[0--84]$^\degree$). \borus\ also includes as a free parameter the torus inclination angle $\theta_{\rm i}$, which is the angle between the observer and the torus axis. The l.o.s. column density is modelled using the $zphabs$ and $cabs$ components.

Finally, as shown in Equations~\ref{eq:myt} and \ref{eq:borus}, our modelling includes three further components for both the \myt\ and the \borus\ analysis. The first one is a cross-normalization constant between the \xmm\ and the \nus\ observations, $C_{\rm NuS}$, to model possible calibration offsets between the two instruments. Indeed, in all observations and with all models we find a cross-normalization $C_{\rm NuS}\sim$1.1--1.15, in agreement with previous results reported in the literature \citep[e.g.,][]{madsen17,osorio20,balokovic21}. In principle, there can be variability even between different cameras on the same instrument (MOS versus pn in \xmm; FPMA versus FPMB in \nus). However, we find that no additional cross-instrument components are required in our analysis, since when included in our fit they are always consistent with 1.

The second component is a secondary power law, $zpo2$, that treats the fraction of emission which is not affected by obscuration: this fractional value is parameterized with the constant $f_{\rm s}$. Finally, NGC 1358 presents diffuse X-ray emission below 1\,keV, which we model with a phenomenological thermal component \texttt{mekal}, where both the gas temperature and metallicity are left free to vary.

\subsubsection{Clumpy torus model}\label{sec:uxclumpy}
\citet{buchner19} presented \uxcl\footnote{Which can be downloaded at the following link: \url{https://github.com/JohannesBuchner/xars/blob/master/doc/uxclumpy.rst}}, an X-ray spectral model which assumes that the AGN obscuration is caused by a clumpy distribution of material. More in detail, in \uxcl\ the obscuring material is axi-symmetric, and the number $N$ of clouds between the observer and the accreting SMBH is
\begin{equation}
    N = N_0\cdot exp\left\{-\left(\frac{\beta}{{\rm TOR}\sigma}\right)^{2}\right\},
\end{equation}
where $N_0$ is the number of clouds on the equatorial plane, $\beta$ is the inclination angle towards the torus pole, and TOR$\sigma$ is the obscuring material angular width, which models the torus scale height and is a free parameter in the model. The clouds angular size distribution is exponential and centered at $\theta_{\rm cloud}$=1$\degree$, and the size of a single cloud is $D$=$d_{\rm BH-cl}$sin($\theta_{\rm cloud}$), where $d_{\rm BH-cl}$ is the distance between the cloud and the SMBH \citep{nenkova08a,nenkova08b}. Notably, \uxcl\ (and more in general models where the obscuring material is assumed to be clumpy) allows one to set up a varying \nhlos\ scenario while keeping the obscurer geometry self-consistent.

Finally, \uxcl\ includes an inner ring of Compton-thick material, whose covering factor CTK is a free parameter in the model. This additional component mimics a ``reflection mirror'' which is needed to model an excess of reprocessed emission observed in some low--$z$, heavily obscured AGNs \citep{buchner19} and can be linked, for example, to a ``warped disk'' obscurer \citep{buchner21}. In \texttt{XSpec}, the model is written as follows:
\begin{equation}\label{eq:uxclumpy}
C_{\rm NuS}*pha*(uxclumpy + f_{\rm S}*uxclumpy\_omni)\\ +mekal,
\end{equation}
where $uxclumpy$ models both the transmitted and the reflected component (including the fluorescent lines), while $uxclumpy\_omni$ models the so-called ``warm mirror emission'', which is the emission scattered, rather than absorbed by the obscuring material. The parameters of $uxclumpy$ and $uxclumpy\_omni$ are linked. $C_{\rm NuS}$, $pha$, $f_{\rm S}$ and $mekal$ are the same components described in Section~\ref{sec:model_uniform}.

\subsection{Evolution with time of the torus properties}\label{sec:multi-epoch}
For consistency with previous works, and to test how joint multi-epoch spectral fitting can tighten the constraints on the spectral parameters, we performed a single-epoch spectral fit for each of the new observations reported in Table~\ref{tab:obs_summary}. We report a detailed description of these fits in Appendix A.

\begin{figure*} 
\begin{minipage}{0.49\textwidth} 
 \centering 
 \includegraphics[width=1.0\textwidth]{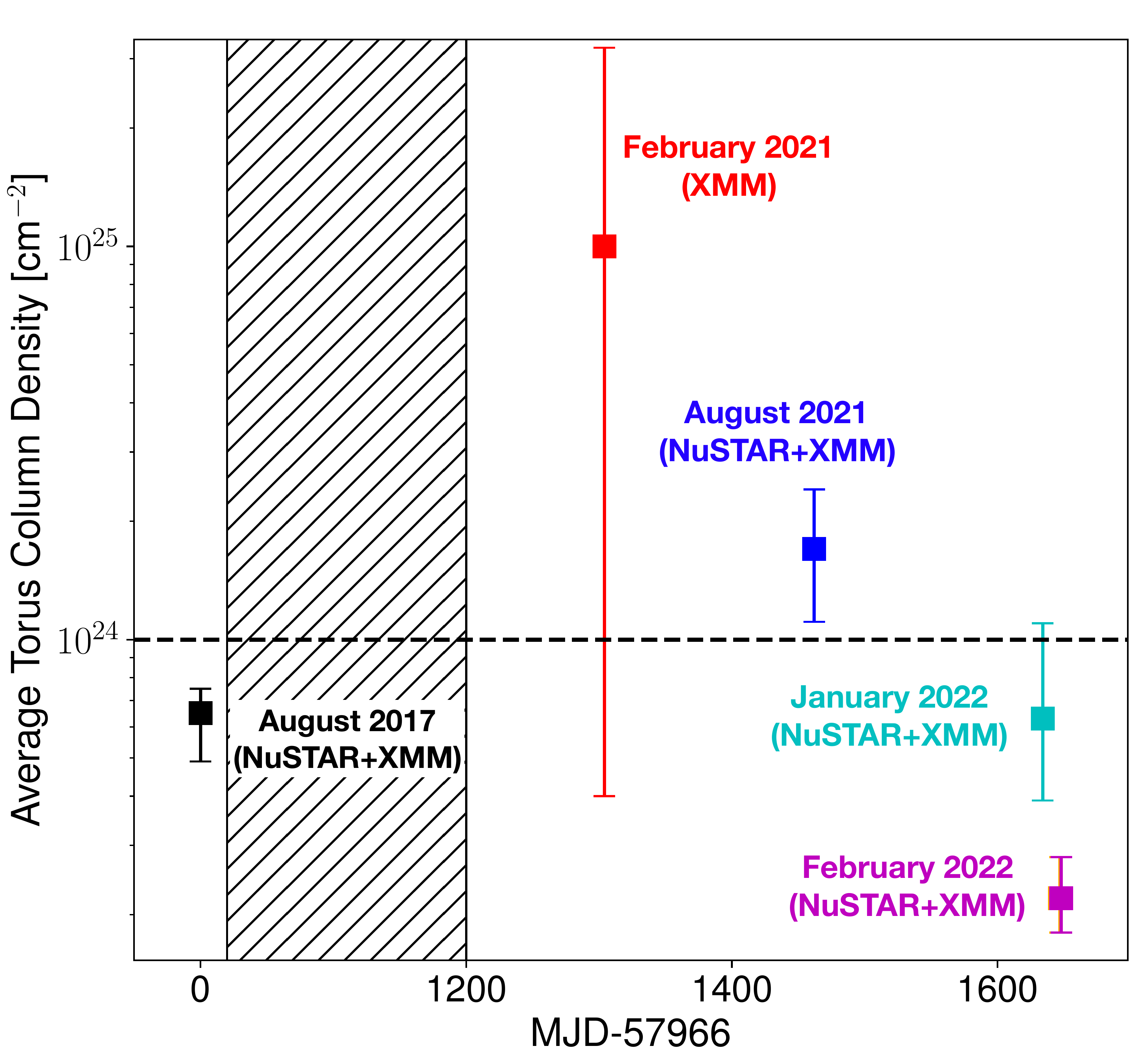} 
 \end{minipage} 
 \begin{minipage}{0.49\textwidth} 
 \centering 
 \includegraphics[trim={0 0 0 1cm},clip,width=1\textwidth]{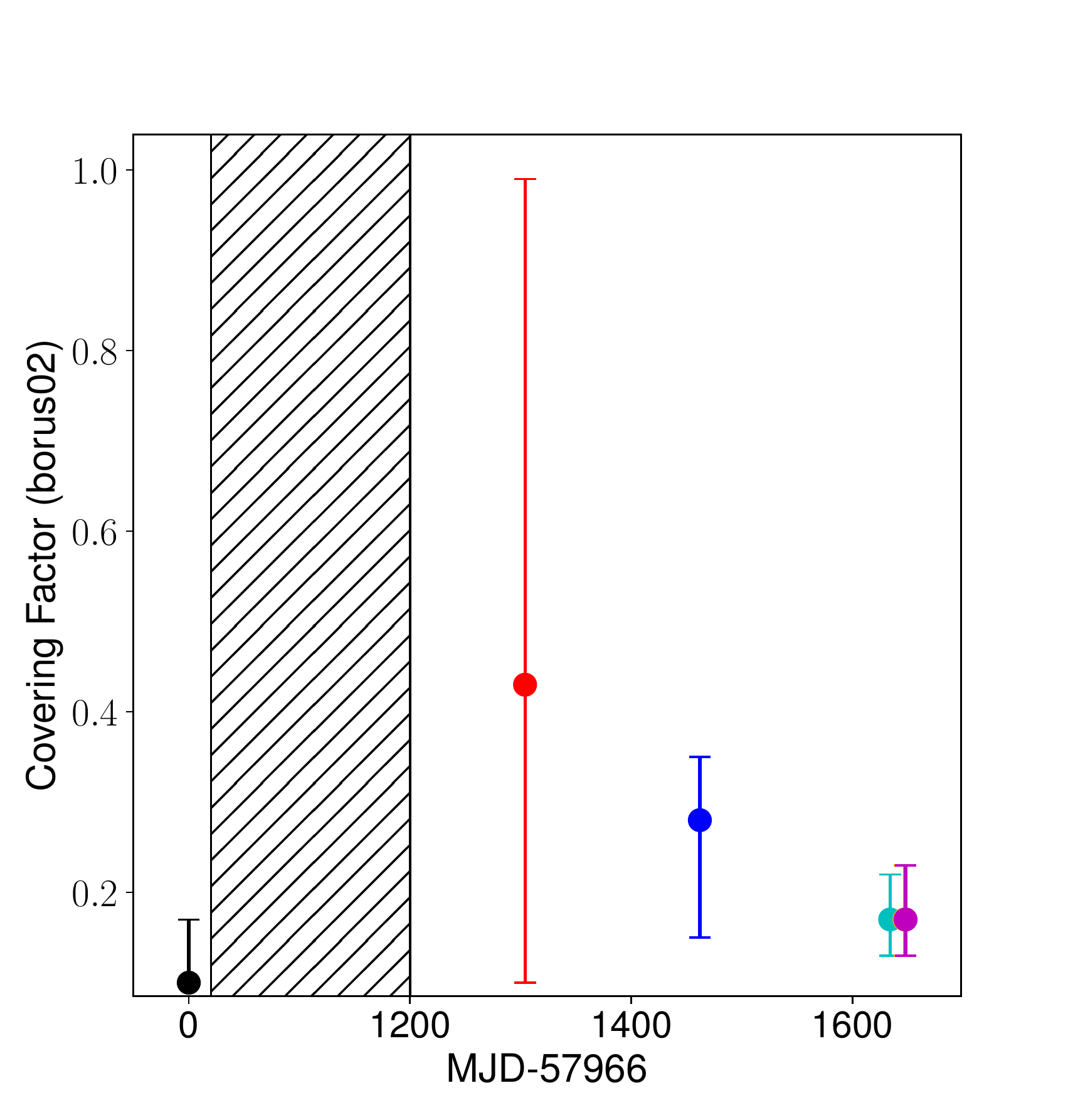} 
 \end{minipage}
\begin{minipage}{0.48\textwidth} 
 \centering 
 \includegraphics[width=1\textwidth]{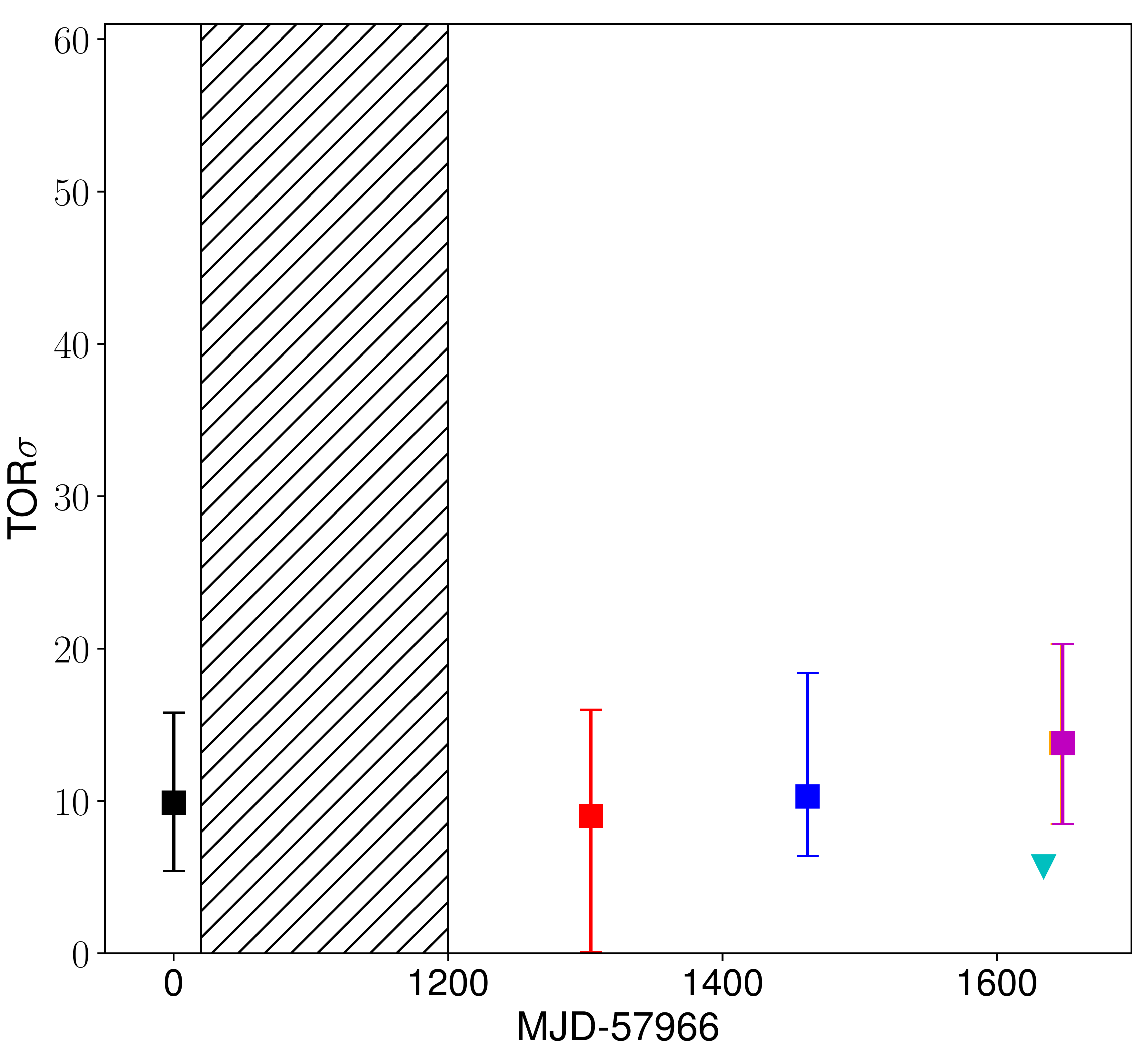}
 \end{minipage}
 \begin{minipage}{0.5\textwidth} 
 \centering 
 \includegraphics[trim={0 0 0 1cm},clip,width=1.0\textwidth]{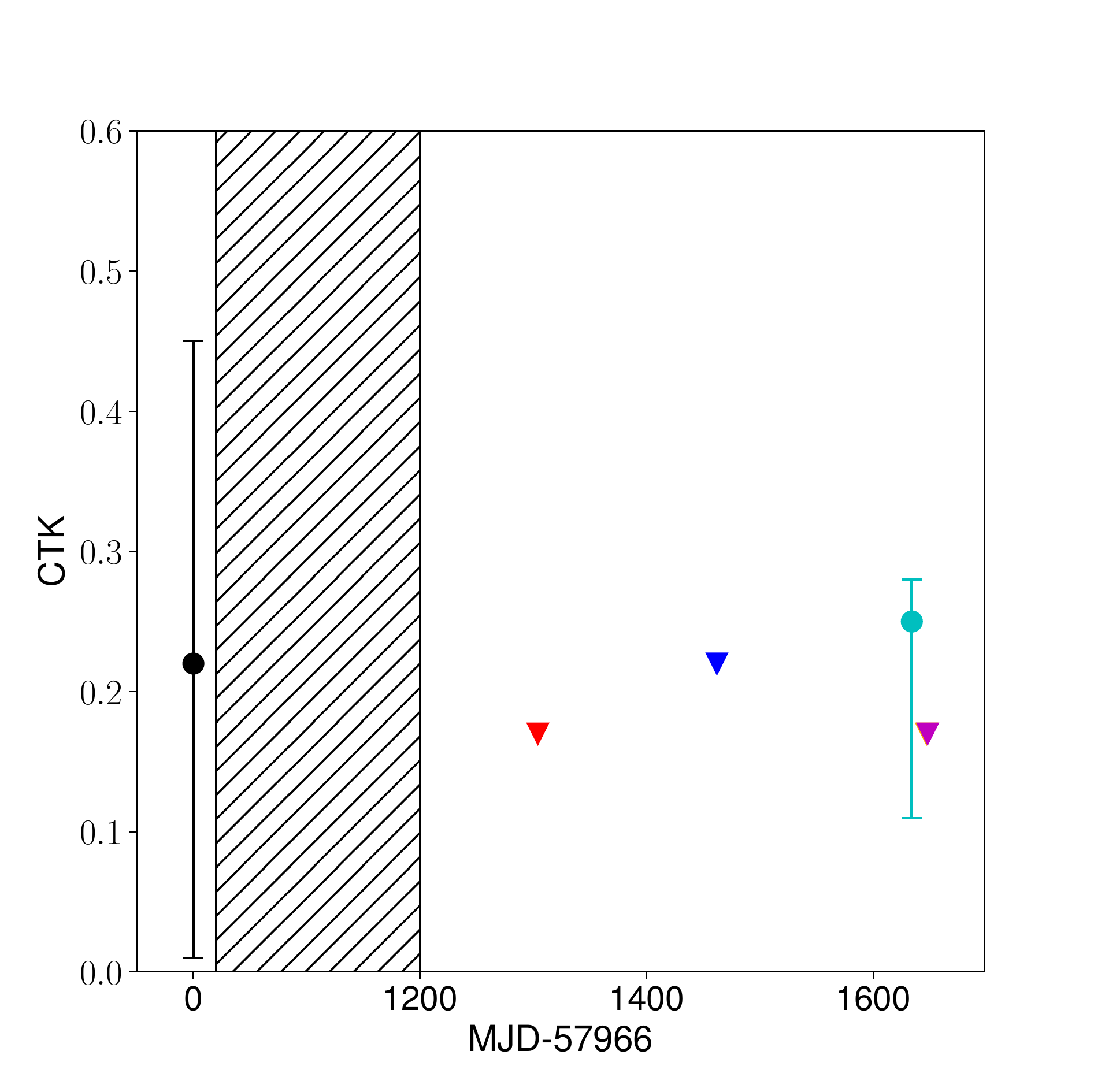} 
 \end{minipage} 
\caption{\normalsize 
Different properties of the obscuring material derived from five single-epoch spectra (we report the results of the spectral analysis in the appendix. Top: Torus average column density (left) and covering factor (right) as computed using \borus. Bottom: cloud vertical height, TOR$\sigma$, and covering factor of the inner ring, CTK, as computed using \uxcl. 
The color-code is the same in all panels; to increase the plot clarity, the first 1200 days (hatched area) are not in scale. In the top left panel, the dashed horizontal line marks the \nhtor=10$^{24}$\,cm$^{-2}$ threshold. 90\,\% confidence upper limits are plotted as downwards triangles.
}\label{fig:nh_tor_fc_single_epoch}
\end{figure*}

As discussed in \citet[][]{balokovic18,balokovic21} and, more recently, in \citet[][]{saha22}, however, the simultaneous fit of multi-epoch X-ray spectra is the most efficient way to reduce the uncertainties on the different spectral parameters and break degeneracies between them. In particular, the multi-epoch fitting approach is key to put tight constraints on the torus covering factor, its average column density, and the inclination angle between the torus axis and the observer, three parameters that can be highly degenerate in single-epoch observations. 

To further clarify the importance of multi-epoch observations, we report in Figure~\ref{fig:nh_tor_fc_single_epoch} four spectral parameters computed in each of the five single-epoch observations performed between August 2017 and February 2022. We also note that the inclination angle $\theta_{\rm i}$ between the observer and the torus axis is loosely constrained, if not fully unconstrained, in all single-epoch observations. Two of the parameters shown in the Figure are computed using \borus: the torus average column density (top left) and covering factor (top right). The other two are computed using \uxcl: the obscuring material scale height (bottom left) and the covering factor of the inner reflector (bottom right).

A first clear evidence is that \xmm\ observations alone are not effective in reliably constraining these parameters. More importantly, while both \uxcl\ parameters are consistent, within the uncertainties, in each of the five epochs, both \borus\ parameters show evidence for variability. 
In particular, the variability in \nhtor\ is found to be fairly large, varying in the range log\nhtor=[23.3--24.2]. Such a disagreement between the single-epoch measurements should not be treated physically, given that the overall amount of material in the obscurer is not expected to vary in time-scales of months. Rather, this result suggests that a uniform torus model is less efficient than a clumpy torus one in modelling the X-ray emission of NGC 1358, a result consistent with the idea that in this source the obscuring material surrounding the accreting SMBH is distributed in highly inhomogeneous clumps.

\begin{table*}
\renewcommand*{\arraystretch}{1.5}
\scalebox{0.75}{
\centering
  \begin{tabular}{ccccccccc}
       \hline
       \hline     
       & \multicolumn{2}{c}{\myt} & \multicolumn{2}{c}{\myt} & \multicolumn{2}{c}{\borus} & \multicolumn{2}{c}{\uxcl} \\
       & \multicolumn{2}{c}{decoupled, 0$\degree$} & \multicolumn{2}{c}{decoupled,  90$\degree$}  \\
       & Best-fit & \% Err$_{\rm M-S}$ &  Best-fit & \% Err$_{\rm M-S}$ &  Best-fit & \% Err$_{\rm M-S}$ &  Best-fit & \% Err$_{\rm M-S}$\\
       \hline
       $\chi^2$/dof & 2520.7/2457 & -- & 2513.8/2457 & -- & 2518.2/2457 & -- & 2547.7/2457 & -- \\
       $\Gamma$ & 1.54$_{-0.04}^{+0.04}$ & 50\,\% & 1.54$_{-0.04}^{+0.04}$ & 50\,\% & 1.45$_{-0.02}^{+0.01}$ & 75\,\% & 1.58$_{-0.03}^{+0.03}$ & 55\,\% \\
       \nhlos\ 2017--08--01 [10$^{24}$\,cm$^{-2}$] & 1.68$_{-0.11}^{+0.15}$ & 45\,\% &  1.87$_{-0.15}^{+0.17}$ & 60\,\% & 1.64$_{-0.03}^{+0.05}$ & 70\,\% & 1.49$_{-0.10}^{+0.09}$ & 60\,\% \\
       \nhlos\ 2021--02--25 [10$^{24}$\,cm$^{-2}$] & 1.21$_{-0.05}^{+0.06}$ & 60\,\% &  1.26$_{-0.06}^{+0.07}$ & 65\,\% & 1.34$_{-0.03}^{+0.04}$ & 60\,\% & 1.02$_{-0.10}^{+0.08}$ & 50\,\% \\
       \nhlos\ 2021--08--02 [10$^{24}$\,cm$^{-2}$] & 0.87$_{-0.04}^{+0.04}$ & 20\,\% &  0.88$_{-0.04}^{+0.04}$ & 50\,\% & 0.87$_{-0.02}^{+0.02}$ & 60\,\% & 0.85$_{-0.06}^{+0.05}$ & 30\,\% \\
       \nhlos\ 2022--01--21 [10$^{24}$\,cm$^{-2}$] & 0.71$_{-0.03}^{+0.03}$ & 30\,\% & 0.71$_{-0.03}^{+0.03}$ & 40\,\% & 0.70$_{-0.01}^{+0.02}$ & 60\,\% & 0.70$_{-0.03}^{+0.03}$ & 30\,\% \\
       \nhlos\ 2022--02--03 [10$^{24}$\,cm$^{-2}$] & 0.74$_{-0.04}^{+0.04}$ & 30\,\% & 0.75$_{-0.04}^{+0.04}$ & 40\,\%  & 0.73$_{-0.01}^{+0.02}$ & 60\,\% & 0.72$_{-0.03}^{+0.03}$ & 25\,\% \\
       \nhlos\ 2022--02--04 [10$^{24}$\,cm$^{-2}$] & 0.76$_{-0.04}^{+0.04}$ & 40\,\% & 0.77$_{-0.04}^{+0.04}$ & 55\,\% & 0.80$_{-0.03}^{+0.02}$ & 55\,\% & 0.74$_{-0.05}^{+0.04}$ & 10\,\% \\
       \nhtor\ [10$^{24}$\,cm$^{-2}$]  & 0.56$_{-0.11}^{+0.19}$ & 80\,\% & 0.36$_{-0.05}^{+0.05}$ & 75\,\%  & 0.35$_{-0.03}^{+0.06}$ & 85\,\% & ... \\
       A$_S$ & 0.21$_{-0.02}^{+0.03}$ & 50\,\% & 0.40$_{-0.05}^{+0.05}$ & 85\,\% & ... & ... & ... & ...\\
       $f_c$ & ...  &... & ...  &... & 0.17$_{-0.02}^{+0.02}$ & 70\,\% & ... & ... \\
       $\theta\rm _{i}$ [$^\circ$]  & ... & ... & ... & ... & 87$^f$ & -- & 90$^f$ & -- \\
       TOR$\sigma$ [$^\circ$]  & ... & ... & ... & ... & ... & ... & 15.3$_{-2.5}^{+2.7}$ & 60\,\% \\
       CTK  & ... & ... & ... & ... & ... & ... & $<$0.10 & 30\,\% \\
       $f_s$ 10$^{-2}$ & 0.10$_{-0.03}^{+0.03}$ & 55\,\% & 0.16$_{-0.03}^{+0.03}$ & 60\,\% & 0.12$_{-0.02}^{+0.01}$ & 75\,\% & 0.28$_{-0.08}^{+0.10}$ & 20\,\% \\
       $kT$ [keV] & 0.63$_{-0.03}^{+0.03}$ & 75\,\% & 0.61$_{-0.03}^{+0.03}$ & 70\,\% & 0.65$_{-0.04}^{+0.03}$ & 45\,\% & 0.65$_{-0.03}^{+0.03}$ & 70\,\% \\
       $Z$/$Z_\odot$ & 0.05$_{-0.02}^{+0.02}$ & 80\,\% & 0.08$_{-0.03}^{+0.05}$ & 95\,\% & 0.05$_{-0.01}^{+0.01}$ & 95\,\% & 0.05$_{-0.02}^{+0.02}$ & 60\,\% \\
       log(L$_{2-10}$)  2017--08--01 [\lu] & 42.90$_{-0.10}^{+0.07}$ & -- & 42.93$_{-0.09}^{+0.08}$ & -- & 42.74$_{-0.09}^{+0.06}$ & -- & 42.75$_{-0.04}^{+0.04}$ & --\\
       log(L$_{10-40}$) 2017--08--01 [\lu] & 43.74$_{-0.32}^{+0.23}$ & -- & 43.38$_{-0.29}^{+0.24}$ & -- & 43.11$_{-0.45}^{+0.33}$ & -- & 42.93$_{-0.19}^{+0.13}$ & --\\
       log(L$_{2-10}$)  2021--02--25 [\lu] & 43.04$_{-0.05}^{+0.04}$ & -- & 42.89$_{-0.04}^{+0.04}$ & -- & 43.02$_{-0.05}^{+0.05}$ & -- & 42.77$_{-0.10}^{+0.08}$ & --\\
       log(L$_{2-10}$)  2021--08--02 [\lu] & 42.96$_{-0.06}^{+0.04}$ & -- & 42.96$_{-0.05}^{+0.04}$ & -- & 42.93$_{-0.07}^{+0.05}$ & -- & 42.91$_{-0.04}^{+0.04}$ & --\\
       log(L$_{10-40}$) 2021--08--02 [\lu] & 43.79$_{-0.21}^{+0.13}$ & -- & 43.67$_{-0.21}^{+0.12}$ & -- & 43.77$_{-0.15}^{+0.10}$ & -- & 43.15$_{-0.13}^{+0.09}$ & --\\
       log(L$_{2-10}$)  2022--01--21 [\lu] & 42.99$_{-0.09}^{+0.07}$ & -- & 42.98$_{-0.04}^{+0.04}$ & -- & 42.91$_{-0.06}^{+0.04}$ & -- & 42.95$_{-0.05}^{+0.03}$ & --\\
       log(L$_{10-40}$) 2022--01--21 [\lu] & 43.86$_{-0.32}^{+0.15}$ & -- & 43.63$_{-0.28}^{+0.22}$ & -- & 43.64$_{-0.26}^{+0.13}$ & -- & 43.18$_{-0.20}^{+0.15}$ & --\\
       log(L$_{10-40}$) 2022--02--03 [\lu] & 43.55$_{-0.29}^{+0.16}$ & -- & 43.45$_{-0.31}^{+0.17}$ & -- & 43.53$_{-0.26}^{+0.14}$ & -- & 43.21$_{-0.19}^{+0.14}$ & --\\
       log(L$_{2-10}$)  2022--02--04 [\lu] & 42.94$_{-0.06}^{+0.05}$ & -- & 42.93$_{-0.04}^{+0.03}$ & -- & 42.93$_{-0.06}^{+0.04}$ & -- & 42.89$_{-0.06}^{+0.05}$ & --\\
       \hline
	\hline
	\vspace{0.02cm}
\end{tabular}
}
\caption{\normalsize Summary of the best-fit results for the joint spectral fit of all the \nus\ and \xmm\ observations taken between August 01, 2017, and February 03--04, 2022. 
$\Gamma$ is the main power law component photon index.
\nhlos\ and \nhtor\ are the line-of-sight and average column density, respectively, in units of cm$^{-2}$. 
A$_S$ is the relative intensity of the reprocessed component with respect to the main one in \myt. 
$f_c$ is the covering factor of the obscuring material as computed by \borus, $f_c$ = cos($\theta_{\rm T}$), where $\theta\rm _{T}$ is the angle (in degrees) between the axis of the torus and the edge of torus. $\theta\rm _{i}$ is the angle (in degrees) between the observer and the torus axis. 
In \uxcl, TOR$\sigma$ is the angular width of the cloud population and CTK is the covering factor of inner Compton-thick ring of clouds. $kT$ and $Z$ are the temperature (in keV) and metallicity (in units of $Z_\odot$) of the thermal \texttt{mekal} component. 
L$_{2-10}$ and L$_{10-40}$ are the intrinsic luminosities in units of erg\,s$^{-1}$ in the 2--10\,keV and 10--40\,keV band, respectively.
For all the parameters we report Err$_{\rm M-S}$, the fractional change of the uncertainties with respect to the corresponding single-epoch observations (i.e., ``30\,\%'' means that the multi-epoch uncertainty on the parameter is 30\,\% smaller than the average of the single-epoch uncertainties). The single-epoch results are reported in the Appendix}.
\label{tab:fit_multi-epoch}
\end{table*}

\subsection{Joint multi-epoch fit}
Given the limitations of a single-epoch fitting that we highlighted in the previous section, we performed a simultaneous fit of the \nus\ and \xmm\ spectra derived from the observations taken between August, 2017, and February, 2022, with the goal of reducing the parameters uncertainties and breaking cross-parameter degeneracies. We included in our models a cross-observation normalization to account for any flux variability not related to \nhlos, and left the l.o.s. column density free to vary in each of the six epochs: we assumed two independent \nhlos\ and normalization values for the \nus\ and \xmm\ observations taken on February 3 and 4. We then fit the spectra assuming no intra-epoch variability for all the other parameters: the power law photon index $\Gamma$, the scattered fraction $f_s$, the average torus column density (in \myt\ and \borus), the torus covering factor (in \borus\ and \uxcl).

\begin{figure*} 
\begin{minipage}{0.32\textwidth} 
 \centering 
 \includegraphics[width=1\textwidth]{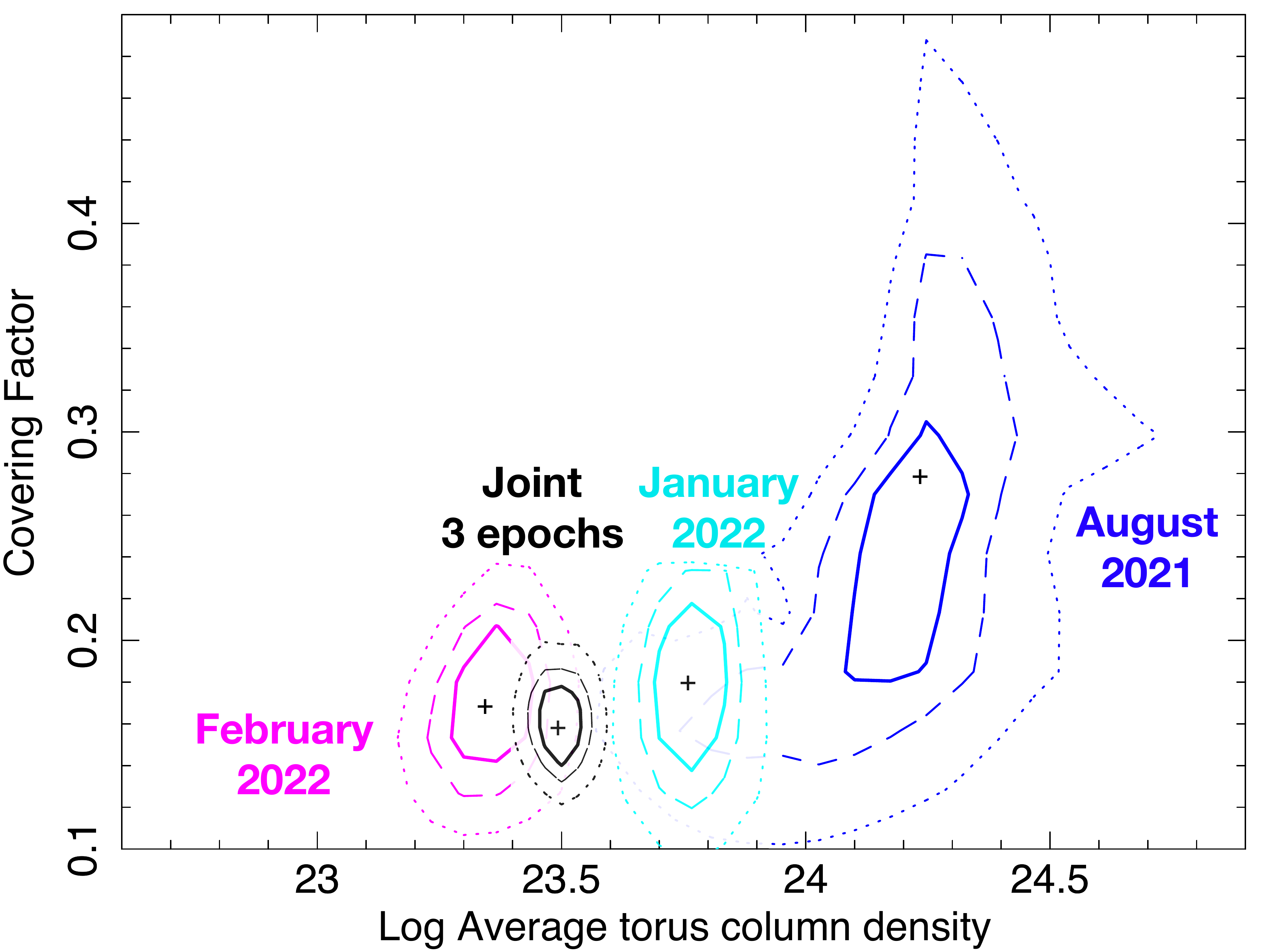} 
 \end{minipage}
 \begin{minipage}{0.32\textwidth} 
 \centering 
 \includegraphics[trim={0 0 0 0 cm},clip,width=1.0\textwidth]{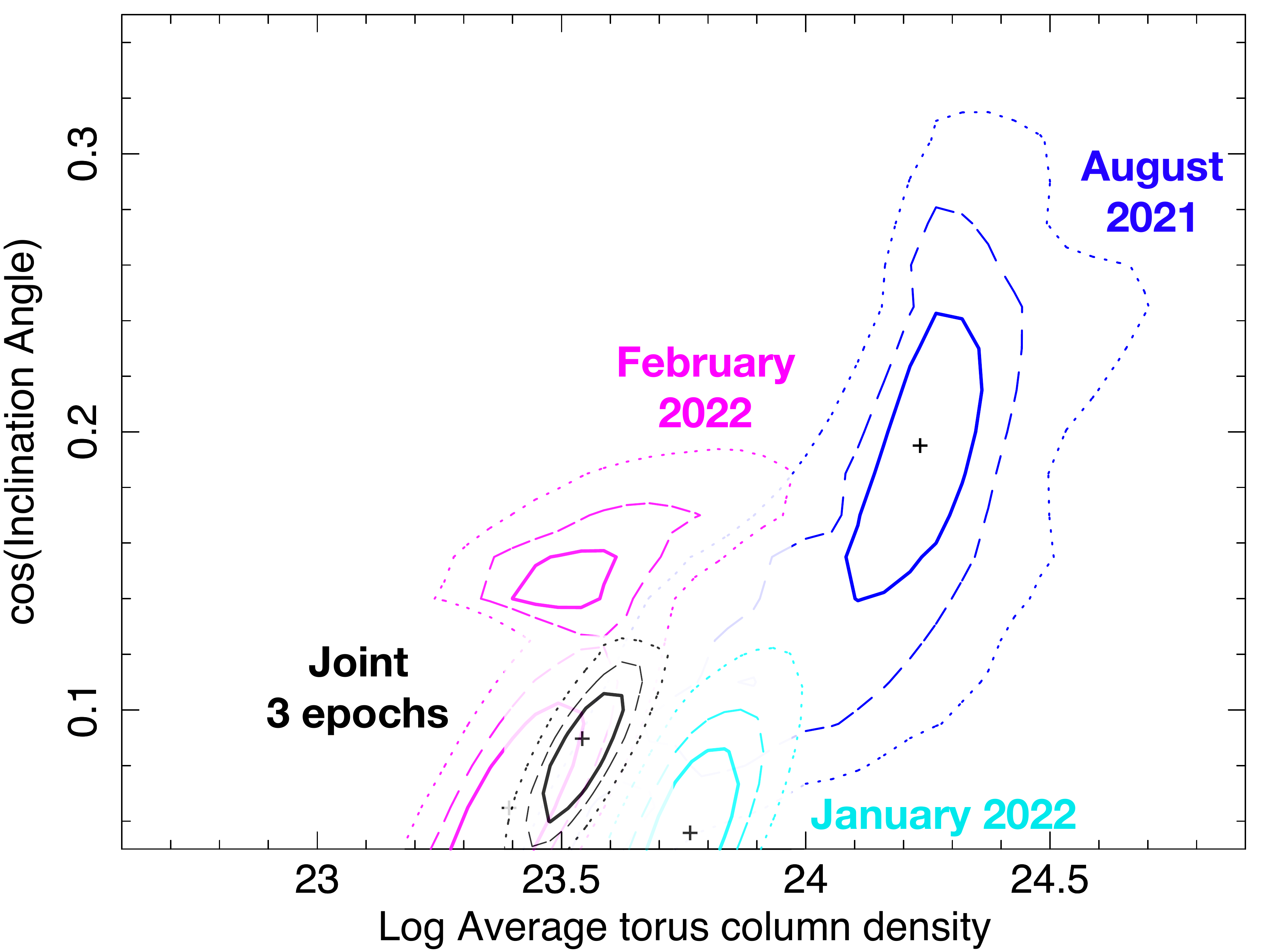} 
 \end{minipage} 
\begin{minipage}{0.32\textwidth} 
 \centering 
 \includegraphics[width=1.0\textwidth]{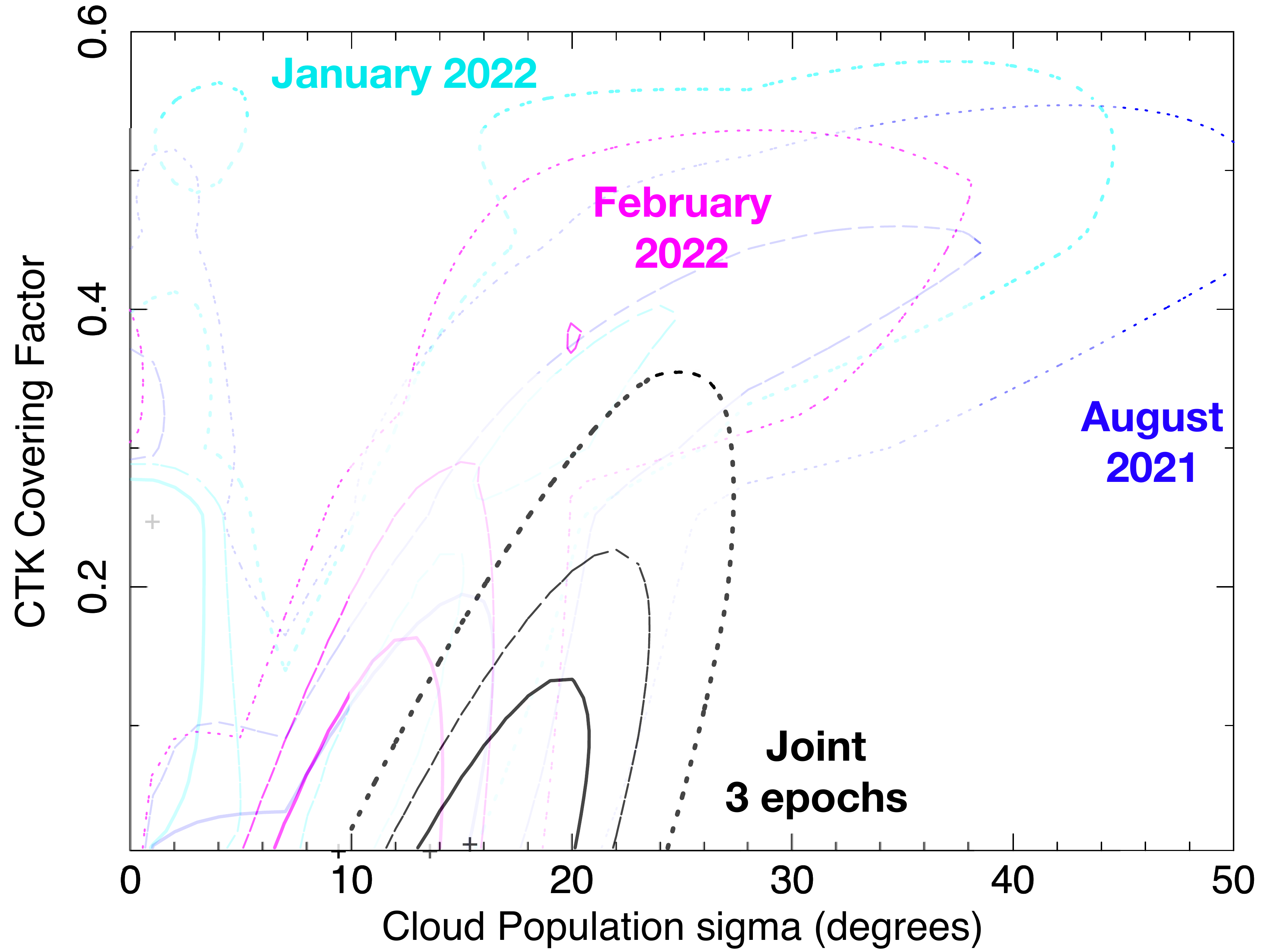} 
 \end{minipage}
\caption{\normalsize 
68, 90 and 99\,\% confidence contours of the covering factor as a function of the average torus column density (left), the cosine of the inclination angle as a function of the average torus column density (center), and the torus CT obscurer covering factor as a function of the cloud population opening angle, TOR$\sigma$ (right). 
The first two plots are obtained using \borus, the third one using \uxcl. The contours obtained from the multi-epoch joint fit are plotted in black, while those obtained fitting the various single-epoch spectra are color-coded as in Figure~\ref{fig:parameters_evolution}.
}\label{fig:contours_multi_epoch}
\end{figure*}

We report in Table~\ref{tab:fit_multi-epoch} the results of the multi-epoch fit, as well as the fractional variation of the uncertainties on the parameters with respect to the single-epoch results. The increase in count statistics and the use of multi-epoch data strongly reduce the uncertainties on all parameters. For example, the errors on the \nhlos\ values measured in the different epochs decrease by 10--50\,\% in \myt\ and \uxcl\ and up to 80\,\% in \borus. We also measure significant reductions on the uncertainties of the average torus column density ($\sim$10-50\,\% in \myt; $\sim$80\,\% in \borus) and in the torus covering factor ($\sim$30-50\,\% in \uxcl; $\sim$70\,\% in \borus). Finally, the uncertainties on the power law photon index decrease by 30--70\,\%. We note that the \borus\ best-fit photon index ($\Gamma$=1.45$_{-0.02}^{+0.01}$) is close to the model lower boundary ($\Gamma$=1.4), another possible indirect evidence for the limitations of a uniform torus model to describe the NGC 1358 obscurer. 

\begin{figure} 
 \centering 
 \includegraphics[width=0.47\textwidth]{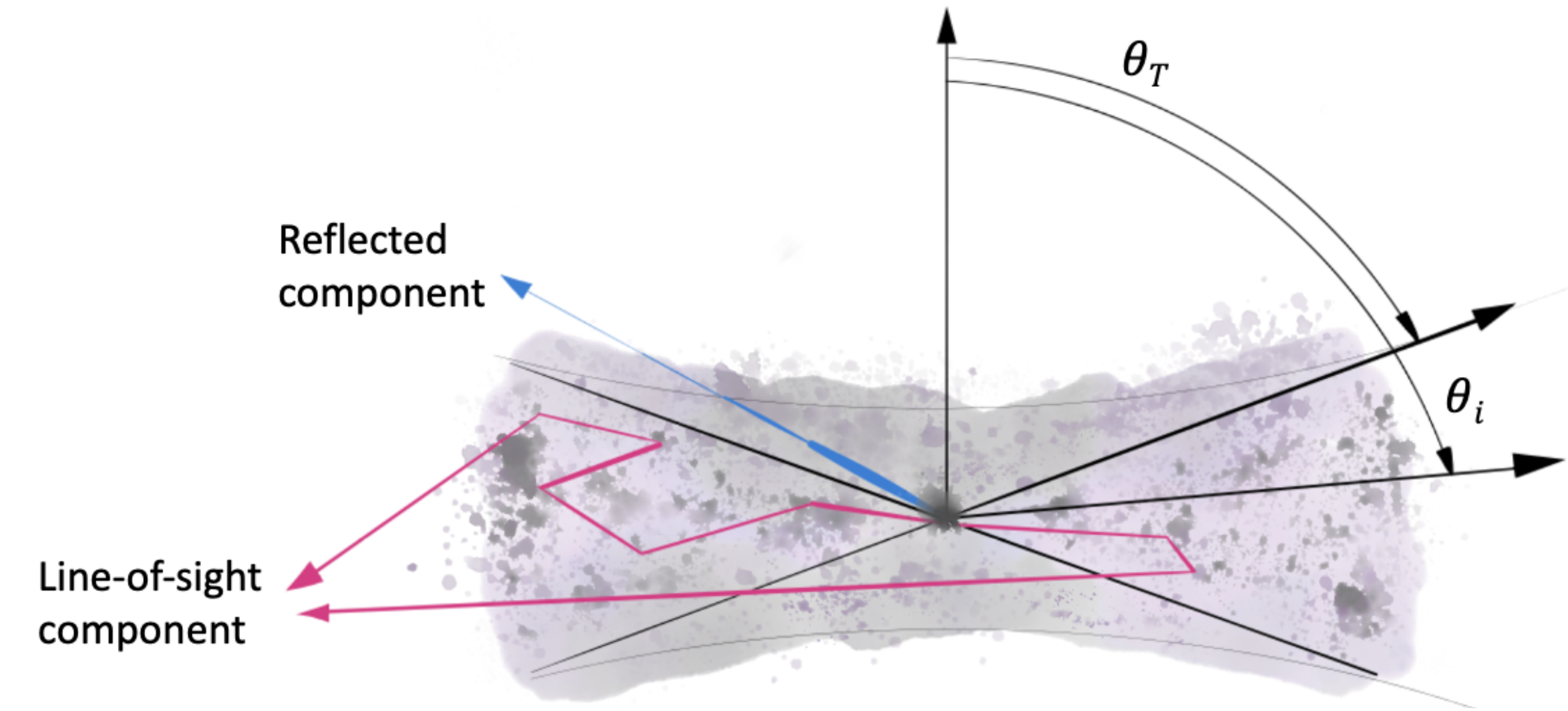} 
\caption{\normalsize 
Sketch of the obscuring material surrounding NGC 1358, based on the best-fit results obtained in this work. $\theta_{\rm T}$ is the torus opening angle, while $\theta_{\rm i}$ is the angle between the observer and the torus axis.
}\label{fig:sketch_torus}
\end{figure}

In Figure~\ref{fig:contours_multi_epoch} we report the comparison between single-- and multi--epoch 68--90--99\,\% confidence contours of the torus covering factor and inclination angle as function of the average torus column density, as measured using \borus, and of the torus vertical extent and inner ring covering factor, as measured with \uxcl. For computational reasons, the multi-epoch contours have been computed using only the joint \nus\ and \xmm\ observations taken between August, 2021, and February, 2022 (i.e., the February, 2021, and August, 2017 observations are not included in the computation of the multi--epoch contours). For consistency, we thus compare these multi-epoch contours with the single-epoch ones obtained in the three set of observations.

For all three pairs of parameters, the multi--epoch fit allows one to break infra-parameter degeneracies and reliably confirm the ``low covering factor, Compton-thin average torus column density'' scenario. Furthermore, the left and central plots once again highlight how single-epoch observations with \borus\ can produce inconsistent \nhtor\ measurements, as we discussed in the previous section. In the multi-epoch fit, instead, we measure with \borus\ a Compton-thin average column density, log\nhtor=23.5$\pm$0.1 and a covering factor $f_c$=0.17$\pm$0.02; the inclination angle is consistent with an ``edge on view'' scenario, being $\theta_{\rm i}>$83$\degree$. We measure the same average torus column density using \myt\ decoupled in its 90$\degree$ configuration: when using the 0$\degree$ configuration, we measure a slightly larger average column density, but the uncertainties are also significantly larger (log\nhtor=24.0$^{+0.4}_{-0.7}$). Finally, with \uxcl\ we measure a cloud vertical extent TOR$\sigma$=15.3$_{-2.5}^{+2.7^\circ}$ and an upper limit on the covering factor of the inner CT ring CTK$<$0.10, once again confirming the low--$f_c$ scenario. We report in Figure~\ref{fig:sketch_torus} a sketch of the NGC 1358 obscuring torus, based on this best-fit results.

\section{Characterizing the obscuring material in NGC 1358}\label{sec:discussion}
\begin{figure}[htbp]
 \centering 
 \includegraphics[width=0.47\textwidth]{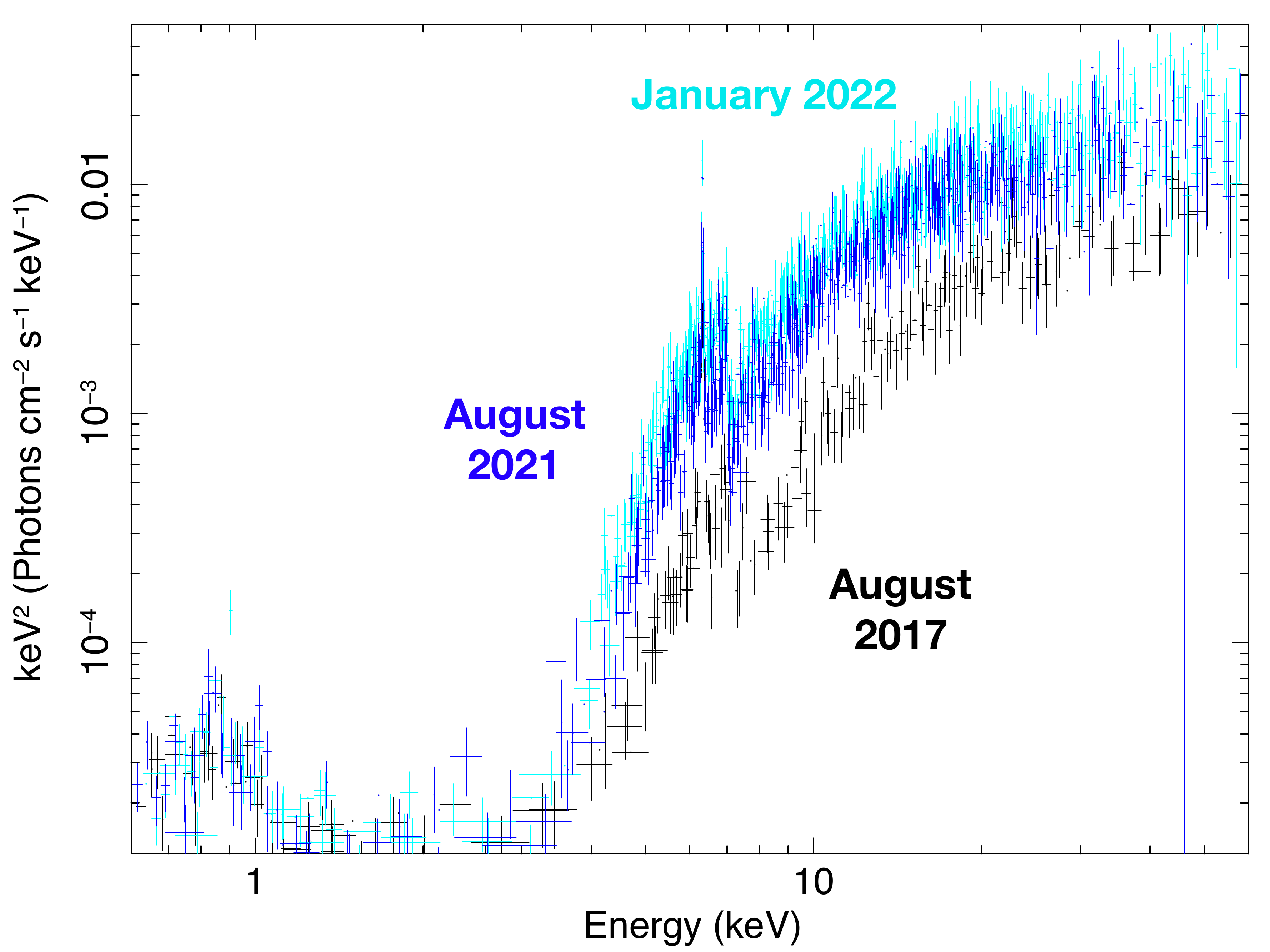} 
\caption{\normalsize 
Unfolded spectra of the August 2017 (black), August 2021 (blue) and January 2022 (cyan) observations of NGC 1358, fitted using the \uxcl\ model.
}\label{fig:spectra_comparison}
\end{figure}

\begin{figure*} 
\begin{minipage}{0.49\textwidth} 
 \centering 
 \includegraphics[width=1\textwidth]{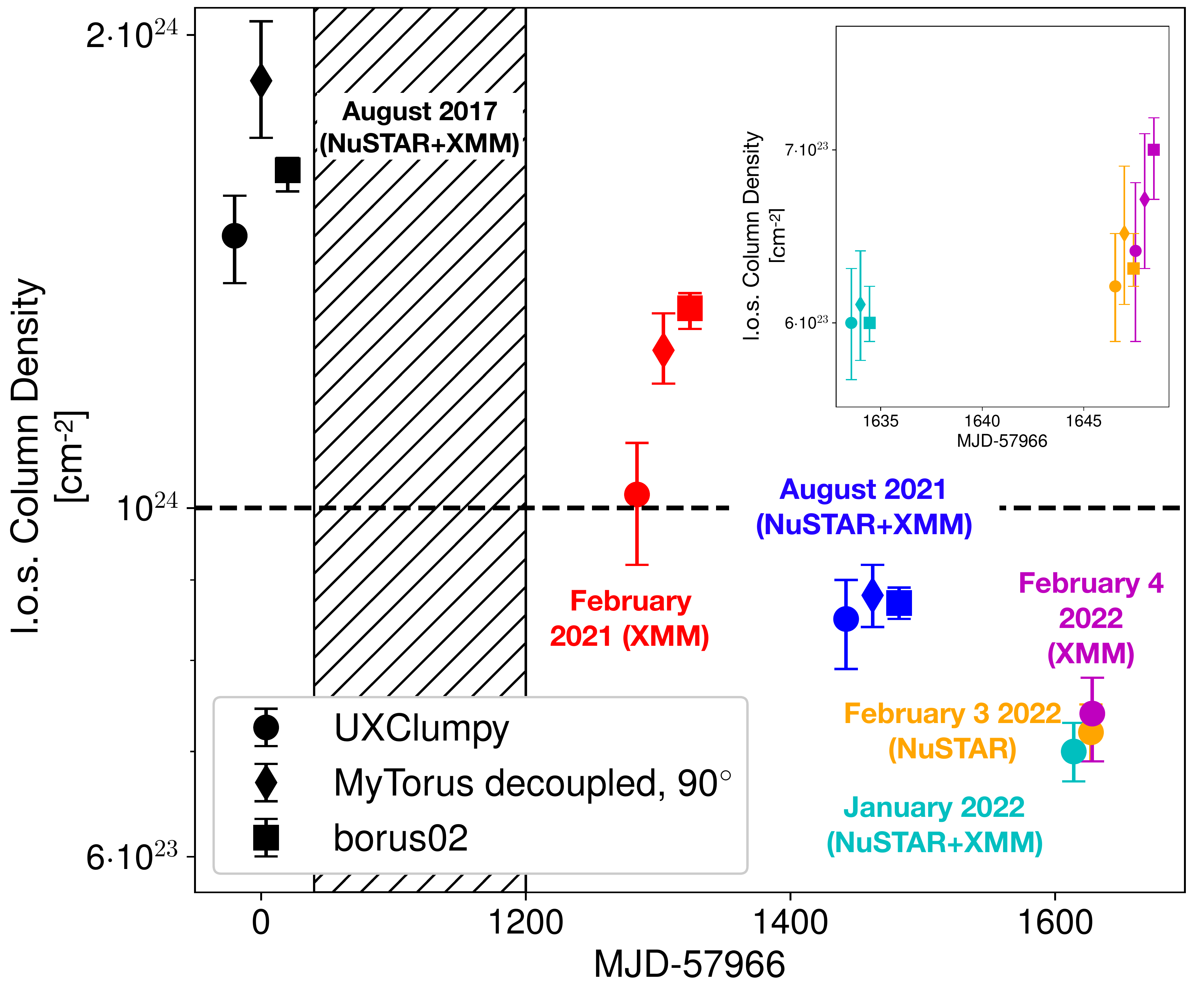} 
 \end{minipage}
 \begin{minipage}{0.49\textwidth} 
 \centering 
 \includegraphics[width=1.0\textwidth]{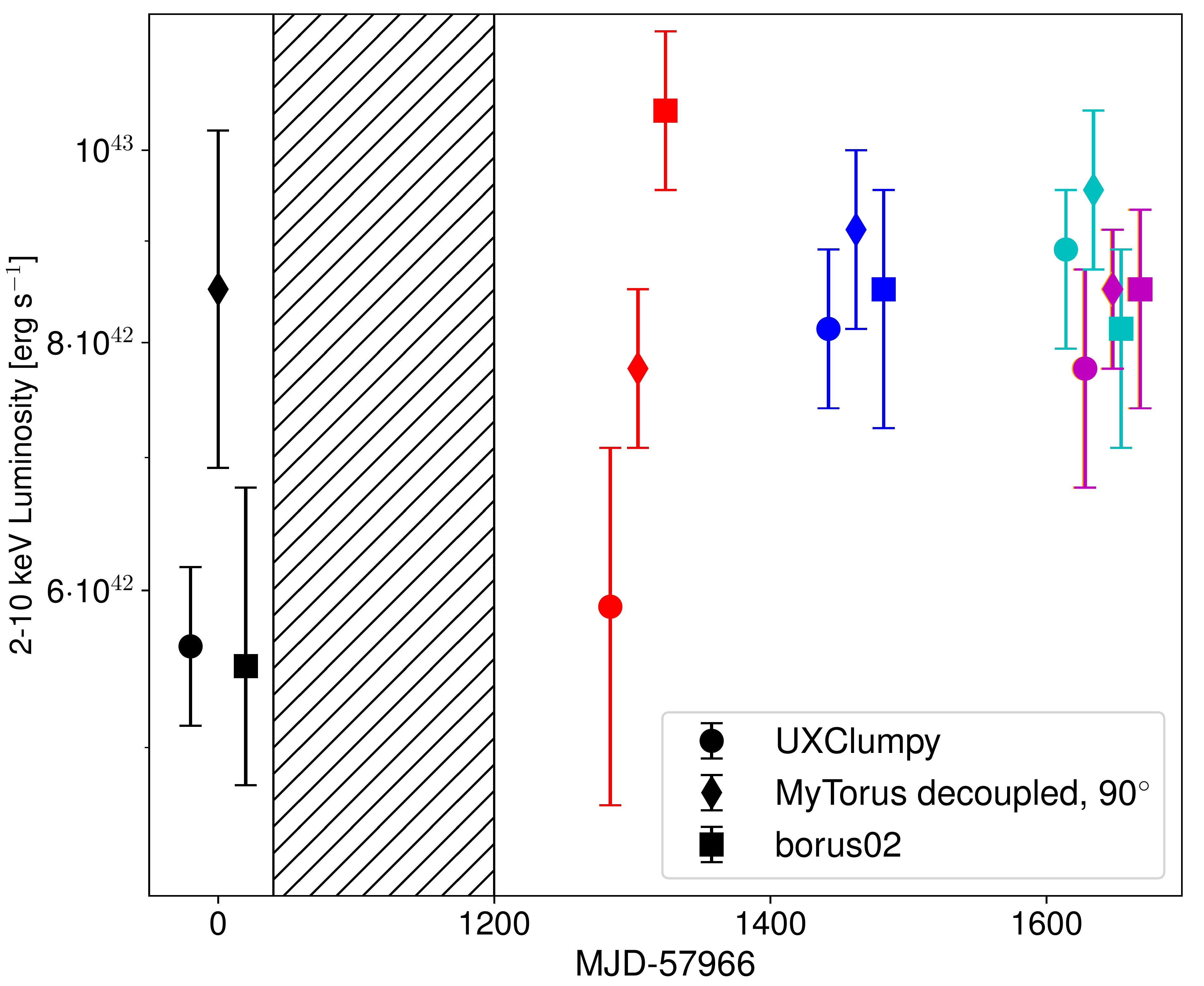} 
 \end{minipage} 
\caption{\normalsize 
Line-of-sight column density (left) and 2--10\,keV intrinsic luminosity (right) of NGC 1358 as obtained using \uxcl\ (circles), \myt\ decoupled in its 90$^\circ$ configuration (diamonds) and \borus\ (squares) in each of the epochs where an X-ray observation was performed, starting from the joint \nus--\xmm\ observation taken in August 2017. No luminosity variability was observed between the February 3 and 4, 2022 observations, so we plot only one data point. In the left panel, we show in an inset the results of the 2022 observations, to avoid overcrowding the plot. The \uxcl\ and \borus\ data-points are shifted by 20 days (0.5 days in the \nhlos\ inset) for visualization purposes. To further increase the plot clarity, the first 1200 days (hatched area) are not in scale.
}\label{fig:parameters_evolution}
\end{figure*}

We report in Figure~\ref{fig:spectra_comparison} the joint \nus\ and \xmm\ spectra of NGC 1358 taken in August 2017, August 2021 and January 2022. The August 2017 spectrum is significantly fainter than the other two over the 2--60\,keV range. A minor, but still notable difference is also observed between the August 2021 and the January 2022 observation, the second being brighter. To better quantify this variability between observations, and break the \nhlos--luminosity degeneracy, we report in Figure~\ref{fig:parameters_evolution} the evolution with time of two main parameters: the line-of-sight column density (left) and the AGN 2--10\,keV intrinsic, absorption--corrected luminosity (right). The best-fit values and uncertainties are those obtained jointly fitting the observations.

As it can be seen, the l.o.s. column density of the material surrounding the accreting SMBH in NGC 1358 is highly variable over different time scales. The first and most significant change in \nhlos\ is the one observed between the 2017 observation and the February 2021 one. In a time-span of $\sim$4 years we observe a drop in \nhlos\ by $\sim$30\,\% (from $\sim$1.5--1.9$\times$10$^{24}$\,cm$^{-2}$ to $\sim$1.0--1.3$\times$10$^{23}$\,cm$^{-2}$). 
We then observe a further $\sim$15--30\,\% reduction in a time-span of about five months (from $\sim$1.0--1.3$\times$10$^{23}$\,cm$^{-2}$ on February 25, 2021, to 8--9$\times$10$^{23}$\,cm$^{-2}$ on August 2, 2021), with a consequent transition from a Compton thick to a Compton thin state;  
a further $\sim$20\,\% decrease, down to 7$\times$10$^{23}$\,cm$^{-2}$ is then observed between August 2021 and January 2022.
Finally, as shown in the inset of Figure~\ref{fig:parameters_evolution}, we measure a tentative, intriguing new increase in the two-weeks time-span between our two most recent observations, although this last result is not confirmed by all models (in particular, this trend is stronger when fitting with \borus). 
Furthermore, we find some tentative evidence for $\sim$1\,day variability thanks to the non-simultaneity of the \nus\ and \xmm\ February 2022 observations. We note that this evidence, while marginal, strengthens the ``new \nhlos\ uprise'' scenario (i.e., the \nhlos\ measured with \xmm\ on February 4, 2022, is larger than the one measured with \nus\ on February 3, 2022).

To better quantify the reliability of this \nhlos\ variability, one needs to check for potential degeneracies between parameters. In particular, it is essential do understand if the AGN intrinsic luminosity experienced any significant change in the time-span covered by our X-ray observations. For this reason, we report in  Figure~\ref{fig:parameters_evolution}, the evolution with time of the 2--10\,keV luminosity as computed from our best-fit models. No significant trend with luminosity is detected, regardless of the model used to fit the data. We note that the 2--10\,keV luminosity value for the 2017 observation decreased by a factor ~2.5 with respect to the one reported in \citet{zhao19a}, and it is now consistent with the values measured in 2021-2022. Such a result suggests that single-epoch measurements are reliable when measuring parameters such as the l.o.s. column density and (provided there is enough statistic above 10\,keV) the torus average column density and covering factor. Single-epoch observations, however, are much less efficient in disentangling the contribution of the primary and reprocessed component to the overall emission, which consequently can lead to incorrect luminosity estimates.

\begin{figure}[htbp]
 \centering 
 \includegraphics[width=0.47\textwidth]{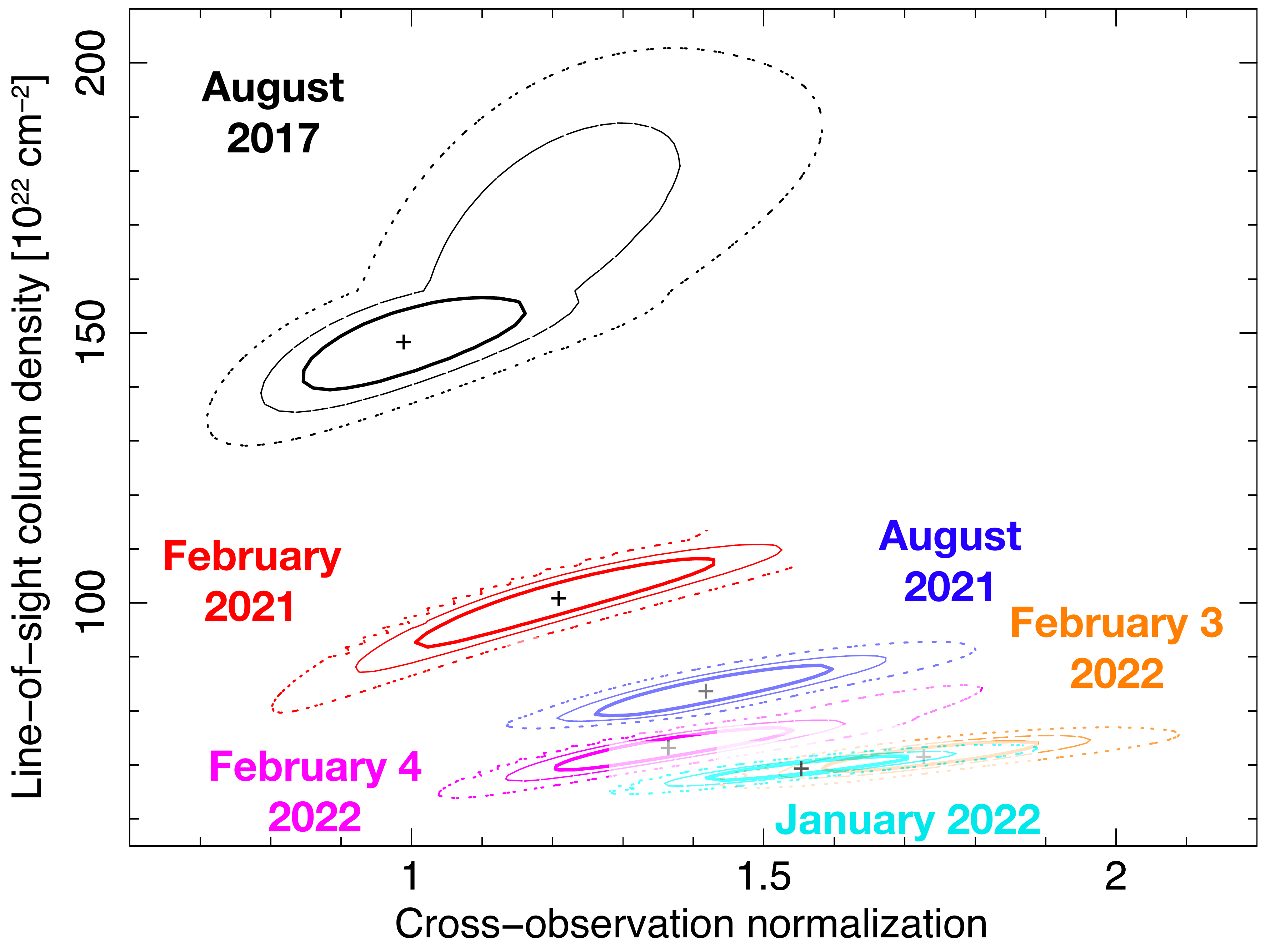} 
\caption{\normalsize 
68, 90 and 99\,\% confidence contours of the l.o.s. column density as measured using \uxcl\ as a function of cross-observation flux normalization. In all contours, the cross-observation normalizations are the \xmm\ ones: the only exception being the February 3 contours, which are computed using the \nus\ normalization for consistency with the fact that \nhlos\ is also measured from the \nus\ data alone. As discussed in Appendix~\ref{sec:feb22_obs}, the \nus\ cross-normalizations are usually $\sim$10\,\% larger than the \xmm\ ones.
}\label{fig:const_NH_contours}
\end{figure}

To further underline that the high-quality \nus\ and \xmm\ data make it possible to break any \nhlos--luminosity degeneracy, we show in Figure~\ref{fig:const_NH_contours} the confidence contours of the l.o.s. column density as a function of the cross-observation flux normalization\footnote{In all contours, the cross-observation normalizations are the \xmm\ ones: the only exception being the February 3 contours, which are computed using the \nus\ normalization for consistency with the fact that \nhlos\ is also measured from the \nus\ data alone.}. 
This parameter takes into account any flux variability which is not related to \nhlos\ variability, and is therefore a good proxy of the 2--10\,keV luminosity. As it can be seen, the \nhlos\ trend is still present and is therefore not significantly affected by AGN luminosity--related degeneracies.

\begin{figure} 
 \centering 
 \includegraphics[width=0.47\textwidth]{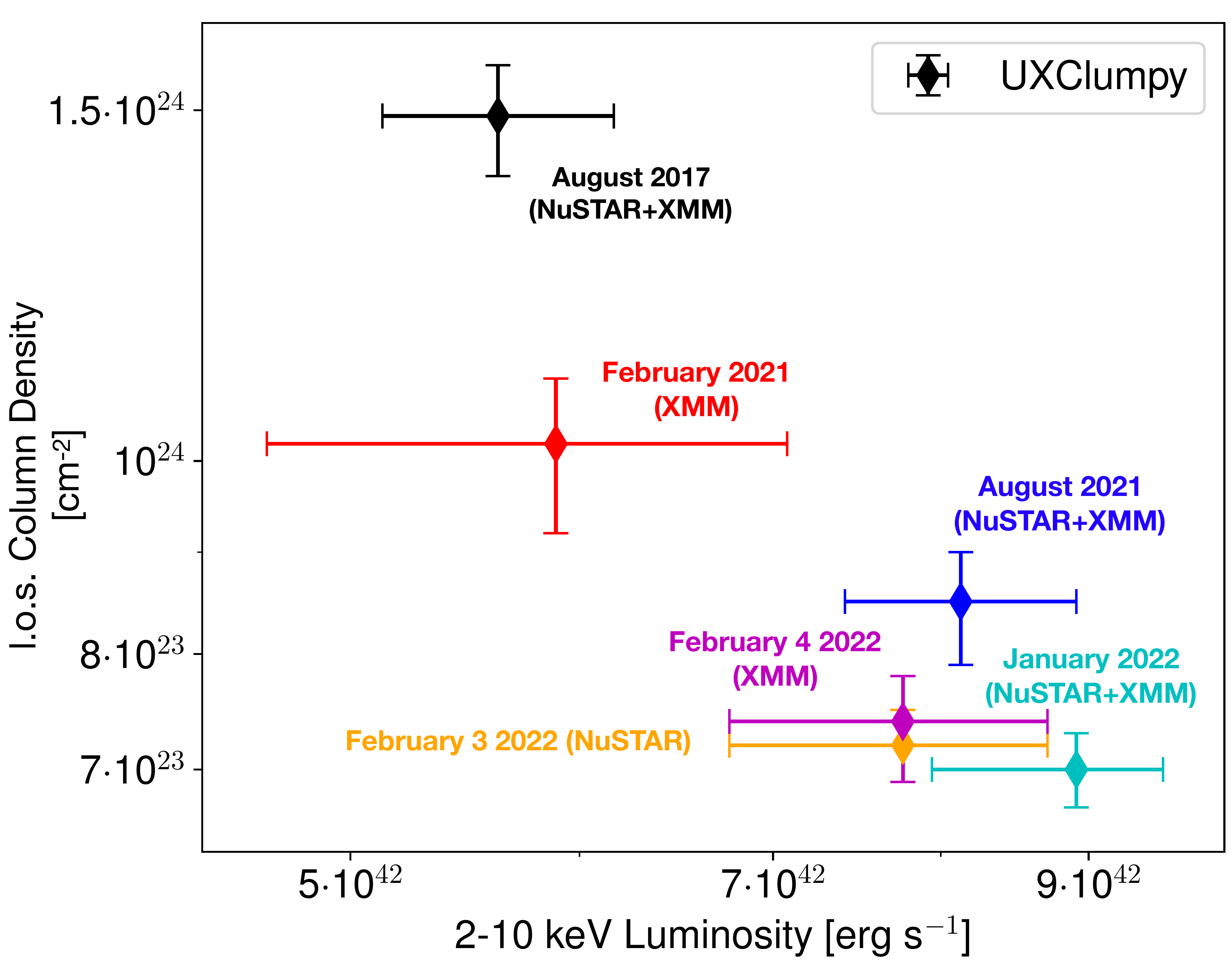} 
\caption{\normalsize 
Line-of-sight column density as computed with \uxcl\ as a function of the 2--10\,keV intrinsic luminosity in NGC 1358. The color-code for each observing epoch is the same used in Figures~\ref{fig:nh_tor_fc_single_epoch}, \ref{fig:parameters_evolution} and \ref{fig:const_NH_contours}.
}\label{fig:correlation_nh_lx_cf}
\end{figure}

In Figure~\ref{fig:correlation_nh_lx_cf}, left panel, we plot the 2--10\,keV luminosity as a function of \nhlos, as computed using \uxcl, to better underline the correlation between these two quantities.

A possible, qualitative explanation for the observed \nhlos\ and X-ray luminosity variability is a self-regulated ``AGN feedback'' scenario (see, e.g., \citealt{gaspari20} for a review). We work under the assumption that the 2--10\,keV luminosity can be used to estimate the AGN bolometric luminosity and can therefore be a proxy for the SMBH Eddington ratio, $\lambda_{\rm Edd}$=$L_{\rm bol}$/$L_{\rm Edd}$. We remark that there are several caveats to take into account with respect to this approach. In particular, there is observational evidence, particularly in Type I AGN, of a lack of correlation between variability in the X-ray continuum and variability in the bolometric luminosity. This suggests that the observed X-ray luminosity variability might be linked to changes in the X-ray corona, particularly on short timescales like those sampled here.

We use Equation~21 from \citet{marconi04} to convert the 2--10\,keV luminosities into bolometric luminosities. We then adopt the correlation between SMBH mass and stellar velocity dispersion reported in \citet{gaspari19} to compute the NGC 1358 SMBH mass, using the stellar velocity dispersion measured by \citet{nelson95}, $\sigma_*$=173$\pm$14\,km\,s$^{-1}$. The SMBH mass is therefore log(M$_{\rm BH}$/M$\odot$)=8.22$\pm$0.15: this value is slightly larger than the one reported by \citet{woo02} using the \citet{tremaine02} correlation (log(M$_{\rm BH}$/M$\odot$)=7.88). Based on these values, we find that the Eddington ratio might have only marginally varied from $\sim$4$\times$10$^{-3}$ in 2017 to $\sim$5$\times$10$^{-3}$ in 2021--2022. We note that the uncertainties on the Eddington ratio measurements can be fairly large, given that the intrinsic scatter in the M$_{\rm BH}$--$\sigma_*$ we used is $\epsilon$=0.36$\pm$0.02 \citep{gaspari19}. Indeed, when using the M$_{\rm BH}$--M$_*$ relation from \citet{suh20}, where M$_*$ is computed from $\sigma_*$=173$\pm$14\,km\,s$^{-1}$ using the \citet{zahid16} relation, we obtain log(M$_{\rm BH}$/M$\odot$)=7.51$\pm$0.75. In a scenario where log(M$_{\rm BH}$/M$\odot$)=7.51, the Eddington ratio of NGC 1358 would have been $\sim$2(4)$\times$10$^{-2}$ in 2017 (2021--2022).

Keeping in mind the above caveats, it is still helpful for the interested reader to discuss at least a qualitative physical interpretation of the retrieved obscuration, in particular in the currently accepted framework of self-regulated AGN feeding/feedback (e.g.~\citealt{gaspari20} for a review).Indeed, the AGN loop experiences a flickering alternation of feeding and feedback events on micro and macro scales over the several Gyr evolution. 
Specifically, higher obscuration phases are associated with stronger CCA rain \citep[e.g.,][]{gaspari13,gaspari17_cca}, in which the feeding-dominated stage is driven by condensing cool clouds that rain down toward the meso-- and ultimately the micro-scale, thus inducing higher \nhlos\ and lower luminosity (as found in Figure \ref{fig:correlation_nh_lx_cf} during 2017). Given the AGN self-regulation, such a process is expected to quickly trigger a feedback event (with higher AGN luminosities and lower \nhlos) as soon as CCA has driven a critical mass inflow near the SMBH horizon. 
Given the uncertainties associated with our measurements, a definitive answer will be achieved by extending the X-ray monitoring campaign.

\subsection{Modeling of the obscuring clouds geometry through l.o.s. column density variability}
Following \citet{risaliti02,risaliti05}, the distance between the obscuring clouds and the SMBH can be computed with the equation

\begin{equation}
    d_{\rm BH}=600\,t_{100}^2\,n_{10}^2\,N_{\rm H,24}^{-2}\,R_{\rm S}, 
\end{equation}
where t$_{100}$ is the variability time in units of 100\,ks, $n_{10}$ is the cloud density in units of 10$^{10}$\,cm$^{-3}$ and N$_{\rm H,24}$ is the column density of a cloud in units of 10$^{24}$\,cm$^{-2}$. As mentioned above, based on the correlation between SMBH mass and stellar velocity dispersion, we measured a SMBH mass log(M$_{\rm BH}$/M$\odot$)=8.22$\pm$0.15, so the Schwarzschild radius is $R_{\rm S}$=$\frac{2 G M_{BH}}{c^2}$=4.9$\times$10$^{13}$\,cm. We then work under the assumption that the variability observed between two consecutive observations is due to a single cloud having column density $N_{\rm H,24}$=|$N_{\rm H,Obs2}$-$N_{\rm H,Obs1}$|/10$^{24}$ passing between the observer and the X-ray emitter corona. Finally, it has been shown using a variety of methods that the X-ray corona size varies in the range $D$=3--15\,$R_{\rm S}$ \citep[e.g.,][]{mchardy05,fabian09,fabian15,chartas16,kamraj18}, so that the cloud density can be computed as $n$=$\frac{|N_{\rm H,Obs2}-N_{\rm H,Obs1}|}{D}$.

We report in Table~\ref{tab:cloud_dist} the range of SMBH-cloud distances we obtain using the \nhlos\ variability and time separation between observations for the five consecutive pair of observations taken between August, 2017, and February, 2022. For consistency with the ``clumpy torus scenario'' that is suggested by our data, we use the \nhlos\ values obtained using \uxcl: the results do not however change significantly when using the \borus\ or \myt\ l.o.s. column density measurements.

Sampling different time separations allows us to sample different SMBH-cloud distances and/or cloud sizes. For example, the tentative \nhlos\ variability observed between the \nus\ February 3, 2022 observation and the \xmm\ February 4, 2022 one, if real, would be caused by material located at only a few hundred gravitational radii from the SMBH. Notably, this is the scale of the accretion disk itself for a SMBH with M$_{\rm BH}\sim$10$^8$\,M$_\odot$, as measured through reverberation mapping \citep[e.g.,][]{jha22}. Time-scales of $\sim$15 days like the one between our January 2022 and February 2022 observations, instead, sample clouds which are located at distances typically associated with the broad line region and the obscuring torus, which is from $\sim$0.02\,pc (for a coronal size $D$=15\,$R_{\rm S}$) to $\sim$1\,pc (assuming $D$=3\,$R_{\rm S}$).  This is the so-called meso-scale for self-regulated AGN feeding/feedback (\citealt{gaspari20}), which is the crucial transitional regime linking the tiny SMBH physics to the macro properties of the host halo.  We also note that $\sim$10$^{-2}$\,pc is the scale of the dust sublimation radius (i.e., the lower limit on the dusty torus inner boundary) for an AGN with 2--10\,keV luminosity $\sim$10$^{43}$\lu\ \citep[see, e.g.,][]{netzer15}. We note, however, that the best-fit results for these last three epochs are consistent with a no-variability scenario.

Finally, observations taken a few months apart (such as our February 2021 and August 2021 ones, pair 2, or the latter and the January 2022 one, pair 3) probe distances of $\sim$1--10$^2$\,pc, once again the looser constraint being associated with the more compact coronal size. 
We also note that the four-year time separation between the 2017 Compton-thick observation and the August 2021 one prevents us from reliably locating the material responsible for the high obscuration reported in \citet{zhao19a}, or even determining if this high-obscuration status was due to a single cloud or to a combination of clouds randomly interjecting our line of sight.

\begin{table}
\renewcommand*{\arraystretch}{1.5}
\scalebox{0.69}{
  \begin{tabular}{ccccc}
       \hline
       \hline     
Obs. range  & t$_{100}$ & $\Delta N_{\rm H,24}$ & $d_{\rm BH,3RS}$ & $d_{\rm BH,15RS}$\\
MM/YYYY--MM/YYYY           & 100\,ks & 10$^{24}$\,cm$^{-2}$ & pc & pc\\
       \hline
08/2017--02/2021 & 1126.7 & 0.61$\pm$0.35 & 5.6$\times$10$^3$ & 224 \\
02/2021--08/2021 & 136.5  & 0.38$\pm$0.15 & 82.2 & 3.3 \\
08/2021--01/2022 & 148.6  & 0.17$\pm$0.10 & 97.5 & 3.9 \\
01/2022--02/2022 & 11.2   & $<$0.14 & 0.56 & 0.02 \\
02/2022--02/2022 & 0.9    & $<$0.13 & 3.3$\times$10$^{-3}$ & 1.3$\times$10$^{-4}$ \\
	\hline
	\vspace{0.02cm}
\end{tabular}
}
\caption{\normalsize Summary of the cloud properties inferred from each pair of X-ray observations, assuming that the \nhlos\ variability measured between two consecutive epochs is due to a single cloud. d$_{\rm BH}$ is the distance between the cloud and the SMBH; t$_{100}$ is the time difference between the two observations, in units of 100\,ks; $\Delta N_{\rm H,24}$ is the difference in l.o.s. Hydrogen column density, in units of 10$^{24}$\,cm$^{-2}$. Finally, $d_{\rm BH,3RS}$ and $d_{\rm BH,15RS}$ are the cloud-SMBH distances computed assuming a coronal size equal to 3 and 15\,$R_{\rm S}$, respectively.}
\label{tab:cloud_dist}
\end{table}

\section{Conclusions}\label{sec:conclusions}
In this paper, we presented the multi-epoch, \nus\ and \xmm\ 0.6--70\,keV monitoring campaign of NGC 1358 which took place between February 2021 and february 2022. The l.o.s. column density of the target was predicted to be highly variable based on the properties of the obscuring material: namely, a small covering factor and a large offset between the CT l.o.s. column density measured in August 2017 and the Compton-thin average torus column density. This evidence made the source an ideal candidate changing-look CT-AGN. We summarize the main results of this work.

\begin{enumerate}
    \item The selection method we proposed to select candidate changing-look CT-AGNs turned out to be highly effective. We find that in 2021--2022 the l.o.s. column density of the material surrounding the AGN in NGC 1358 decreased by a factor $\sim$3 with respect to the 2017 observation, and the source transitioned from Compton-thick to Compton-thin.
    This result opens the way for a more extended \nus--\xmm\ campaign to target the rest of the candidate changing-look CT-AGNs population and further characterize the properties of the obscuring material surrounding accreting SMBHs.
    \item We found NGC 1358 to be l.o.s. column density-variable over a wide range of time-scales: these results suggest that the obscuring material is distributed in clouds of different \nhlos\ located at distances from the accreting SMBH as small as a few hundreds of gravitational radii and as a large as tens to hundreds of pc (depending on the cloud sizes). In such a scenario, a clumpy torus model offers a more self-consistent explanation to the \nhlos\ variability than a uniform torus one.
    \item Multi-epoch X-ray observations with \nus\ and \xmm\ are, as of today, one of the most efficient methods to reliably measure geometrical properties of the torus such as its covering factor and clumpiness (i.e., difference between average and l.o.s. column density). The 3-epoch fit we performed led to a reduction on the fit parameters uncertainties, with respect to the single-epoch fits, which vary from 20 to 80\,\%.
    \item The high spectral data quality of the \nus\ and \xmm\ observations enables simultaneous measurements of \nhlos\ and 2--10\,keV intrinsic luminosity (and therefore derive an estimate of the SMBH Eddington ratio) in each of the epochs we analyzed. 
    \item The anti-correlation between column density and luminosity (Fig.~\ref{fig:correlation_nh_lx_cf}) can be understood in the framework of a self-regulated AGN feeding and feedback cycle driven via CCA raining clouds (e.g., \citealt{gaspari20}). 
    However, only by continuing to monitor with \nus\ and \xmm\ such a remarkable AGN, we will able to better constrain the current variability/obscuration trends and probe the detailed self-regulation in NGC\,1358.
\end{enumerate}

\section*{Acknowledgments}
We thank the referee for the useful suggestions, which helped us in improving the paper. We thank Mauro Dadina for granting us access to the Sbuccia cluster to compute the multi-epoch confidence contours.
This research has made use of the NuSTAR Data Analysis Software (NuSTARDAS) jointly developed by the ASI Space Science Data Center (SSDC, Italy) and the California Institute of Technology (Caltech, USA). 
SM acknowledges funding from the INAF ``Progetti di Ricerca di Rilevante Interesse Nazionale'' (PRIN), Bando 2019 (project: ``Piercing through the clouds: a multiwavelength study of obscured accretion in nearby supermassive black holes'').
MG acknowledges partial support by NASA Chandra GO9-20114X and HST GO-15890.020/023-A, and the {\it BlackHoleWeather} program. FS acknowledges funding from the INAF mainstream 2018 program ``Gas-DustPedia: A definitive view of the ISM in the Local Universe''. This work makes use of Matplotlib \citep{hunter07} and NumPy \citep{harris20}.

\bibliographystyle{apj}
\bibliography{NGC1358_monitoring}

\begin{appendix}
\section{Single-epoch spectral fits of the \nus\ and \xmm\ observations analyzed in this work}
In this Appendix we report the results of the single-epoch spectral fits performed on the \nus\ and \xmm\ observations taken between February, 2021, and February, 2022. We also report here the images of the best-fit models obtained using \myt, \borus\ and \uxcl\ for the four sets of observations analyzed in this paper.

\subsection{\xmm\ observation, 2021-02-25}
\begin{figure*}[ht]
\begin{minipage}{0.49\textwidth} 
 \centering 
 \includegraphics[width=0.76\textwidth,angle=-90]{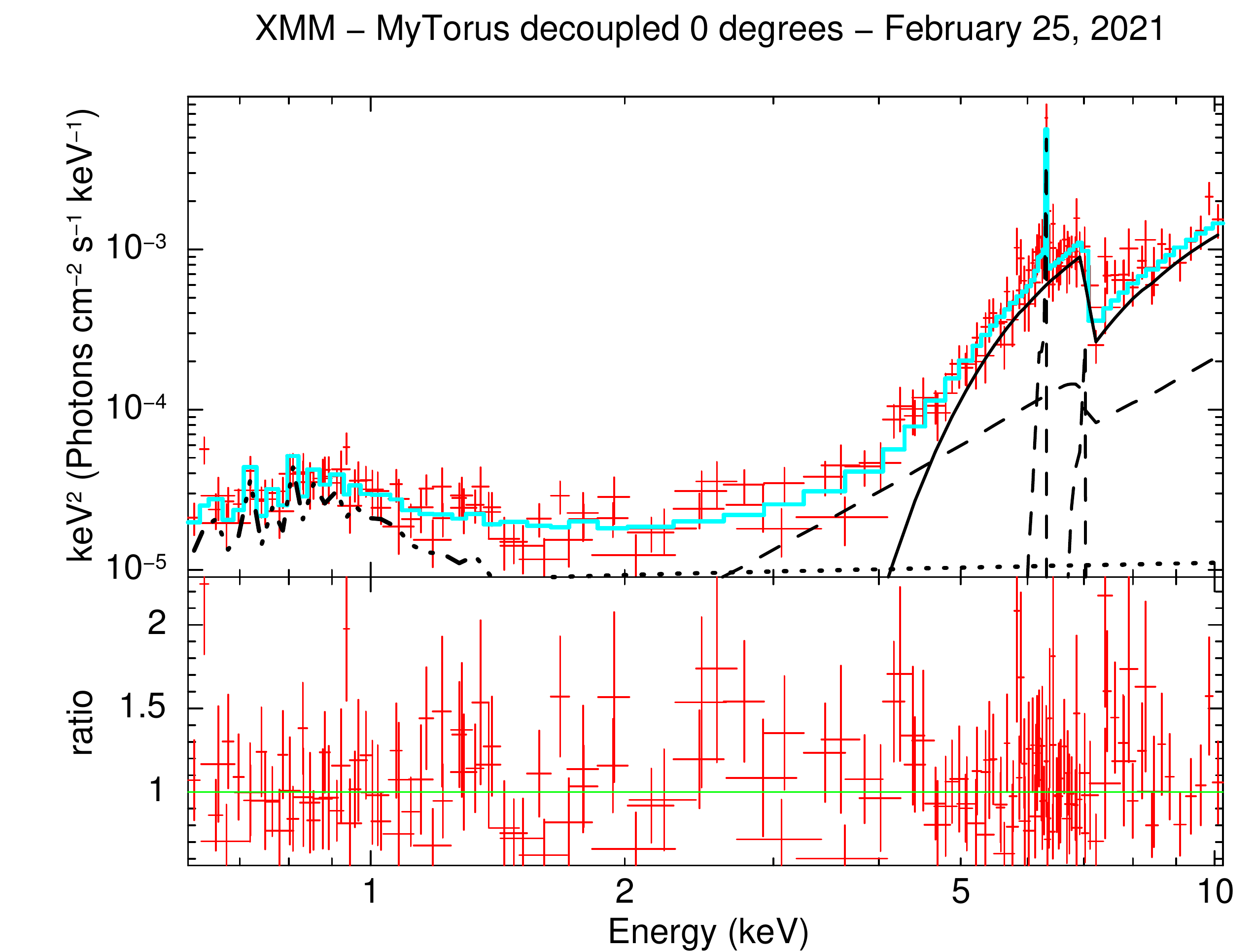} 
 \end{minipage} 
 \begin{minipage}{0.49\textwidth} 
 \centering 
 \includegraphics[width=0.76\textwidth,angle=-90]{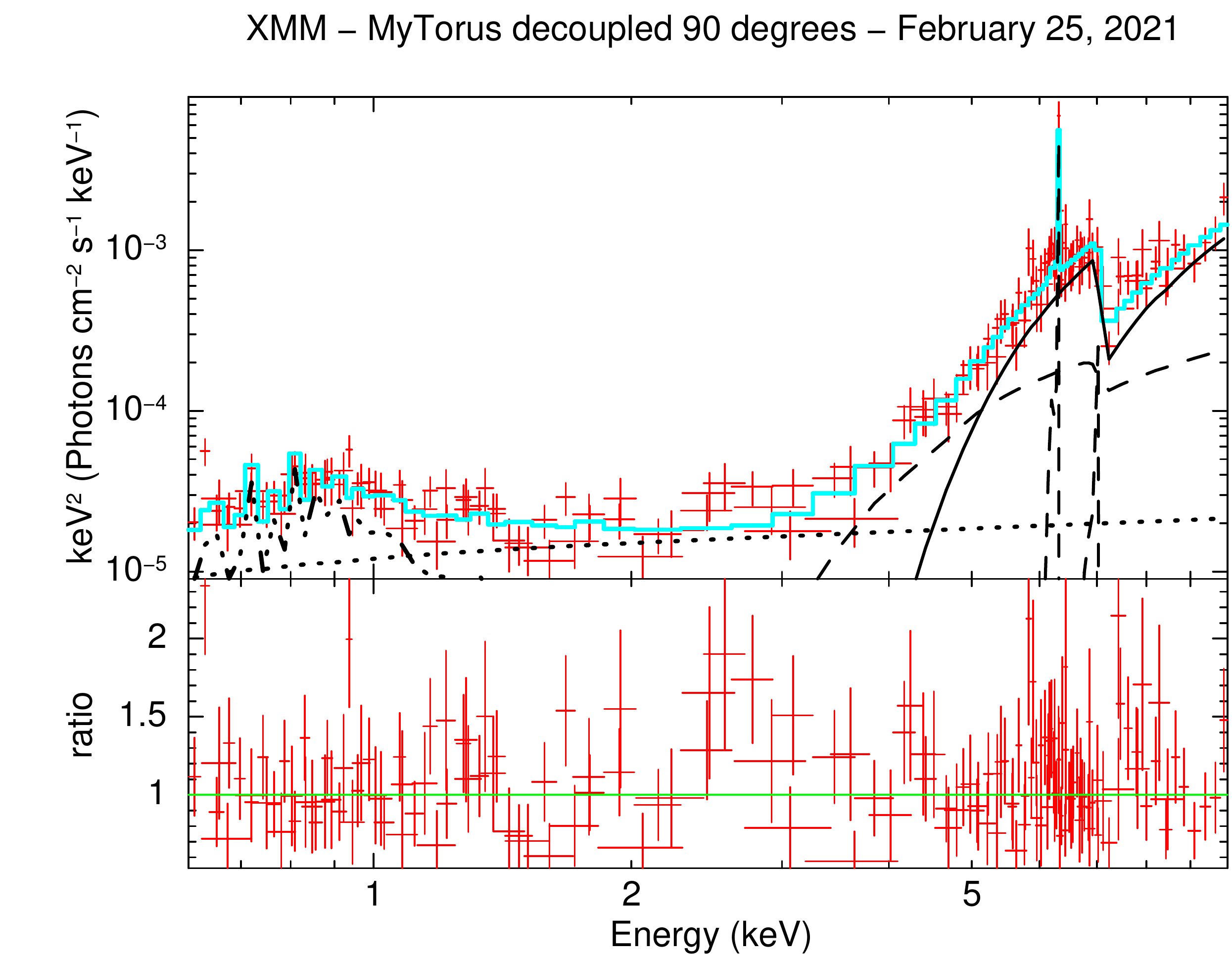} 
 \end{minipage} 
\begin{minipage}{.49\textwidth} 
 \centering 
 \includegraphics[width=0.76\textwidth,angle=-90]{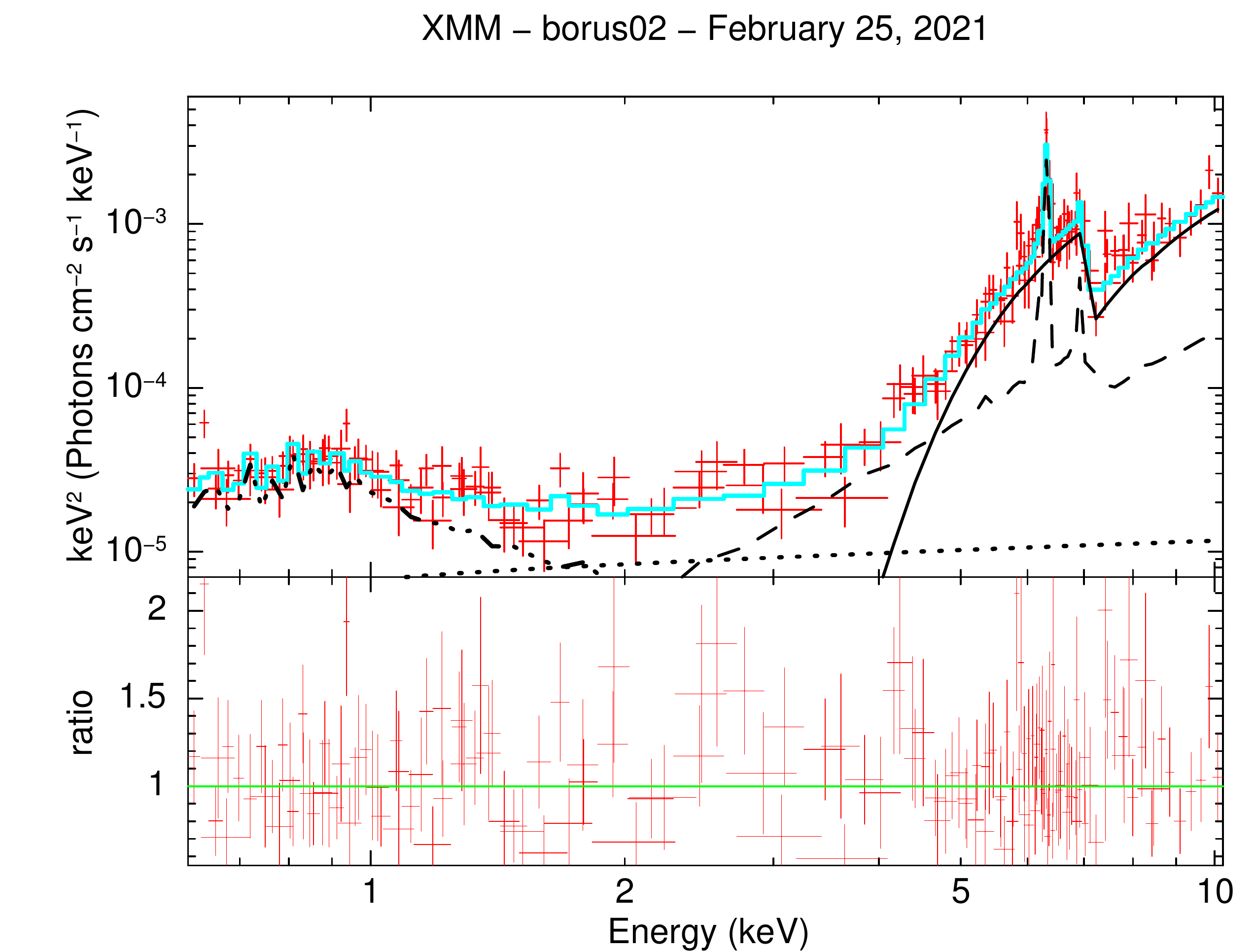} 
 \end{minipage} 
 \begin{minipage}{.49\textwidth} 
 \centering 
 \includegraphics[width=0.76\textwidth,angle=-90]{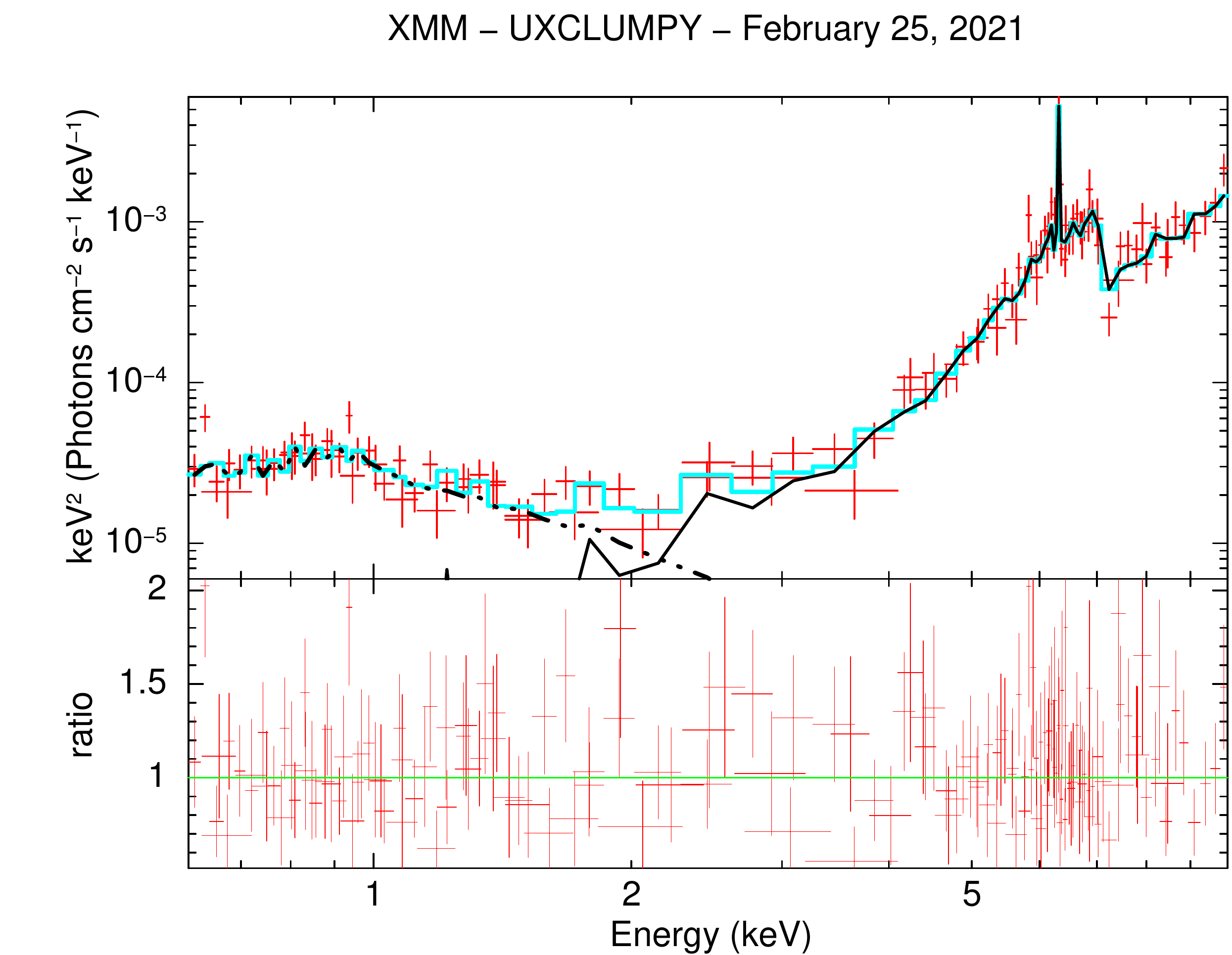} 
 \end{minipage} 
\caption{\normalsize 
Unfolded \xmm\ spectrum of the February 25, 2021 observation of NGC 1358. On the top panel we show the best fits obtained using \myt\ in its decoupled, $\theta$=0$\degree$ (left) and $\theta$=90$\degree$ configuration. In the bottom panel, we report the \borus\ (left) and \uxcl\ (right) best fit models. In all panels, the overall model is plotted as a solid cyan line, the absorbed main power law component is plotted as a solid black line, the reprocessed emission as a dashed black line, the scattered component as dashed black line, and the thermal \texttt{mekal} component as a dash-dotted black line.
}\label{fig:spec_20210225}
\end{figure*}

The first observation we analyze is the only one without simultaneous \nus\ data and was taken by \xmm\ on February 25, 2021. We report the results of our analysis in Table~\ref{tab:fit_2021} and the best-fit spectra obtained with \myt\ and \borus\ in Figure~\ref{fig:spec_20210225}. There is a general good agreement between the results obtained with \myt\ (either assuming $\theta_{S}$=90$\degree$ or $\theta_{S}$=0$\degree$), \borus\ and \uxcl. In particular, we measure a typical AGN photon index $\Gamma$=1.8--1.9 (although with fairly large uncertainties, $\Delta\Gamma\sim$0.4) and a line-of-sight column density around the Compton-thick threshold, \nhlos$\sim$10$^{24}$\,cm$^{-2}$.

The lack of \nus\ data prevented us from constraining other parameters, such as the average column density and the covering factor, when fitting the data with \myt\ in one of its two configurations or with \borus. This is due to the fact that variations of \nhtor\ or $f_c$ mostly affect the $>$5--10\,keV spectrum \citep[see, e.g., Figure~A1 in][]{zhao20}.
The \uxcl\ fit, which is also the one with the best reduced $\chi^2$ ($\chi$/d.o.f.=123.7/148; as a reference, the \borus\ fit has $\chi$/d.o.f.=130.3/149), supports instead the ``low covering factor'' scenario. Indeed, we measure a cloud vertical dispersion TOR$\sigma$=9.0$_{-8.9}^{+7.0\circ}$ and a covering factor of the inner ring CTK$<$0.17.

\begin{table*}
\renewcommand*{\arraystretch}{1.5}
\scalebox{0.75}{
\centering
  \begin{tabular}{ccccc|cccc}
       \hline
       \hline     
       Date & \multicolumn{4}{c|}{2021-02-25 (\xmm\ only)} & \multicolumn{4}{c}{2021-08-02 (\nus\ + \xmm)} \\
       Model & \myt & \myt & \borus & \uxcl & \myt & \myt & \borus & \uxcl \\
       & decoupled, 0$\degree$ & decoupled,  90$\degree$ & & & decoupled, 0$\degree$ & decoupled, 90$\degree$  \\
       \hline
       $\chi^2$/dof & 132.1/150 & 136.6/150 & 130.3/149 & 123.7/148 & 660.5/599 & 662.0/599 & 659.9/598 & 658.1/598\\
       $C_{\rm NuS}$ & N/A & N/A & N/A & N/A & 1.13$_{-0.06}^{+0.07}$ & 1.12$_{-0.06}^{+0.07}$ & 1.13$_{-0.06}^{+0.07}$ & 1.11$_{-0.06}^{+0.07}$\\
       $\Gamma$ & 1.90$_{-0.50*}^{+0.29}$ & 1.79$_{-0.39*}^{+0.35}$ & 1.75$_{-0.35*}^{+0.47}$ & 1.94$_{-0.39}^{+0.25}$ & 1.61$_{-0.10}^{+0.11}$ & 1.60$_{-0.10}^{+0.10}$ & 1.60$_{-0.09}^{+0.07}$ & 1.63$_{-0.07}^{+0.09}$\\
       $norm$ 10$^{-2}$ & 0.59$_{-0.46}^{+0.74}$ & 0.64$_{-0.49}^{+1.00}$ & 0.40$_{-0.28}^{+1.16}$ & 0.80$_{-0.49}^{+0.78}$ & 0.51$_{-0.13}^{+0.22}$ & 0.56$_{-0.19}^{+0.28}$ & 0.45$_{-0.10}^{+0.14}$ & 0.68$_{-0.18}^{+0.29}$\\
       \nhlos\ [10$^{24}$\,cm$^{-2}$] & 0.94$_{-0.20}^{+0.14}$ & 1.10$_{-0.20}^{+0.20}$ & 0.92$_{-0.10}^{+0.17}$ & 1.09$_{-0.05}^{+0.35}$ & 0.86$_{-0.05}^{+0.06}$ & 0.94$_{-0.09}^{+0.09}$ & 0.83$_{-0.05}^{+0.06}$ & 0.91$_{-0.08}^{+0.09}$\\
       \nhtor\ [10$^{24}$\,cm$^{-2}$] & 10.00$_{-*}^{+*}$ & 0.28$_{-0.08}^{+0.10}$ & 10.05$_{-9.62}^{+21.57*}$ & ... & 1.98$_{-1.50}^{+1.39}$ & 0.37$_{-0.10}^{+0.20}$ & 1.74$_{-0.59}^{+0.71}$ & ...\\
       A$_S$ & 0.59$_{-0.21}^{+0.36}$ & 0.47$_{-0.19}^{+0.54}$ & ... & ... & 0.24$_{-0.08}^{+0.07}$ & 0.40$_{-0.11}^{+1.38}$ & ... & ...\\
       $f_c$ & ...  &... & 0.43$_{-0.33*}^{+0.56}$ & ... & ...  &... & 0.28$_{-0.13}^{+0.07}$ & ...\\
       $\theta\rm _{i}$ [$^\circ$] & ... & ... & 67.3$_{-42.0}^{+17.2}$ & 65.9$_{-44.3}^{+17.0}$ & ...  &... & 78.7$_{-5.6}^{+4.2}$ & 87.8$_{-9.8}^{+2.2*}$\\
       TOR$\sigma$ [$^\circ$]  & ... & ... & ... & 9.0$_{-8.9}^{+7.0}$ & ...  &... & ... & 10.3$_{-3.9}^{+8.1}$\\
       CTK  & ... & ... & ... & $<$0.17 & ...  &... & ... & $<$0.22\\
       $f_s$ 10$^{-2}$ & 0.15$_{-0.12}^{+0.46}$ &  0.21$_{-0.11}^{+0.34}$ & 0.19$_{-0.16}^{+0.47}$ & $<$0.28 & 0.13$_{-0.05}^{+0.06}$ & 0.16$_{-0.06}^{+0.15}$ & 0.15$_{-0.09}^{+0.10}$ & 0.12$_{-0.12*}^{+0.07}$\\
       $kT$ [keV] & 0.64$_{-0.11}^{+0.09}$ & 0.63$_{-0.12}^{+0.08}$ & 0.68$_{-0.10}^{+0.11}$ & 0.74$_{-0.11}^{+0.08}$ & 0.59$_{-0.10}^{+0.06}$ & 0.58$_{-0.10}^{+0.06}$ & 0.60$_{-0.10}^{+0.06}$ & 0.62$_{-0.08}^{+0.07}$\\
       $Z$/$Z_\odot$ & 0.05$_{-0.03}^{+0.13}$ & 0.08$_{-0.06}^{+3.31}$ & 0.04$_{-0.02}^{+0.06}$ & 0.02$_{-0.01}^{+0.05}$ & 0.18$_{-0.12}^{+0.32}$ & 0.44$_{-0.37}^{+1.93}$ & 0.14$_{-0.10}^{+1.18}$ & 0.08$_{-0.05}^{+0.10}$\\
       log(F$_{2-10}$) [\flu]  & --12.06$_{-0.74}^{+0.02}$ & --12.05$_{-0.29}^{+0.01}$ & --12.05$_{-0.36}^{+0.23}$ & --12.05$_{-0.36}^{+0.01}$ & --11.83$_{-0.10}^{+0.01}$ & --11.83$_{-0.07}^{+0.01}$ & --11.83$_{-0.08}^{+0.03}$ & --11.83$_{-0.33}^{+0.04}$\\
       log(F$_{10-40}$) [\flu]  & N/A & N/A & N/A & N/A & --10.68$_{-0.08}^{+0.01}$ & --10.68$_{-0.05}^{+0.01}$ & --10.68$_{-0.10}^{+0.01}$ & --10.68$_{-0.49}^{+0.03}$\\
       log(L$_{2-10}$) [\lu]  & 42.84$_{-0.05}^{+0.05}$ & 42.95$_{-0.05}^{+0.05}$ & 42.80$_{-0.11}^{+0.10}$ & 42.85$_{-0.25}^{+0.12}$ & 42.97$_{-0.05}^{+0.05}$ & 43.02$_{-0.05}^{+0.04}$ & 42.92$_{-0.09}^{+0.07}$ & 42.99$_{-0.13}^{+0.16}$ \\
       log(L$_{10-40}$) [\lu]  & N/A & N/A & N/A & N/A & 43.17$_{-0.03}^{+0.04}$ & 43.22$_{-0.03}^{+0.04}$ & 43.10$_{-0.05}^{+0.06}$ & 43.16$_{-0.10}^{+0.16}$ \\
       \hline
	\hline
	\vspace{0.02cm}
\end{tabular}
}
\caption{\normalsize Summary of the best-fit results for the spectral fit of the observations taken on February 25, 2021 (\xmm\ only) and on August 02, 2021 (\nus\ and \xmm). $C_{\rm NuS}$ is the cross-normalization between \NuSTAR\ and \XMM. $\Gamma$ and $norm$ are the main power law component photon index and normalization at 1\,keV in photons\,keV$^{-1}$\,cm$^{-2}$\,s$^{-1}$, respectively.
\nhlos\ and \nhtor\ are the line-of-sight and average column density, respectively, in units of cm$^{-2}$. 
A$_S$ is the relative intensity of the reprocessed component with respect to the main one in \myt. 
$f_c$ is the covering factor of the obscuring material as computed by \borus, $f_c$ = cos($\theta_{\rm T}$), where $\theta\rm _{T}$ is the angle (in degrees) between the axis of the torus and the edge of torus. $\theta\rm _{i}$ is the angle (in degrees) between the observer and the torus axis. 
In \uxcl, TOR$\sigma$ is the angular width of the cloud population and CTK is the covering factor of inner Compton-thick ring of clouds. $kT$ and $Z$ are the temperature (in keV) and metallicity (in units of $Z_\odot$) of the thermal \texttt{mekal} component. 
F$_{2-10}$ and F$_{10-40}$ are the observed fluxes in units of erg\,cm$^{-2}$\,s$^{-1}$ in the 2--10\,keV and 10--40\,keV, respectively. 
L$_{2-10}$ and L$_{10-40}$ are the intrinsic luminosities in units of erg\,s$^{-1}$ in the 2--10\,keV and 10--40\,keV, respectively. Upper and lower limits flagged with $^*$ have reached the parameter boundary.}
\label{tab:fit_2021}
\end{table*}

\subsection{\nus\ and \xmm\ observations, 2021-08-02}
\begin{figure*}[ht]
\begin{minipage}{0.49\textwidth} 
 \centering 
 \includegraphics[width=0.76\textwidth,angle=-90]{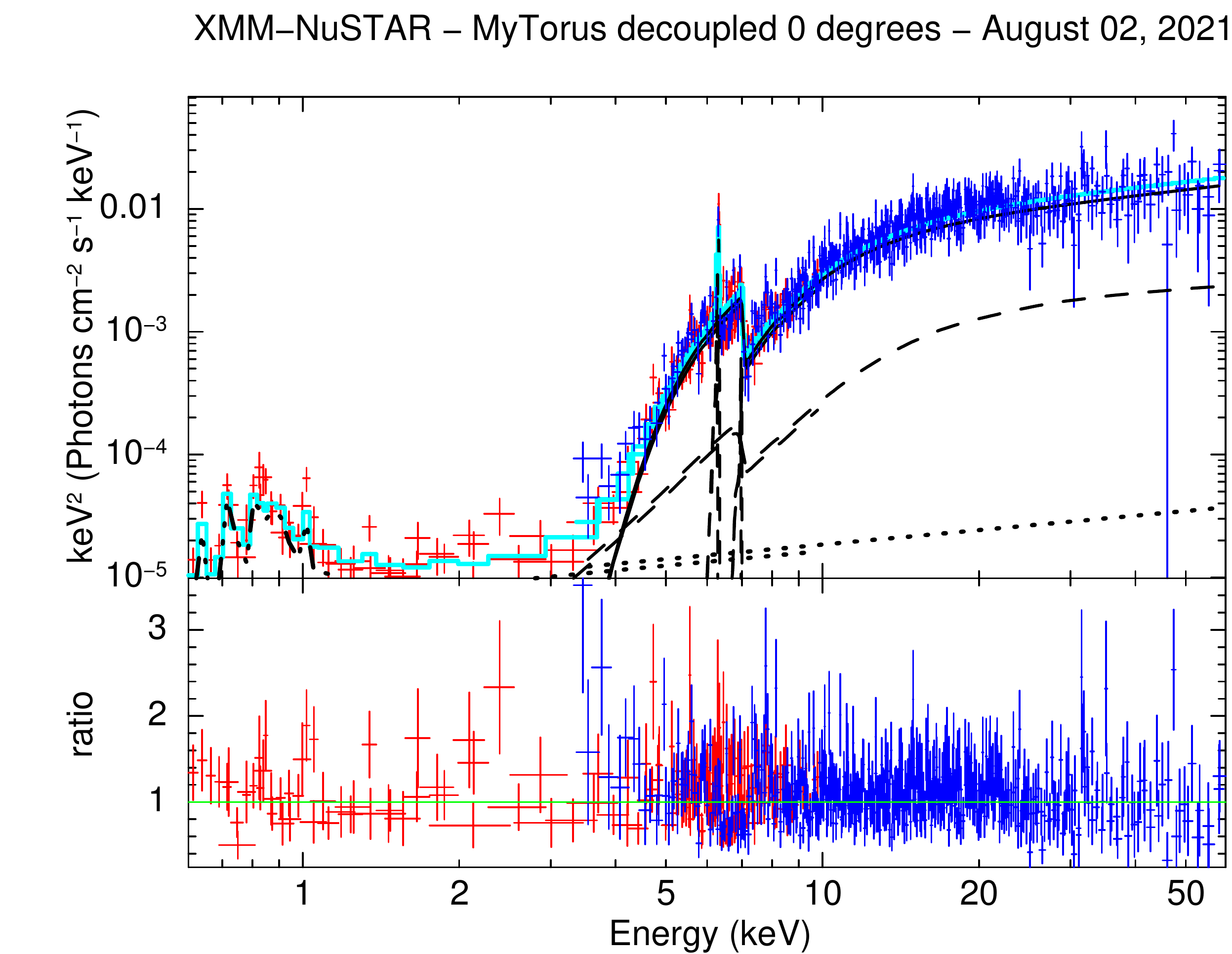} 
 \end{minipage} 
 \begin{minipage}{0.49\textwidth} 
 \centering 
 \includegraphics[width=0.76\textwidth,angle=-90]{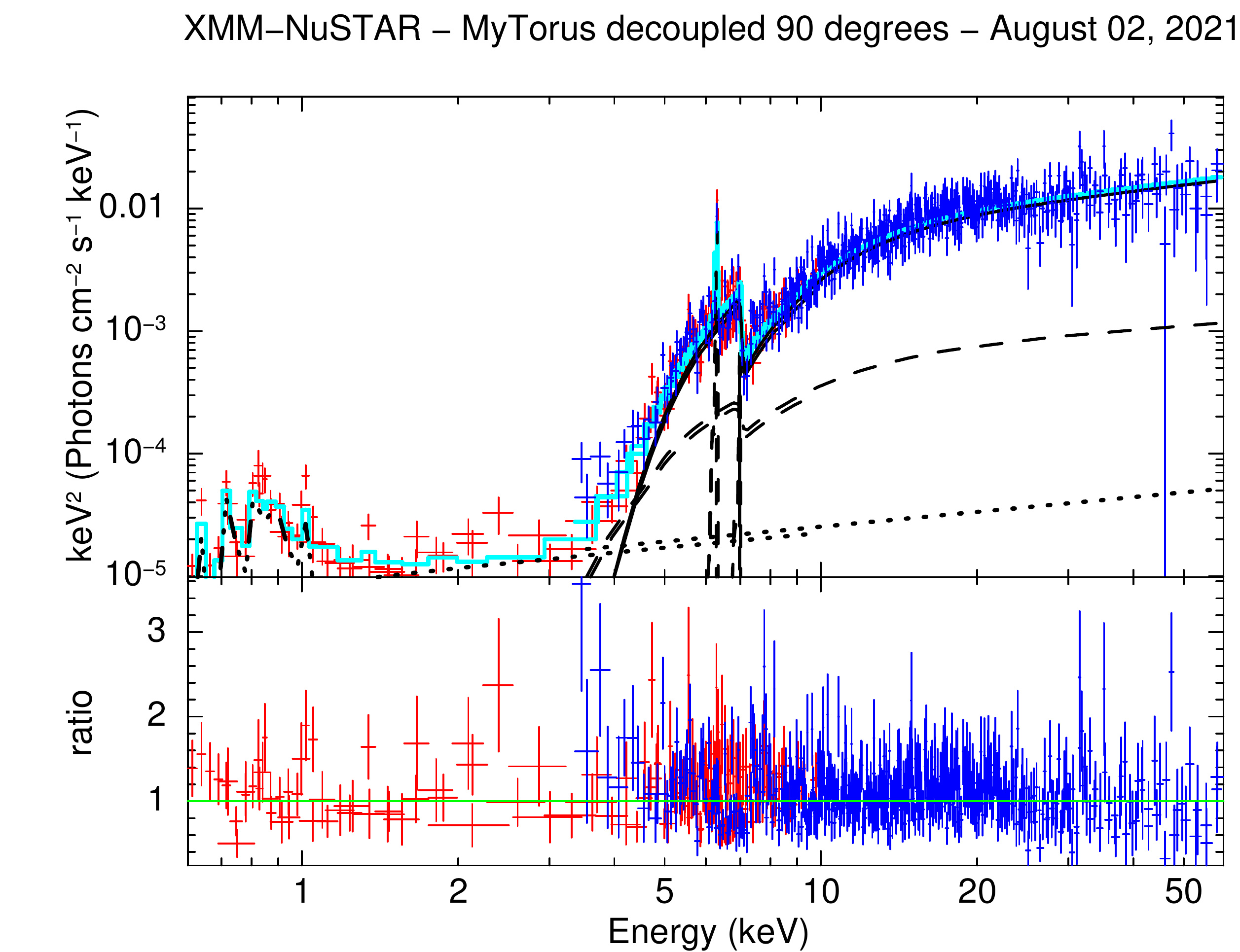} 
 \end{minipage} 
\begin{minipage}{.49\textwidth} 
 \centering 
 \includegraphics[width=0.76\textwidth,angle=-90]{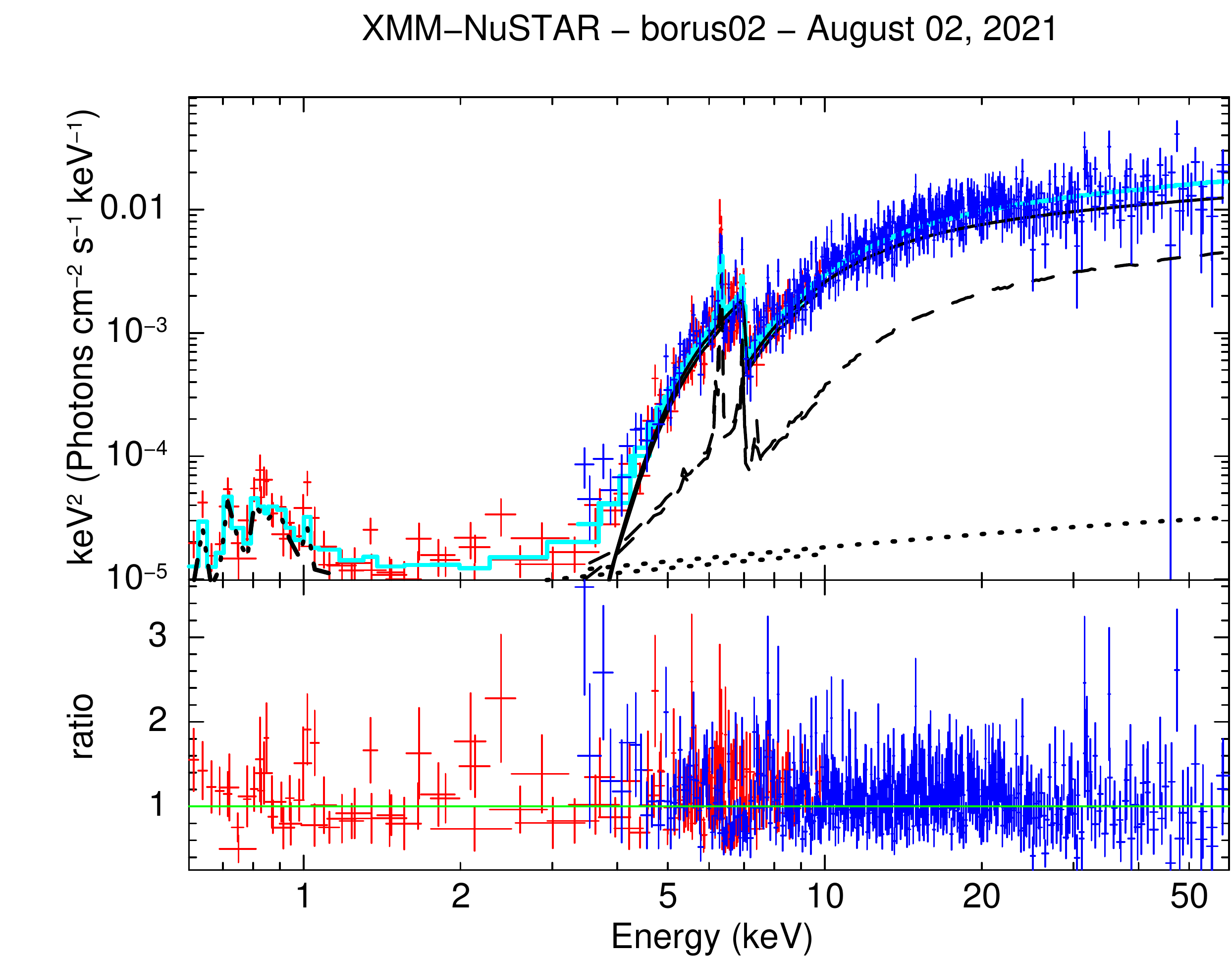} 
 \end{minipage} 
  \begin{minipage}{.49\textwidth} 
 \centering 
 \includegraphics[width=0.76\textwidth,angle=-90]{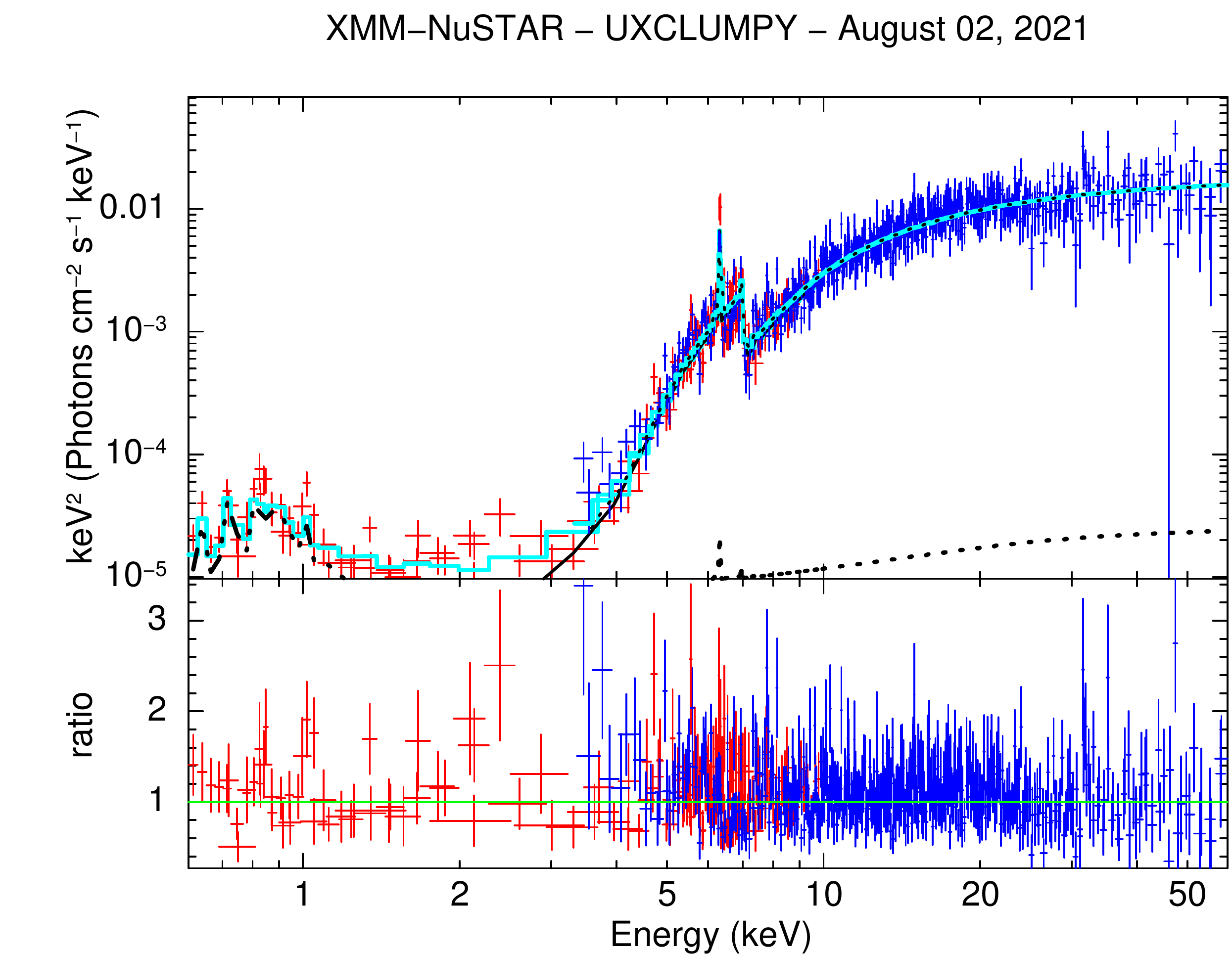} 
 \end{minipage} 
\caption{\normalsize 
Unfolded \xmm\ (red) and \nus\ (blue) spectra of the August 02, 2021 observation of NGC 1358. On the top panel we show the best fits obtained using \myt\ in its decoupled, $\theta$=0$\degree$ (left) and $\theta$=90$\degree$ configuration. In the bottom panel, we report the \borus\ (left) and \uxcl\ (right) best fit models. In all panels, the overall model is plotted as a solid cyan line, the absorbed main power law component is plotted as a solid black line, the reprocessed emission as a dashed black line, the scattered component as dashed black line, and the thermal \texttt{mekal} component as a dash-dotted black line.
}\label{fig:spec_20210802}
\end{figure*}

The second observation analyzed in this work was taken quasi-simultaneously by \nus\ and \xmm\ on August 02, 2021. We report the results in Table~\ref{tab:fit_2021}, while the best-fit spectra are shown in Figure~\ref{fig:spec_20210802}. 
There is a general excellent agreement between the results obtained with \myt\ (either assuming $\theta_{S}$=90$\degree$ or $\theta_{S}$=0$\degree$), \borus\ and \uxcl, and all four models have almost identical reduced $\chi^2$. We measure a typical AGN photon index $\Gamma\sim$1.6$\pm$0.1 (slightly harder, although consistent within the uncertainties, than the one measured in the February 2021 observation) and a line-of-sight column density just below the Compton-thick threshold, \nhlos$\sim$8--9$\times$10$^{23}$\,cm$^{-2}$.

As for the other properties of the obscuring material, the average column density measured using \myt\ decoupled in its ``0$\degree$'' configuration is consistent with the one we measure with \borus, log\nhtor$\sim$24.2--24.3. Such a value is slightly larger than the one found by \citet[][log\nhtor$\sim$23.8]{zhao19a}, although the 2017 measurement is in agreement with the August 2021 one at the 90\,\% confidence level. We also note that the \nhtor\ obtained using \myt\ decoupled in its ``90$\degree$'' configuration is much lower (log\nhtor=23.6$\pm$0.2) and in even closer agreement with the \citet{zhao19a} one. 
Finally, the covering factor we measure using \borus\ ($f_c$=0.28$_{-0.13}^{+0.08}$) is in agreement with the one reported by \citet[][$f_c<$0.17]{zhao19a}, and a similarly low covering factor is found when using \uxcl\ (TOR$\sigma$=10.3$_{-3.9}^{+8.1\circ}$ and CTK$<$0.22).

\subsection{\nus\ and \xmm\ observations, 2022-01-21}
\begin{figure*}[ht]
\begin{minipage}{0.49\textwidth} 
 \centering 
 \includegraphics[width=0.75\textwidth,angle=-90]{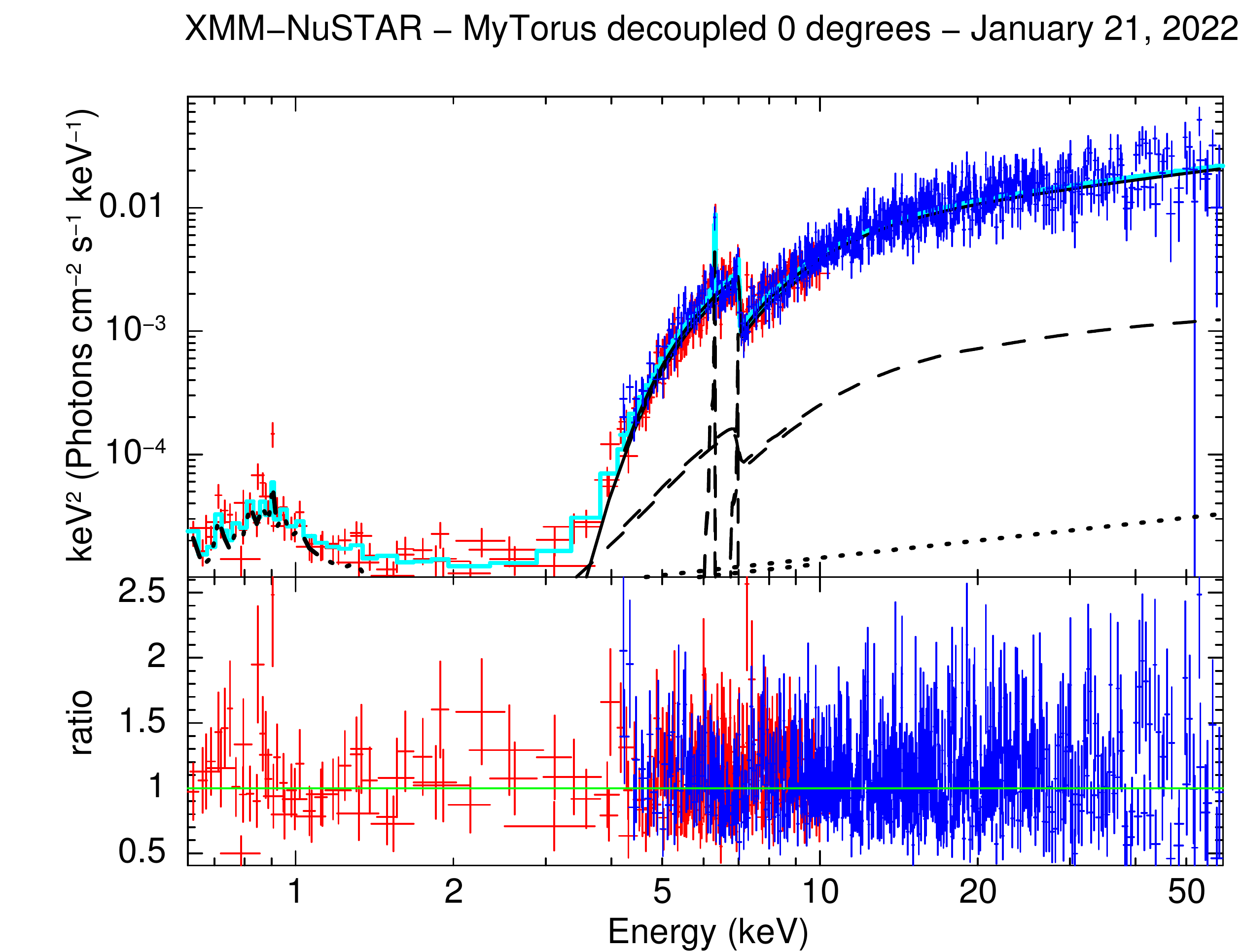} 
 \end{minipage} 
 \begin{minipage}{0.49\textwidth} 
 \centering 
 \includegraphics[width=0.75\textwidth,angle=-90]{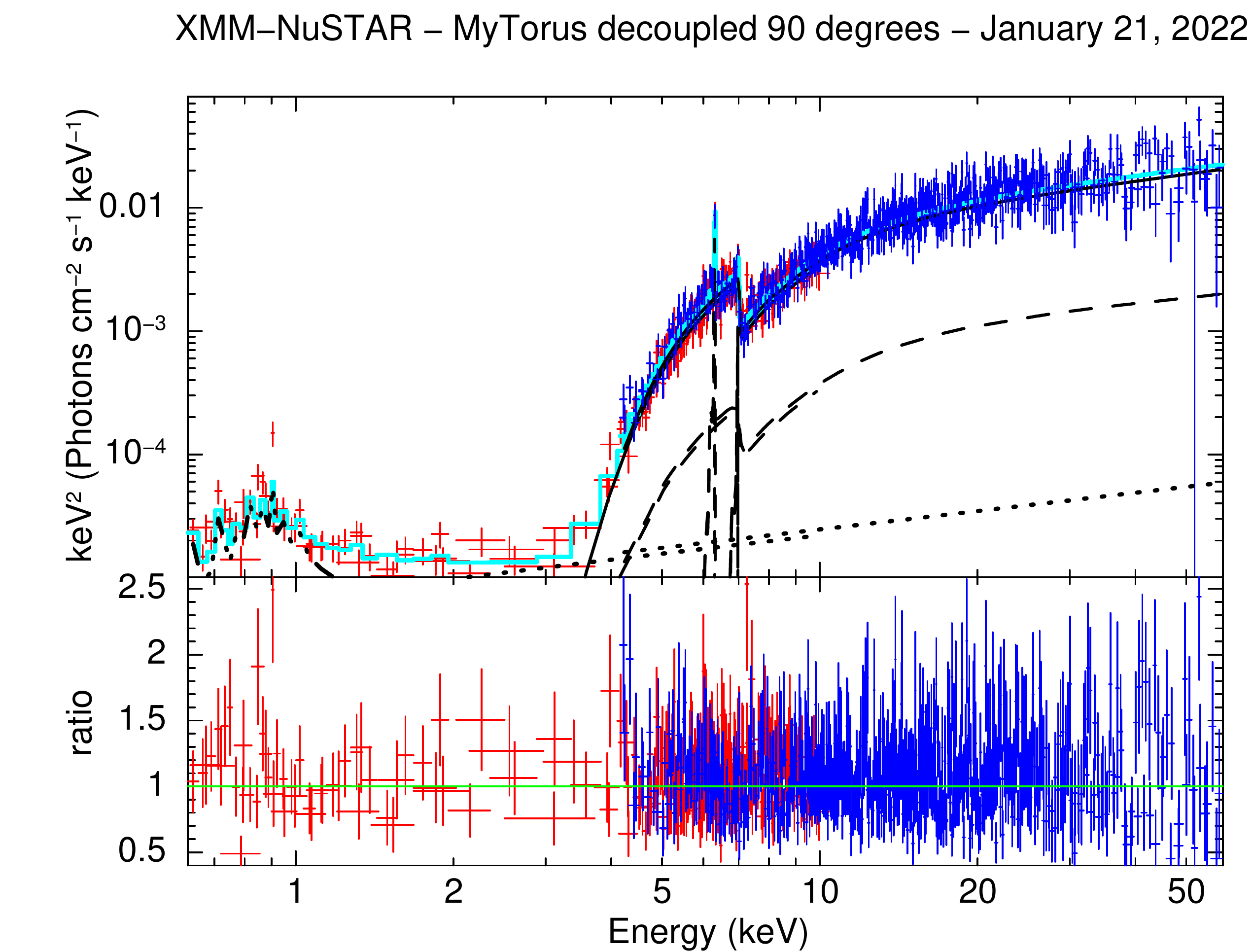} 
 \end{minipage} 
\begin{minipage}{.49\textwidth} 
 \centering 
 \includegraphics[width=0.75\textwidth,angle=-90]{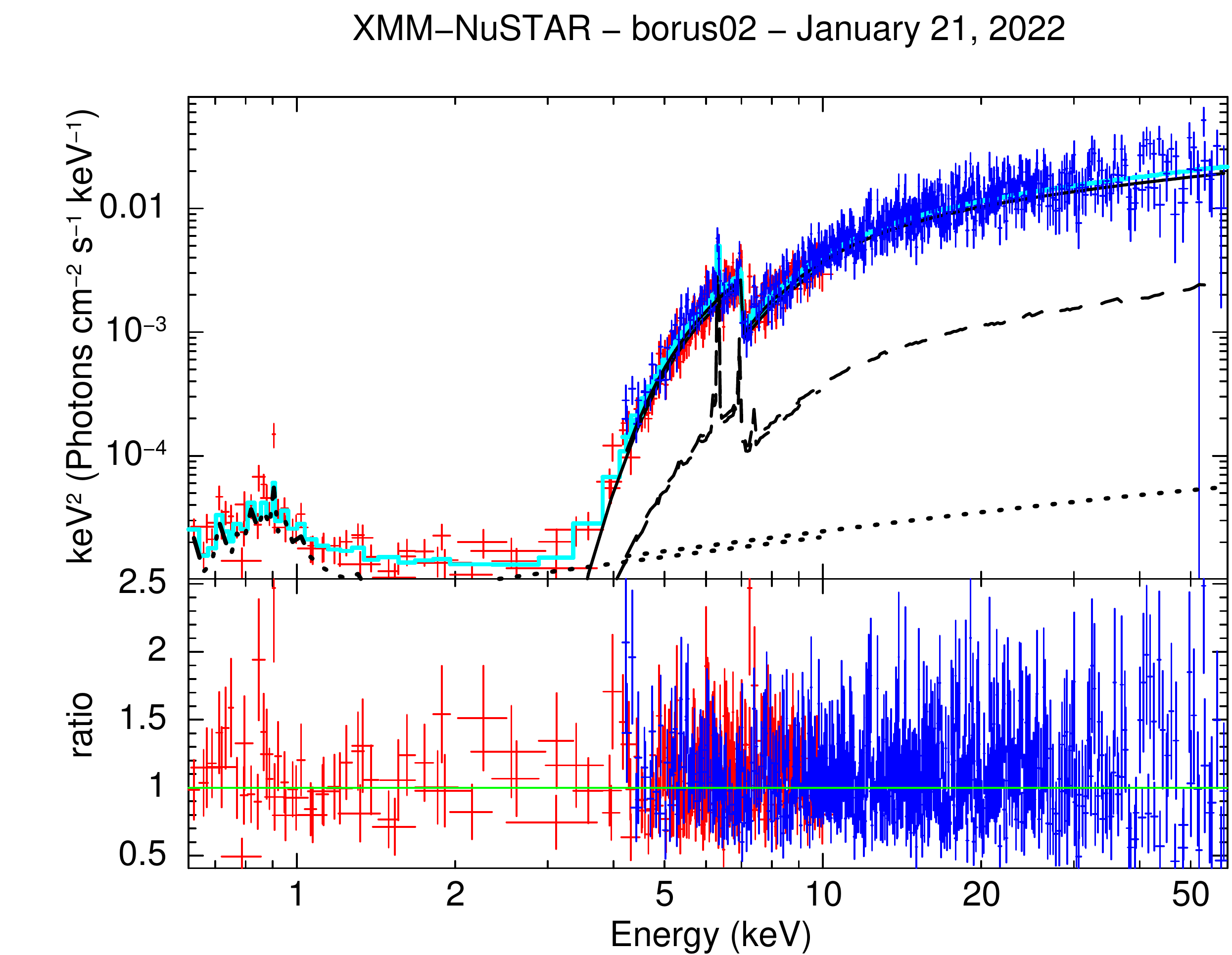} 
 \end{minipage} 
  \begin{minipage}{.49\textwidth} 
 \centering 
 \includegraphics[width=0.75\textwidth,angle=-90]{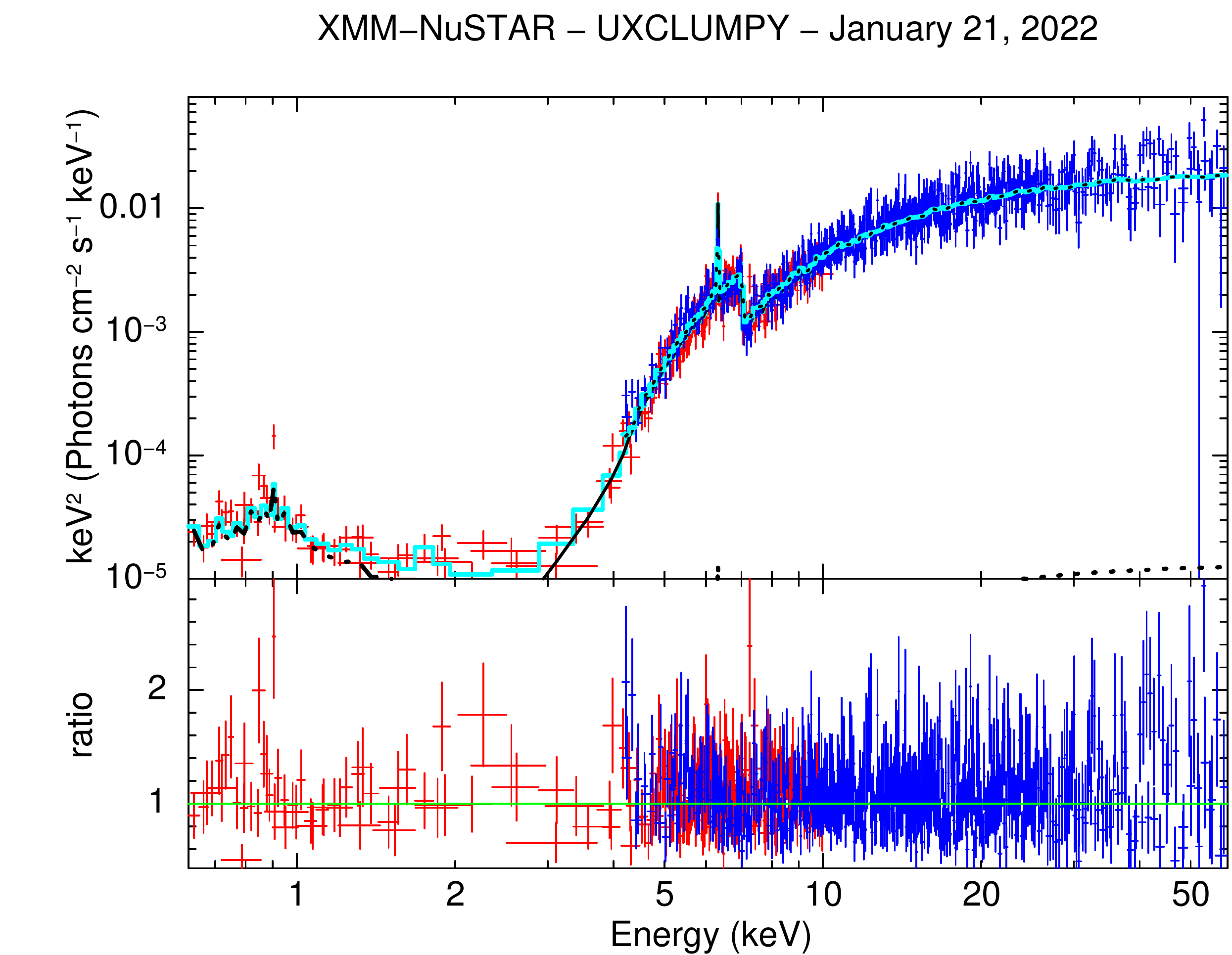} 
 \end{minipage} 
\caption{\normalsize 
Unfolded \xmm\ (red) and \nus\ (blue) spectra of the January 21, 2022 observation of NGC 1358. On the top panel we show the best fits obtained using \myt\ in its decoupled, $\theta$=0$\degree$ (left) and $\theta$=90$\degree$ configuration. In the bottom panel, we report the \borus\ (left) and \uxcl\ (right) best fit models. In all panels, the overall model is plotted as a solid cyan line, the absorbed main power law component is plotted as a solid black line, the reprocessed emission as a dashed black line, the scattered component as dashed black line, and the thermal \texttt{mekal} component as a dash-dotted black line.
}\label{fig:spec_20220121}
\end{figure*}

The third set of observations of NGC 1358 was performed quasi-simultaneously by \nus\ and \xmm\ on January 21, 2022. The results of the spectral analysis are reported in Table~\ref{tab:fit_2022}, while the spectra are shown in Figure~\ref{fig:spec_20220121}. The line-of-sight column density further decreased by $\sim$20\,\% with respect to the observations performed 5.5 months earlier, being $\sim$7$\times$10$^{23}$\,cm$^{-2}$; such a result is model-independent.

The agreement between the four models extends to all the other spectral parameters. In particular, both \borus\ ($f_c$=0.17$_{-0.04}^{+0.05}$) and \uxcl\ (TOR$\sigma$<5.7$^\circ$ and CTK=0.26$_{-0.14}^{+0.03}$) once again favor a low--covering factor scenario, in agreement with our previous findings. The average torus column density is found to be log\nhtor$\sim$23.8 using both \borus\ and \myt\ decoupled in its 90$\degree$ configuration. As mentioned before, such a value is the same reported in \citep[][]{zhao19a}: furthermore, in this observation \nhlos\ is basically identical to \nhtor. 

\subsection{\nus\ and \xmm\ observations, 2022-02-03/04}\label{sec:feb22_obs}
The fourth and final set of \nus\ and \xmm\ observations was taken on February 3 and 4, 2022. As reported in Table~\ref{tab:obs_summary}, the \nus\ observation ended $\sim$13\,hours before the start of the \xmm\ observation. This small temporal offset allowed us to search for short-time scale variability. We first assumed that no \nhlos\ variability occurred between the two observations, and only allowed for flux variability which we parameterize with the usual cross-instrument constant $C_{\rm Nus}$. For all models, we find  $C_{\rm Nus}$=1.31--1.32$\pm$0.07; as a comparison, in both the August 2021 and January 2022 observations we measure a cross-normalization $C_{\rm Nus}\sim$1.1. For this reason, we decided to leave \nhlos\ free to vary between the \nus\ and \xmm\ observations: with this new model, we obtain a cross-normalization $C_{\rm Nus}\sim$1.1, in close agreement with the values obtained in the other epochs. Furthermore, all models favor a scenario where \nhlos\ increased by 4--9$\times$10$^{22}$\,cm$^{-2}$ between the February 3 \nus\ observation and the \xmm\ observation taken half a day later. We also note, however, that the two \nhlos\ values are always in agreement within their 90\,\% confidence uncertainties.

Regardless of the model we used to perform the spectral fit, the \nhlos\ variability scenario was always statistically favored with respect to the normalization--luminosity variability one, therefore we report in Table~\ref{tab:fit_2022} the results obtained with the double--\nhlos\ model. As mentioned above, there is a tentative evidence for an increase in \nhlos\ within the two observations, and with respect to the January observation, particularly in the \myt\ and \borus\ models ($\Delta$\nhlos$\sim$10$^{23}$\,cm$^{-2}$), while the discrepancy is milder when fitting the data with \uxcl.

As for the average torus properties, all models support a low--$f_c$ scenario with an average column density \nhtor$\sim$2--3\,$\times$10$^{23}$\,cm$^{-2}$.

\begin{table*}
\renewcommand*{\arraystretch}{1.5}
\scalebox{0.75}{
\centering
  \begin{tabular}{ccccc|cccc}
       \hline
       \hline     
       Date & \multicolumn{4}{c|}{2022-01-21 (\nus\ + \xmm)} & \multicolumn{4}{c}{2022-02-04 (\nus\ + \xmm)} \\
       Model & \myt & \myt & \borus & \uxcl & \myt & \myt & \borus & \uxcl \\
       & decoupled, 0$\degree$ & decoupled,  90$\degree$ & & & decoupled, 0$\degree$ & decoupled, 90$\degree$ & & \\
       \hline
       $\chi^2$/dof & 796.9/792 & 796.2/792 & 795.1/792 & 795.6/792 & 634.7/650 & 634.9/650 & 631.8/650 & 634.6/650 \\
       $C_{\rm NuS}$ & 1.12$_{-0.05}^{+0.05}$ & 1.11$_{-0.05}^{+0.05}$ & 1.11$_{-0.03}^{+0.04}$ & 1.11$_{-0.05}^{+0.05}$ & 1.14$_{-0.22}^{+0.27}$ & 1.10$_{-0.25}^{+0.28}$ & 1.07$_{-0.23}^{+0.04}$ & 1.20$_{-0.17}^{+0.21}$ \\
       $\Gamma$ & 1.53$_{-0.09}^{+0.07}$ & 1.51$_{-0.06}^{+0.07}$ & 1.44$_{-0.04*}^{+0.09}$ & 1.59$_{-0.04}^{+0.08}$ & 1.50$_{-0.08}^{+0.08}$ & 1.51$_{-0.08}^{+0.08}$ & 1.43$_{-0.03*}^{+0.08}$ & 1.55$_{-0.08}^{+0.05}$ \\
       $norm$ 10$^{-2}$ & 0.46$_{-0.09}^{+0.11}$ & 0.40$_{-0.08}^{+0.12}$ & 0.34$_{-0.06}^{+0.02}$ & 0.64$_{-0.09}^{+0.15}$ & 0.42$_{-0.11}^{+0.14}$  & 0.44$_{-0.12}^{+0.17}$ & 0.36$_{-0.09}^{+0.01}$ & 0.46$_{-0.09}^{+0.09}$ \\ 
       \nhlos\ J [10$^{24}$\,cm$^{-2}$] & 0.69$_{-0.04}^{+0.04}$ & 0.67$_{-0.04}^{+0.06}$ & 0.67$_{-0.04}^{+0.04}$ & 0.70$_{-0.03}^{+0.03}$ & ... & ... & ... & ... \\
       \nhlos\ X [10$^{24}$\,cm$^{-2}$] & ... & ... & ... & ... & 0.79$_{-0.07}^{+0.08}$  & 0.83$_{-0.08}^{+0.09}$ & 0.81$_{-0.08}^{+0.04}$ & 0.73$_{-0.05}^{+0.04} $\\
       \nhlos\ N [10$^{24}$\,cm$^{-2}$] & ... & ... & ... & ... & 0.73$_{-0.06}^{+0.06}$ & 0.75$_{-0.07}^{+0.07}$ & 0.72$_{-0.05}^{+0.02}$ & 0.69$_{-0.03}^{+0.04}$ \\
       \nhtor\ [10$^{24}$\,cm$^{-2}$]  & 1.00$_{-0.52}^{+1.51}$ & 0.60$_{-0.28}^{+0.50}$  & 0.63$_{-0.24}^{+0.47}$ & ... & 0.30$_{-0.12}^{+3.37}$ & 0.21$_{-0.06}^{+0.09}$  & 0.22$_{-0.04}^{+0.06}$  & ... \\
       A$_S$ & 0.18$_{-0.05}^{+0.06}$ & 0.47$_{-0.16}^{+0.29}$ & ... & ... & 0.28$_{-0.09}^{+0.09}$ & 0.48$_{-0.13}^{+0.15}$ & ... & ... \\
       $f_c$ & ...  &... & 0.17$_{-0.04}^{+0.05}$ & ... & ...  & ... & 0.17$_{-0.04}^{+0.06}$ & ... \\
       $\theta\rm _{i}$ [$^\circ$]  & ... & ... & 87$^f$ & 90$^f$ & ...  &... & 87$^f$ & 90$^f$ \\
       TOR$\sigma$ [$^\circ$]  & ... & ... & ... & $<$5.7  & ...  &... & ... & 13.8$_{-5.3}^{+6.5}$ \\
       CTK  & ... & ... & ... & 0.25$_{-0.14}^{+0.03}$  & ...  &... & ... & $<$0.17 \\
       $f_s$ 10$^{-2}$ & 0.10$_{-0.05}^{+0.05}$ & 0.18$_{-0.06}^{+0.05}$ & 0.19$_{-0.03}^{+0.03}$ & $<$0.14 & $<$0.20 & 0.10$_{-0.08}^{+0.10}$ & 0.12$_{-0.03}^{+0.11}$ & 0.19$_{-0.11}^{+0.12}$ \\
       $kT$ [keV] & 0.68$_{-0.06}^{+0.07}$ & 0.66$_{-0.06}^{+0.06}$ & 0.68$_{-0.05}^{+0.06}$ & 0.71$_{-0.06}^{+0.08}$ & 0.64$_{-0.14}^{+0.07}$ & 0.63$_{-0.15}^{+0.10}$ & 0.65$_{-0.05}^{+0.06}$ & 0.65$_{-0.12}^{+0.11}$ \\
       $Z$/$Z_\odot$ & 0.07$_{-0.03}^{+0.08}$ & 0.10$_{-0.06}^{+0.22}$ & 0.08$_{-0.01}^{+0.01}$ & 0.05$_{-0.02}^{+0.05}$ & 0.03$_{-0.02}^{+0.05}$ & 0.04$_{-0.03}^{+0.08}$ & 0.04$_{-0.04}^{+0.01}$ & 0.04$_{-0.03}^{+0.07}$ \\
       log(F$_{2-10}$) [\flu] & --11.66$_{-0.06}^{+0.01}$ & --11.66$_{-0.09}^{+0.01}$ & --11.66$_{-0.08}^{+0.05}$ & --11.65$_{-0.03}^{+0.01}$ & --11.75$_{-0.05}^{+0.01}$ & --11.75$_{-0.04}^{+0.01}$ & --11.75$_{-0.10}^{+0.06}$ & --11.75$_{-0.03}^{+0.01}$ \\
       log(F$_{10-40}$) [\flu] & --10.60$_{-0.03}^{+0.02}$ & --10.60$_{-0.06}^{+0.04}$ & --10.60$_{-0.09}^{+0.07}$ & --10.60$_{-0.02}^{+0.01}$ & --10.61$_{-0.04}^{+0.01}$ & --10.61$_{-0.04}^{+0.01}$ & --10.61$_{-0.13}^{+0.09}$ & --10.61$_{-0.03}^{+0.01}$ \\
       log(L$_{2-10}$) [\lu] & 42.98$_{-0.06}^{+0.06}$ & 42.93$_{-0.03}^{+0.04}$ & 42.92$_{-0.07}^{+0.06}$ & 42.98$_{-0.07}^{+0.10}$ & 42.94$_{-0.08}^{+0.08}$ & 42.97$_{-0.09}^{+0.09}$ & 42.95$_{-0.13}^{+0.10}$ & 42.88$_{-0.09}^{+0.09}$ \\ 
       log(L$_{10-40}$) [\lu] & 43.21$_{-0.04}^{+0.04}$ & 43.19$_{-0.03}^{+0.03}$ & 43.18$_{-0.12}^{+0.09}$ & 43.20$_{-0.07}^{+0.09}$ & 43.22$_{-0.02}^{+0.02}$ & 43.23$_{-0.02}^{+0.02}$ & 43.23$_{-0.04}^{+0.03}$ & 43.18$_{-0.09}^{+0.08}$ \\
       \hline
	\hline
	\vspace{0.02cm}
\end{tabular}
}
\caption{\normalsize Summary of the best-fit results for the spectral fit of the observations taken on January 21, 2021 and on February 03--04, 2022: in the second data-set, the \xmm\ observation was performed on February 03 and the \nus\ one on February 4 (see the text for more details). \nhlos\ J is computed assuming no variability between the \nus\ and \xmm\ observations, while \nhlos\ X and \nhlos\ N are computed separately from the \xmm\ and \nus\ data, respectively. The other parameters are the same reported in Table~\ref{tab:fit_2021}. The inclination angle in \borus\ and \uxcl\ was frozen to the best-fit value because it was otherwise unconstrained if left free to vary. Upper and lower limits flagged with $^*$ have reached the parameter boundary.}
\label{tab:fit_2022}
\end{table*}

\begin{figure*}[ht]
\begin{minipage}{0.49\textwidth} 
 \centering 
 \includegraphics[width=0.74\textwidth,angle=-90]{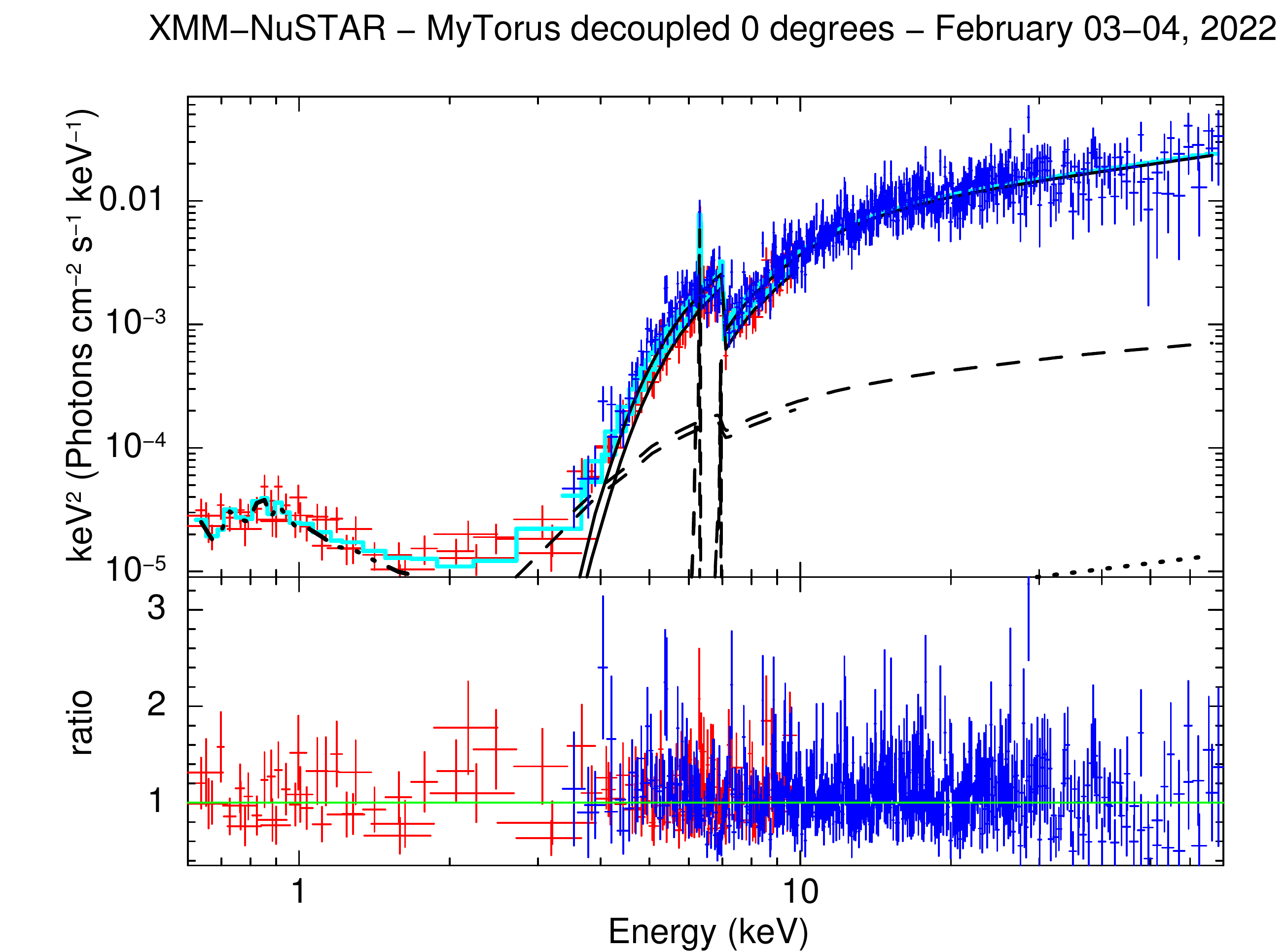} 
 \end{minipage} 
 \begin{minipage}{0.49\textwidth} 
 \centering 
 \includegraphics[width=0.73\textwidth,angle=-90]{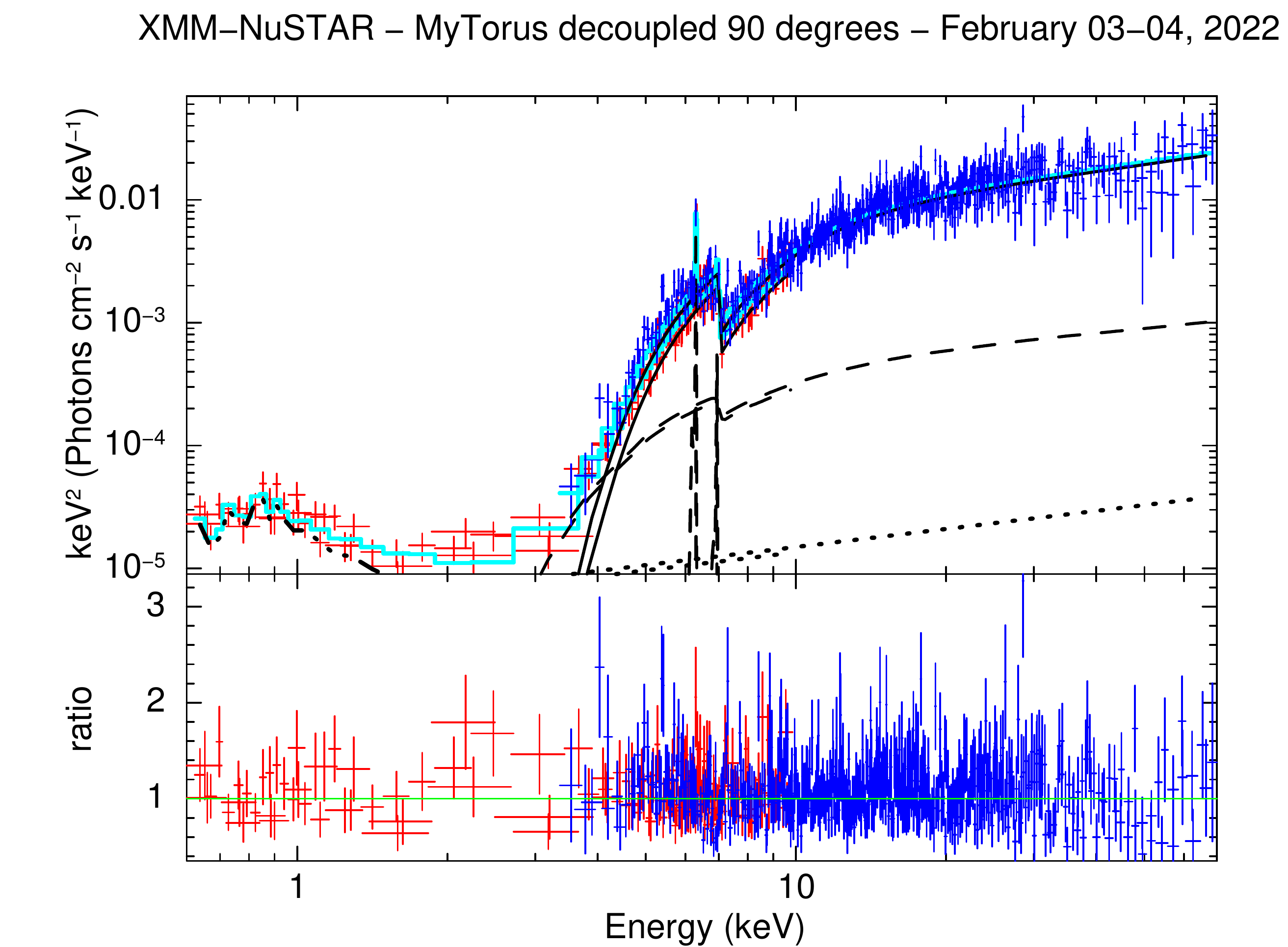} 
 \end{minipage} 
\begin{minipage}{.49\textwidth} 
 \centering 
 \includegraphics[width=0.75\textwidth,angle=-90]{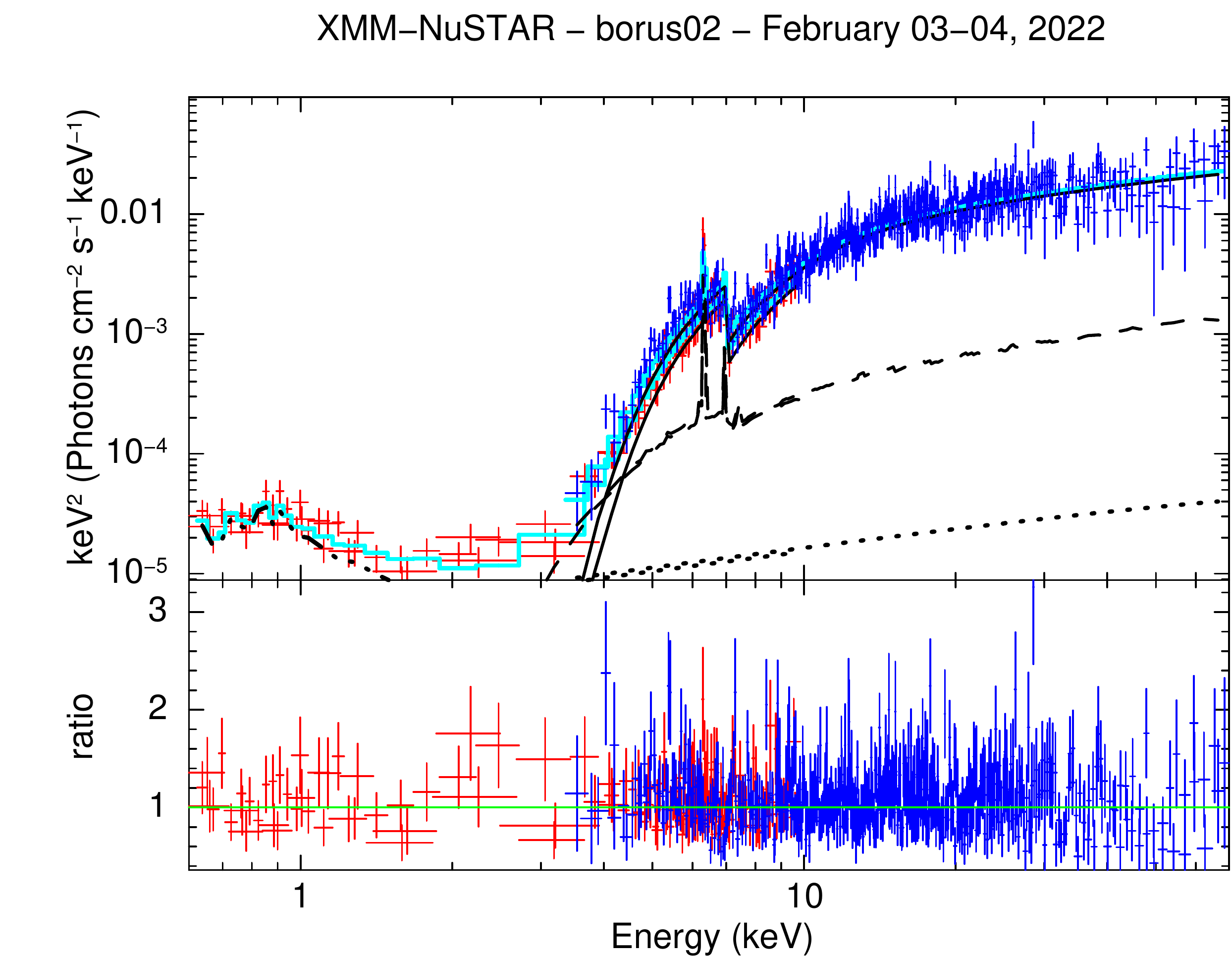} 
 \end{minipage} 
  \begin{minipage}{.49\textwidth} 
 \centering 
 \includegraphics[width=0.75\textwidth,angle=-90]{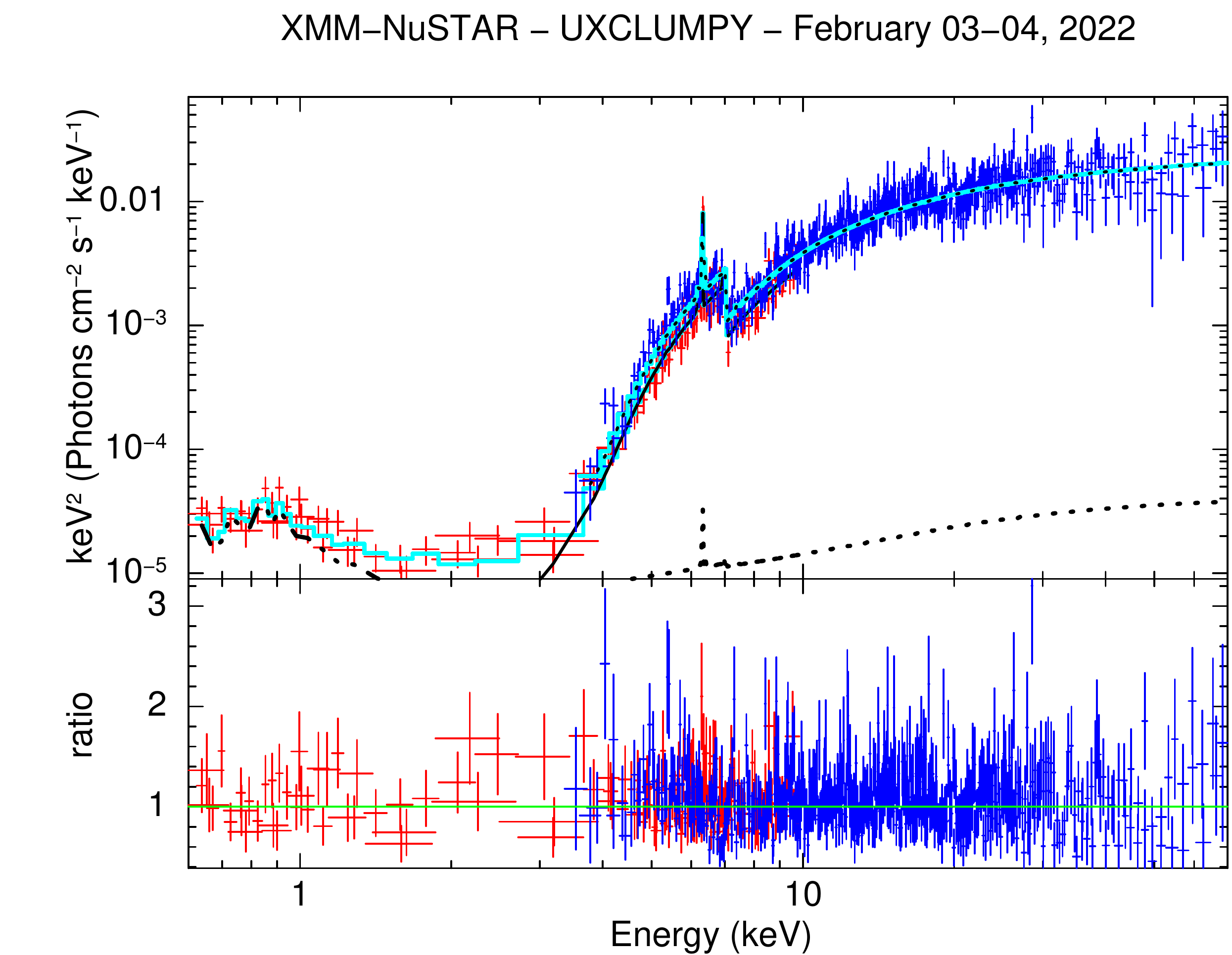} 
 \end{minipage} 
\caption{\normalsize 
Unfolded \xmm\ (red) and \nus\ (blue) spectra of the February 03--04, 2022 observation of NGC 1358. On the top panel we show the best fits obtained using \myt\ in its decoupled, $\theta$=0$\degree$ (left) and $\theta$=90$\degree$ configuration. In the bottom panel, we report the \borus\ (left) and \uxcl\ (right) best fit models. In all panels, the overall model is plotted as a solid cyan line, the absorbed main power law component is plotted as a solid black line, the reprocessed emission as a dashed black line, the scattered component as dashed black line, and the thermal \texttt{mekal} component as a dash-dotted black line.
}\label{fig:spec_20220204}
\end{figure*}

\end{appendix}

\end{document}